\documentclass[12pt,preprint]{aastex}
\shorttitle{Temperature Scale of O-type Stars}
\shortauthors{Massey et al.}
\begin{document}

\title{The Physical Properties and Effective Temperature Scale of
O-type Stars as a Function of Metallicity. II. Analysis of 20 More
Magellanic Cloud Stars, and Results from the Complete
Sample\altaffilmark{1,}\altaffilmark{2}}

\author{Philip Massey\altaffilmark{3}}

\affil{Lowell Observatory, 1400 W. Mars Hill Road, Flagstaff, AZ 86001;
Phil.Massey@lowell.edu}

\author{Joachim Puls and
A. W. A. Pauldrach}
\affil{Universit\"{a}ts-Sternwarte M\"{u}nchen,
Scheinerstrasse 1, 81679, Munich, Germany;
uh101aw@usm.uni-muenchen.de,uh10107@usm.uni-muenchen.de}

\author{Fabio Bresolin, 
Rolf P. Kudritzki, and Theodore Simon}
\affil{Institute for Astronomy, University of Hawaii, 2680 Woodlawn
Drive, Honolulu, HI 96822;
bresolin@ifa.hawaii.edu,kud@ifa.hawaii.edu,tsimon@ifa.hawaii.edu}

\altaffiltext{1}{Based on observations made with the NASA/ESA Hubble
Space Telescope, obtained at the Space Telescope Telescope Science
Institute, which is operated by the Association of Universities
for Research
in Astronomy (AURA), Inc., 
under NASA contract NAS 5-26555.  These observations are
associated with programs 6417, 7739, and  9412.}

\altaffiltext{2}{Based on observations made with the NASA-CNES-CSA {\it
Far Ultraviolet Spectroscopic Explorer,} operated for NASA by John
Hopkins University under NASA contract NAS5-32985.
These observations are
associated with program C002.}

\altaffiltext{3}{
Visiting astronomer, Cerro Tololo Inter-American Observatory (CTIO),
a division of the National Optical Astronomy Observatory, which is
operated by the AURA, Inc.,
under cooperative agreement with the National Science Foundation.}

\begin{abstract}

In order to determine the physical properties of the hottest and most
luminous stars, and understand how these properties change as a function
of metallicity, we have analyzed {\it HST}/UV and high S/N optical 
spectra of an additional
20 Magellanic Cloud stars, doubling the sample presented in the first
paper in this series. Our analysis uses NLTE line-blanketed models that
include spherical extension and the hydrodynamics of the stellar wind.
In addition, our dataset includes {\it FUSE}
observations of OVI and {\it HST} near-UV He~I and He~II lines to test
for consistency of our derived stellar properties for a few stars.
The results from the complete sample
are as follows:
(1) We present
an effective temperature scale for O stars
as a function of metallicity. We find that 
the SMC O3-7 dwarfs are 4000 K
hotter than Galactic stars of the same spectral type.  The difference is
in the sense expected due to the decreased significance of line-blanketing
and wind-blanketing at the lower metallicities that characterize the SMC.
The temperature difference between the SMC and Milky Way O dwarfs decreases
with decreasing temperature, becoming negligible by spectral type B0, 
in accord with the decreased effects of stellar winds at lower temperatures
and luminosities. The temperatures of the LMC stars appear to be intermediate
between that of the Milky Way and SMC, as expected based on their metallicities.
Supergiants show a similar effect, but are roughly 3000-4000 K cooler than
dwarfs for early O stars, also with a negligible difference by B0.  The giants
appear to have the same effective temperature scale as dwarfs, consistent
with there being little difference in the surface gravities. 
When we compare our scale to other recent modeling efforts, we find good
agreement with some CMFGEN results, while other CMFGEN studies are discordant,
although there are few individual stars in common.
WM-Basic modeling by others have resulted in significantly cooler effective
temperatures than what we find, as does the 
recent TLUSTY/CMFGEN study of
stars in the NGC~346 cluster,  but our results lead to a far more coeval
placement of stars in the H-R diagram for this cluster.
(2) We find that the wind momentum of these stars scale with
luminosity {\it and} metallicity in the ways predicted by radiatively-driven
wind theory, supporting the use of photospheric analyses of hot luminous stars
as a distance indicator for galaxies with resolved massive star populations.
(3) A comparison
of the spectroscopic masses with those derived from stellar evolutionary
theory shows relatively good agreement for stars with effective temperatures
below 45000 K; however, stars with higher temperatures all show a significant
mass discrepancy, with the spectroscopic masses a factor of 2 or more smaller
than the evolutionary masses.  This problem may in part be due to unrecognized
binaries in our sample, but the result suggests a possible systematic problem
with the surface gravities or stellar radii derived from our models.
(4) Our sample contains a large number of 
stars of the earliest O-types, including those of the newly proposed O2
subtype. We provide the first quantitative descriptions of their
defining spectral characteristics
and investigate whether the new types are a legitimate extension of the
effective temperature sequence.  We find that the NIII/NIV emission line
ratio used to define the new classes does not, by itself, serve
as an effective temperature indicator within a given luminosity class:
there are O3.5~V stars
which are as hot or hotter than O2~V stars.   However, the He~I/He~II ratio
does not fare much better for stars this hot, as we find that
He~I $\lambda 4471$/ He~II $\lambda 4542$, 
usually taken primarily as a temperature indicator, 
becomes sensitive to both the mass-loss rate and surface gravities for 
the hottest stars.
This emphasizes the need to rely upon {\it all} of the
spectroscopic diagnostic lines, and not simply N~III/N~IV or even He~I/He~II,
for these extreme objects.
(5) The two stars with the most discordant radial velocities in our sample
happen to be O3 ``field stars", i.e., found from the nearest OB associations.
This provides the first compelling observational evidence as to the origin
of the field O stars in the Magellanic Clouds; i.e., that these are 
classic runaway OB stars, ejected from their birth places.
\end{abstract}


\section{Introduction}
\label{Sec-intro}

An accurate knowledge of the stellar effective temperature scale provides
the means of converting the observed properties of a star
into its physical properties.  The issue is particularly
important---and challenging---for the massive O-type stars, as the
bolometric corrections (BCs) are a very steep function of the assumed
effective temperature ($T_{\rm eff}$) at the extremely high $T_{\rm eff}$'s
that characterize these objects.  For these hot stars,
a 10\% error in $T_{\rm eff}$ results in in an error of 30\% in the derived
bolometric
luminosity of a star (starting from the absolute visual magnitude; see
Massey 1998),
compromising any attempt to use stellar evolutionary
tracks to determine distances, initial mass functions (IMFs), and ages
of clusters (see, for example, 
Massey 1998, Slesnick et al.\ 2002).  In addition,
a 10\% uncertainty in $T_{\rm eff}$ results in a factor of 2 or more 
uncertainty in the Lyman continuum flux, affecting our understanding of
the ionization balance of H~II regions and the porosity of the interstellar
medium in general (see, for example, Oey \& Kennicutt 1997, Oey 2004).

At the same time,
the determination of the effective temperature scale of O-type stars is
complicated by the fact that the stellar atmospheres of these stars
are physically complex (see, for example, Kudritzki 1998). 
The strong spectral
features are formed under non-local thermodynamic equilibrium (NLTE)
conditions (Auer \& Mihalas 1972), while stellar winds provide
a significant source of heating of the photosphere 
through the backscattering of radiation 
(Hummer 1982; Abbott \& Hummer 1985).  Modern 
stellar atmosphere models for O-type stars  include NLTE, spherical
extension, the hydrodynamics of the the stellar winds, and the effects
of line blanketing; see, for example,
Pauldrach et al.\ (2001), Hillier et al.\ (2003), and Puls et al.\ (2003, 2005)
for descriptions of the stellar atmosphere programs
WM-BASIC, CMFGEN, and FASTWIND, respectively. 

The most recent improvement in these codes is an improved treatment of
line blanketing.  The results of fitting these improved models to 
stars have suggested that the previous effective temperature scale 
for Galactic O stars may be too hot by as much as
$\sim 20\%$ (see, for example, Martins et al.\ 2002, Bianchi \& Garcia 2002, 
Garcia \& Bianchi 2004, and Repolust et al.\ 2004), while the results
for the Magellanic Clouds are mixed 
(compare Crowther et al.\ 2002,  Bouret et al.\ 2003, Hillier et al.\ 2003, 
and Martins et al.\ 2004).

We are engaged in a new determination of the effective
temperature scale of O-type stars, with an emphasis on how the conversion
from spectral type to $T_{\rm eff}$ depends upon the initial composition
of the gas out of which the star was formed; i.e., how do the physical
properties of an O7.5~V star differ in the SMC, LMC, Milky Way, and the
Andromeda Galaxy, systems which span a range of 10 in metallicity.  In a 
previous paper (Massey et al.\ 2004, hereafter Paper~I) we analyzed
the spectra of a sample of 20 SMC and LMC O-type stars using FASTWIND.
The same code has recently been used to analyze a
sample of Galactic O-type stars (Repolust et al.\ 2004), obtained at
a similar signal-to-noise ratio (S/N) and
spectral resolution (Herrero et al.\ 1992). Our study suggested
that the Magellanic Cloud sample is 3,000-4,000~K hotter than
their Galactic counterparts for the early through mid-Os, although the sample
size precluded a more definitive statement.  

Here we extend
these studies to an additional sample of 20 Magellanic Cloud O-type stars,
and consider the results from the complete sample of 40 stars.  In 
Sec.~\ref{Sec-data} we describe the space- and ground-based data
of the new sample, along with our reduction procedures.  In 
Sec.~\ref{Sec-analysis} we provide the model fits to these spectra,
determining physical parameters, and describe the spectrum of each
star in turn (Sec.~\ref{Sec-stars}). 
Our data includes very high S/N
spectra of stars of the earliest types, including the newly proposed
``O2" type (Walborn et al.\ 2002), and in Sec.~\ref{Sec-O3s} we provide
the first quantitative classification criteria for such stars.  Our analysis
also allows us to investigate whether the physical properties of such
stars are well correlated with the new spectral subtypes (i.e., are O2~V
stars necessarily hotter than stars classified as O3~V or O3.5~V?).
Our analysis of the complete sample of 40 stars then allows us to compare
the physical properties of such stars as a function of metallicity
(Sec.~\ref{Sec-results}), including deriving effective temperature scales
for the SMC, LMC, and Milky Way (Sec.~\ref{Sec-Teff}) O stars, comparing the
Wind-Momentum Luminosity relationship (Kudritzki et al.\ 1995)
to that expected from
radiatively-driven wind theory (Sec.~\ref{Sec-WLR}), and investigating
whether or not the derived spectroscopic masses agree well with the
masses obtained from stellar evolutionary models (Sec.~\ref{Sec-Mass}).
We summarize our conclusions and lay out the need
for future work in Sec.~\ref{Sec-sum}.

We emphasize here that we realize that the model atmospheres are constantly
improving: that just as the models today are considerably more sophisticated
than the first NLTE models of Auer \& Mihalas (1970) on which the first
modern effective temperature scale of O stars were derived (Conti \& Alschuler
1971), so will the models
30 years hence will include better physics.  Furthermore, even the models
of today that purport to include similar physics do not necessarily yield
the same results. So, we accept that ours will not be the final word
on the absolute effective temperature scale.
However, by
using the same models and techniques on a large sample of objects
in the SMC and LMC that have been used on Galactic stars, we can make the
first cut at an effective temperature scale of hot stars that includes
the effects of metallicity.  We believe that the relative differences of
the scales we derive are accurate and significant.

\section{Observations and Reductions}
\label{Sec-data}

The stars are listed in
Table~\ref{tab:stars}.  As discussed in Paper I,
good photometry provides the key to an accurate value of the absolute
visual magnitude M$_V$, needed in order to constrain the stellar radius.
Unless otherwise noted in the table,
the photometry comes from the CCD {\it UBV} measurements by
Massey (2002) or Massey et al.\ (2000),  except for the R136 stars, which
comes from the WFPC2 photometry of Hunter et al.\ (1997).  Conversion
to M$_V$ for the non-R136 stars was done by using the color excesses
determined from the spectral types (see Paper~I), 
adopting distance moduli of 18.9 and
18.5 for the SMC and LMC respectively, following Westerlund (1997) and
van den Bergh (2000).  The M$_V$ values for the R136 stars were derived
in a similar manner by Massey \& Hunter (1998).

The spectroscopic observations are
summarized in Tables~\ref{tab:journal1} and \ref{tab:journal2}.  In
general, our modeling efforts require data in the UV in order to
determine the terminal velocity $v_\infty$ of the stellar wind.  These
data we have obtained with Space Telescope Imaging Spectrograph (STIS)
on {\it HST} using the FUV-MAMA detector.  Data in the blue-optical
region is used to constrain both the surface gravity $g$ (from the
wings of the Balmer absorption lines) and the effective temperatures
$T_{\rm eff}$ (from the He~I and He~II absorption lines).  Since the
detection of very weak (20-50m\AA) He~I $\lambda 4471$ is needed for
our modeling of the hottest stars in our sample, we chose to obtain
long-slit data at high S/N.  The long-slit observatioins allow the
subtraction of nebular emission in this line, and we aimed for for a
S/N that would allow accurate measurement and detection of this
important He~I line.  For all but the R136 stars, we obtained
1.4\AA\ resolution data with the CTIO 4-m telescope and RC
spectrograph, and a S/N of 400-500 was obtained with a great deal of
care to flat-fielding (see Paper~I).  The choice of spectral resolution
was a reasonable match to the expected line widths.  Spectra of the
R136 stars come from CCD observations with {\it HST}/STIS, obtained (in
general) at a S/N of 100 per 0.4\AA\ spectral resolution element,
providing a sensitivity about half as great\footnote{Others recently have made
different choices in obtaining optical data for their modeling of MC O
stars:  Bouret et al.\ (2003) and Heap et al.\ (2005) rely upon the
echelle spectra described by Walborn et al.\ (2000), obtained at higher
dispersion than ours, but with a short slit that precluded easy sky
(nebular) subtraction.  The VLT-Flames survey by Evans et al.\ (2005)
uses fibers on an 8.2-m telescope to sample many stars, with only
modest compromise for nebular subtraction. We consider these approaches
complementary to ours, each with its advantages and disadvantages.}.
Finally, data at H$\alpha$ is needed to determine the mass-loss rate
$\dot{M}$.  For this we relied both upon ground-based CTIO and/or STIS
observations.  Full details can be found in Paper~I.  We note that in
general our stars were isolated compared to the slit width 
(Table~\ref{tab:journal1}.  Although one cannot in general exclude
the possibility of an unresolved companion on the slit, these can
be revealed by the sort of detailed analysis we undertake here.

In addition, we also obtained data in two other wavelength regions for
a few of our stars to provide some independent check on the
modeling---how well do the data in these regions, not used for the model
fits, agree with our results?  One of
these regions is the 3020-3300\AA\ ``near-UV" (NUV) region, obtained
with {\it HST}/STIS-CCD in 
order to observe the He~I $\lambda 3187$ and He~II $\lambda
3203$ lines.  We found relatively good agreement in Paper~I for the two
stars that were observed in the region.  Of more critical interest is
the far-ultraviolet (FUV) region.  We obtained data in the wavelength
range 905\AA-1187\AA\ with the {\it Far Ultraviolet Spectroscopic
Explorer (FUSE)} satellite (Moos et al.\ 2000).  
This spectral region contains the high excitation
O~VI $\lambda \lambda 1032, 1038$ doublet, which provides a useful 
check on the
determination of $v_\infty$ (see Figs~ 5 and 6 of Taresch et
al.\ 1997).  The large aperture (30" x 30"; see Table ~\ref{tab:journal1}) 
necessitated the selection
of isolated targets.

The reduction procedure for both the {\it HST} and CTIO data are given
in Paper~I, and are not repeated here other than to note that the
CCD spectra (both STIS and CTIO) were reduced with
IRAF\footnote{IRAF is distributed by the National Optical Astronomy
Observatories, which are operated by the Association of Universities
for Research in Astronomy, Inc., under cooperative agreement with the
National Science Foundation.} using the optimal extraction algorithms.
For the {\it HST}/STIS CCD data, this led to a substantial improvement in
S/N over 
that delivered by the pipeline (see Fig.~1 of Paper 1).  For the STIS
UV MAMA observations we used the standard pipeline reductions, as we found
that reanalysis of these data
made no improvement to that of the pipeline.

The {\it FUSE} spacecraft and its instrumentation are described
by Moos et al.\ (2000) and Sahnow et al.\ (2000).  The 
{\it FUSE} spectra were extracted from the two-dimensional
raw spectral images using CalFUSE version 2.1.6, which 
uses an optimal extraction algorithm developed by 
S. Lacour (see http://fuse.pha.jhu.edu/analysis/calfuse.html).  
The  observations consisted of from 2 to 6 individual sub-exposures,
which were extracted, calibrated in flux, aligned in wavelength, 
and, if required, scaled in signal level to create a co-added 
spectrum of each star.  Three of our stars (Sk$-67^\circ$22, 
AV~476, and AV~177) were observed in July or August of 2002, 
following the discovery by the {\it FUSE} project 
of an error in the flux calibration procedure used by CalFUSE 
(see http://fuse.pha.jhu.edu/analysis/fcal\_bug/fcal\_bug.html).
We have confirmed that the effects of the reported error
are entirely negligible ($<2$\%) for the three spectra at the 
wavelengths of the O~VI lines, which are of primary
interest to us here.     

These {\it FUSE} data are obtained in 8 channels (LiF1a, LiF1b, LiF2a, LiF2b,
SiC1a, SiC1b, SiC2a, and SiC2b).  For four of the stars, the agreement
between the channels was good in the regions of overlap, and we
averaged the data to produce a higher S/N spectrum, binned at 0.1~\AA.
For one of the stars, AV~177, the agreement between the 8 channels of
{\it FUSE} was poor, with only two of the channels (LiF1a and LiF1b)
showing the expected flux levels and a normal spectrum.  The
observation was repeated at our request, but the new data suffered from
an even worse problem involving data drop out,
 and we obtained useful data only from Li1a and LiF2b.
We combined these four observations for our treatment of AV~177.

\section{Analysis}
\label{Sec-analysis}

Following the procedures of Paper~I, we first determined the star's 
terminal velocity $v_\infty$ from radiative transfer fits of the 
P~Cygni profile of the C~IV $\lambda 1550$ doublet, and (where possible)
Si~IV $\lambda 1400$ and N~V $\lambda 1240$, although the Si~IV line is usually
too weak, and the N~V line is often contaminated by strong interstellar
Ly$\alpha$ absorption.  The fitting is done based upon the SEI method 
(see Lamers et al.\ 1987, Haser 1995, and Haser et al.\ 1995).  We list the
resultant values in Table~\ref{tab:termvels}.  We estimate our uncertainty
in this procedure as 50-100 km s$^{-1}$, with larger
uncertainties (up to 200 km s$^{-1}$) indicated by ``:".  Two stars
had even less certain measurements, and we estimate the uncertainties
as 300-500 km s$^{-1}$.

In running the FASTWIND models, we adopted these values of $v_\infty$ and
began with the assumption that $\beta$ (a measure of the steepness of the
velocity law) has a value of 0.8 following Puls et al.\ (1996).
In addition, we began by adopting
a He/H number ratio of 0.1, and adjusted it if needed.  The absolute visual
magnitudes M$_V$ were drawn from Table~1, and constrained the stellar radius
of each model.  We assumed a metallicity $Z/Z_\odot$ of 0.2 for the SMC stars
and 0.5 LMC stars, as argued in Paper~I.  Following Repolust et al.\ (2004),
we adopted a micro-turbulence velocity of 10 km s$^{-1}$ for the models
with effective temperatures of 36,000~K and below, and 0 km s$^{-1}$ for the
hotter models.

For each star, we ran a series of grids, allowing then the values for 
$T_{\rm eff}$, $\log g$, and $\dot{M}$ to vary, until we had satisfactory
fits (as judged by eye) between the synthetic and observed spectra.
In a number of cases, a good fit to the strengths of the He~I and He~II lines
necessitated a slight increase in the He/H ratio, as described below.
The lines examined included the H$\delta$, H$\gamma$, H$\beta$, and H$\alpha$
Balmer lines, the He~I $\lambda 4387$ singlet line,
the He~I 
$\lambda 4471$ triplet line (and
occasionally the 
He~I $\lambda 4713$ triplet line), 
and the He~II $\lambda 4200, 4542$ lines.
We examined, but did not give much weight, to the fit of the wind-sensitive
He~II $\lambda 4686$ line in determining the fits, but illustrate the
agreement between the models and the observed spectra.

As discussed both by Repolust et al.\ (2004)
and in Paper~I, FASTWIND
does not produce a strong enough He~I $\lambda 4471$ line for giants
and supergiants of spectral type O6 and later.  Fortunately the He~I
$\lambda 4387$ singlet line  
can be used for these spectral types, with
simultaneous good fits to this line and the He~II absorption.
A detailed comparison of the output of FASTWIND and CMFGEN
by Puls et al.\ (2005) has demonstrated that the
same problem should be encountered with CMFGEN (since the agreement of the
He~I triplet lines, including He~I $\lambda 4471$, 
in both codes is remarkable),
consistent with our statement
in Paper~I that this is an example of a long-standing problem with
stellar atmosphere codes (see Voels et al.\ 1989).  This so-called
``generalized dilution effect" is yet to be understood\footnote{Puls et al.\ 
(2005) does reveal 
very significant differences between the strengths of the He~I
{\it singlet} lines 
predicted by CMFGEN and FASTWIND in the temperature regime
36000 to 41000 (for dwarfs) and 31000-35000 (for supergiants).
This disagreement between the behavior of He~I singlets may well
account for some of the systematics between our effective temperature
scale and those of some CMFGEN studies, as well as the strong differences
in results on similar stars found by different CMFGEN studies; see 
the discussion in Sec.~\ref{Sec-Teff}.}.

Below we discuss the individual stars in our sample.  In determining
the spectral types, we first consider the visual impression of the
spectra, following the premises of Walborn \& Fitzpatrick (1990).  We
then measured the equivalent widths of He~I $\lambda 4471$ and He~II
$\lambda 4542$, the principle classification lines, and determined
$\log W'=\log W(4471) - \log W(4542)$, and compared that to that used
to differentiate the spectral subtypes by Conti \& Alschuler (1971).
In practice, there was no disagreement between these two methods.  For
the earliest type stars, we also make reference to the new
classification criteria proposed by Walborn et al.\ (2002, 2004), based
upon the strengths of the N~IV $\lambda 4058$ and N~III $\lambda
\lambda 4534,42$ emission lines and the N~V $\lambda 4603,19$
absorption lines.  Walborn et al.\ (2002) used this to attempt to break
the degeneracy of the O3 spectral class and introduced types O2
(N~IV$>>$N~III), and redefined O3 as ``weak He~I with N~IV$>$N~III".
In Sec.~\ref{Sec-O3s} we attempt to provide a slightly more
quantitative description.  One needs to keep in mind, however, that
there is yet to be a demonstration that stars classified as ``O2" are
necessarily hotter than stars classified as ``O3", although this is
clearly the implication of the scheme; we also examine this in
Sec~\ref{Sec-O3s}.  Furthermore, as emphasized in Paper~I, there is no
{\it a priori} basis for assuming that the strengths of the
``Of" characteristics (He~II
$\lambda 4686$ and N~III $\lambda \lambda 4634, 42$ emission) will
scale with luminosity in the same way in the SMC than in the Milky Way for
stars of the same effective temperatures, 
given the factor of 5 difference in metallicity between these two
systems, and the corresponding differences in stellar winds.  N~III
$\lambda \lambda 4634,42$ emission is a complex NLTE effect and its
size mainly due to effective temperature (Mihalas \& Hummer 1973,
Taresch et al.\ 1997), while He~II $\lambda 4686$ will depend upon the
stellar wind properties, such as density and temperature.  Accordingly,
we will note any discrepancies between these traditional luminosity
indicators and the actual M$_V$ of the stars.

The model fits give us a measure of the {\it effective} surface gravity
$g_{\rm eff}$ which will be the combination of the  
{\it true} surface gravity $g_{\rm true}$ offset by 
centrifugal acceleration due to the rotation of the star.  This correction is
estimated (in a statistical sense) as simply the square of the
projected rotational velocity divided by the stellar radius:
\begin{equation}
g_{\rm true} = g_{\rm eff} + \frac{(v \sin i)^2}{6.96 R}
\end{equation}
(Repolust et al.\ 2004), where the numerical factor allows the use
of the usual units; i.e., km s$^{-1}$ for $v \sin i$ and solar units
for $R$.   We have a good estimate of $v \sin i$ from comparing the
model lines with the observed spectra, except that our spectral resolution
results in a minimum ``rotational" velocity of (generally) 110 km s$^{-1}$.
However, for these stars with minimal rotational velocities the centrifugal
correction is tiny (0.01 in $\log g$), and insignificant compared to the
typical fitting uncertainty of 0.1 in $\log g$ 
(Paper~I and Sec.~\ref{Sec-results}).  However, for the fast-rotators in
our sample the correction can be marginally significant, 
amounting to 0.05~dex in
several cases, and  0.13~dex in one case (AV~296, from Paper~I).  
The spectroscopic masses should be determined from the true surface
gravities; i.e., $M_{\rm spect} = (g_{\rm true}/g_\odot) R^2$.

The final values for the fits are given in Table~\ref{tab:results}.
If we were only able to determine a lower limit to T$_{\rm eff}$, then the
values for $\log g$, $R$, $\dot{M}$, and the derived $M_{\rm spect}$
are uncertain.  Note that in Paper~I the stars
with only a lower limit on $T_{\rm eff}$ had their stellar radii
incorrectly listed as lower limits.

\subsection{Comments on Individual Stars}
\label{Sec-stars}

\subsubsection{SMC}

\begin{figure}
\epsscale{0.6}
\plotone{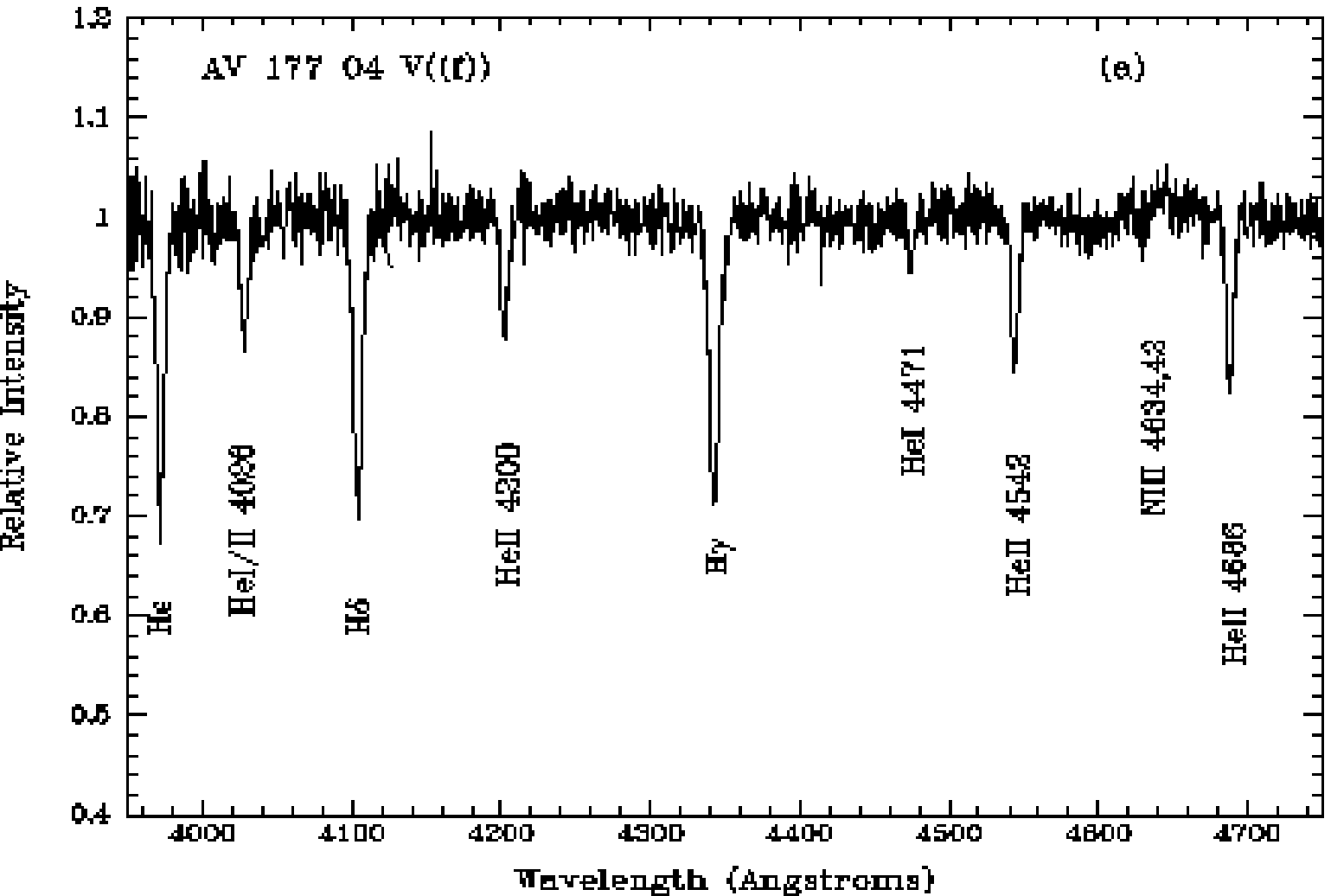}
\plotone{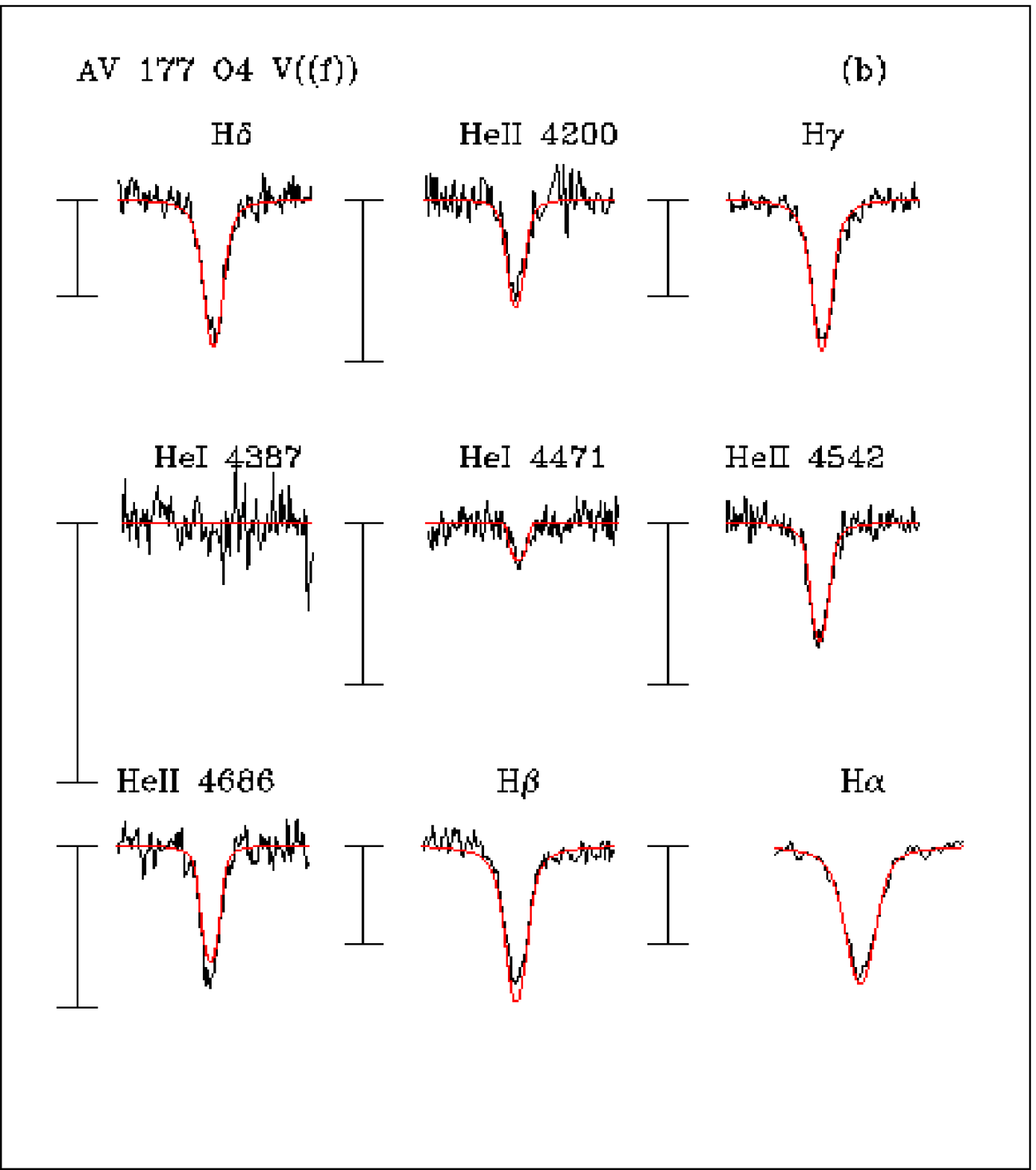}

\caption{\label{fig:av177} AV 177.  (a) A portion of the
blue-optical spectrum of AV~177 is shown with the major lines identified.
(b) Selected spectral lines (black) are shown compared to the model fits 
(red in the on-line version; thin black line in the printed version).
The bar to the left of each line shows a change of 20\% intensity relative
to the continuum, and the top of the bar denotes the 
continuum level. A radial velocity
of 100  km s$^{-1}$ and a rotational
broadening $v \sin i$ of 220 km s$^{-1}$ was used in making this comparison.
}
\end{figure}

{\it AV 177.}---This star was previously classified as O5~V by
Crampton \& Greasley (1982).  Visually, we would classify the star
as an O4~V((f)) (Fig.~\ref{fig:av177}a).  This is confirmed by our
measurement $\log W'=-0.75$.  The strength of He~I $\lambda 4471$
(EW=170m\AA) would preclude it being classified as an O3.  The luminosity
class ``V" based upon the strong He~II $\lambda 4686$ absorption is
consistent with the star's absolute visual magnitude,
M$_V=-4.8$.   There is a trace of emission
at N~III $\lambda \lambda 4634, 42$ (EW$\sim-250$m\AA)
leading to the ``((f))" designation; there is no sign of 
N~IV $\lambda 4058$ emission.
We obtained satisfactory fits
with a slightly increased He/H number ratio (0.15), as shown in 
Fig.~\ref{fig:av177}b.  The mass-loss rate
is low ($< 5 \times 10^{-7} M_\odot$ yr$^{-1}$), and only an upper
limit can be established. (A value of $3 \times 10^{-7} M_\odot$ yr$^{-1}$
is used in the fitting.)
The H$\alpha$ profile obtained from the ground-based data agrees well
with the {\it HST} data; we use the former, since its S/N is greater.

\begin{figure}
\epsscale{0.6}
\plotone{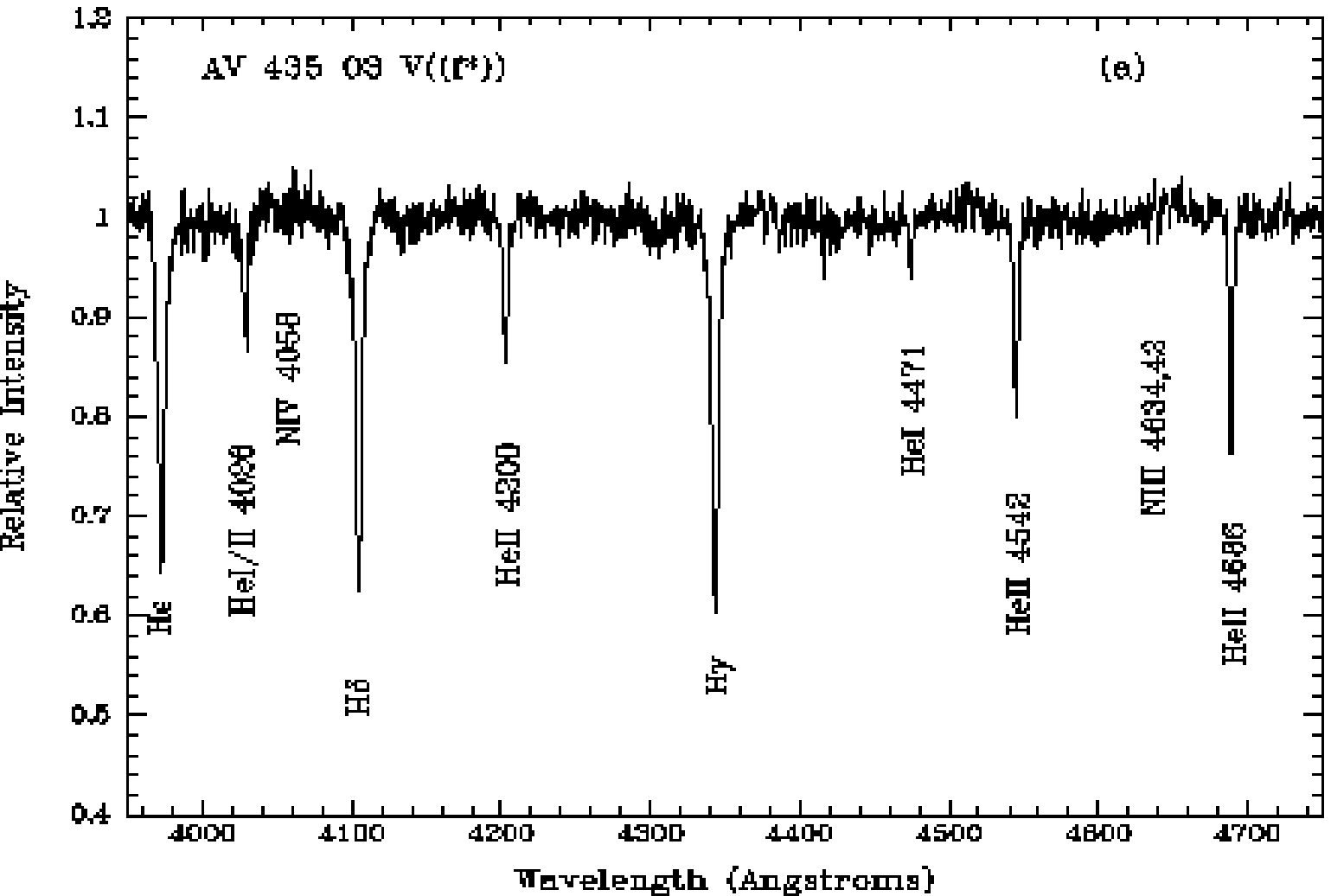}
\plotone{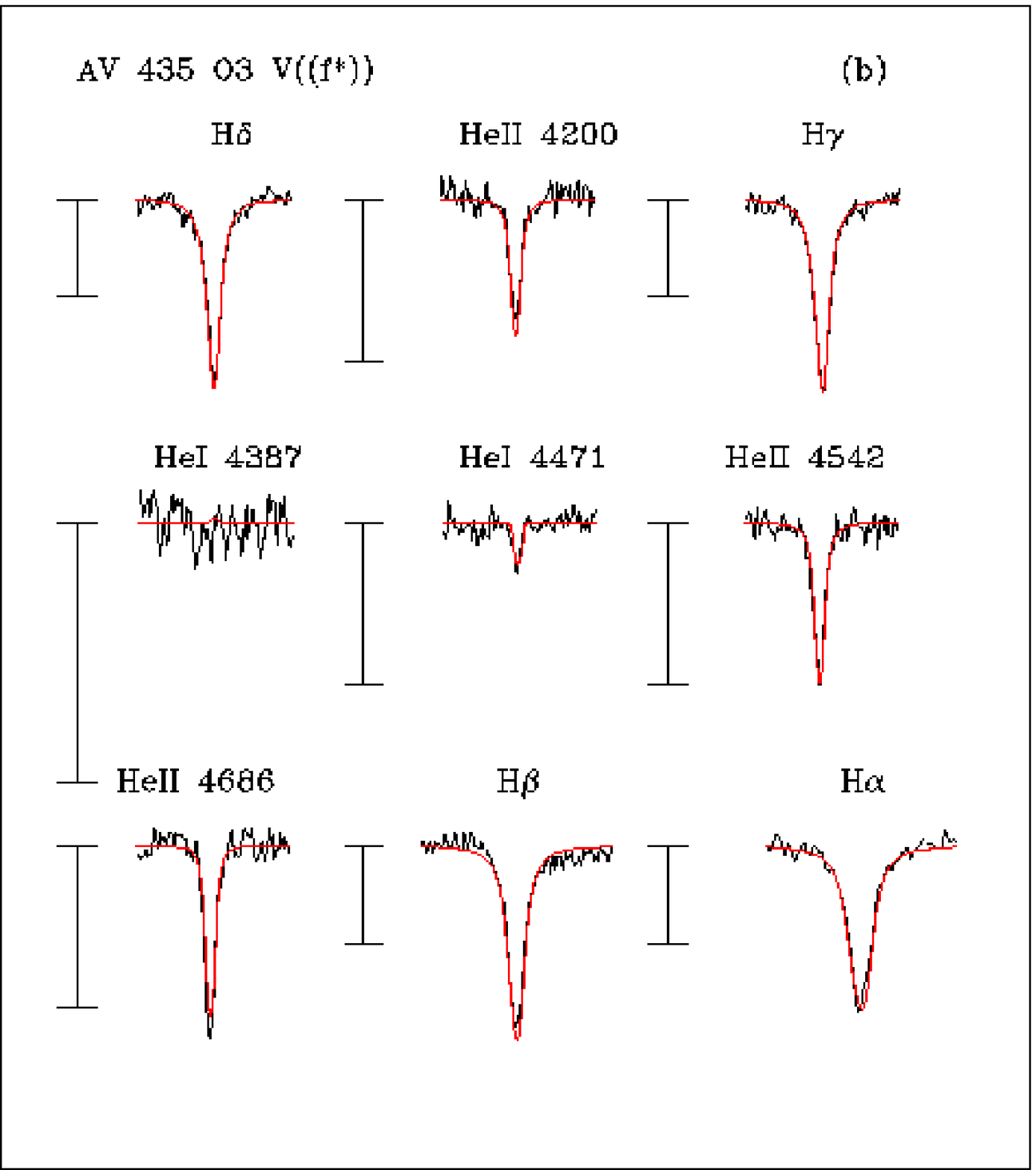}

\caption{\label{fig:av435} AV 435.  Same as Fig.~\ref{fig:av177}.  A 
radial velocity of 200 km s$^{-1}$ and a rotational
broadening $v \sin i$ of 110 km s$^{-1}$ was used in making this comparison.
}
\end{figure}

{\it AV 435.}---This star had been previously classified as O4~V by
Massey et al.\ (1995).  Here we would call it an O3~V.  Although He~I
$\lambda 4471$ is weakly present (Fig.~\ref{fig:av435}a), its equivalent
width is only 125m\AA.  This is comparable to or smaller
the strength seen in the
original O3~V prototypes HD93128 and HDE~303308 (Simon et al.\ 1983),
as well as the other star used by Walborn (1971a) to
define the O3 class, HD~93250 (Kudritzki 1980).  Walborn et al.\ (2002)
has proposed revising the early O-type classes based upon the strength
of selective emission lines, e.g., N~IV $\lambda 4058$ and N~III
$\lambda 4334,42$.  Since N~IV $\lambda 4058$ is (weakly) present (EW=-80m\AA), with
N~III emission weaker (EW$>-60$m\AA), this would also lead to an O3~V((f*))
classification according to these revised criteria.  The ((f*)) simply
refers to the fact that N~IV $\lambda 4058$ is in emission and equal or
stronger
than any N~III emission.  The absolute visual magnitude M$_V=-5.8$
would be consistent with either the ``V" or ``III" luminosity class;
see Conti (1988). The fits we obtained (Fig.~\ref{fig:av435}b) are a
good match to the observed spectrum. Again we found the agreement between
the ground-based H$\alpha$ profile and that obtained with {\it HST} is
good, and
show the former in the figure.  

\begin{figure}
\epsscale{0.6}
\plotone{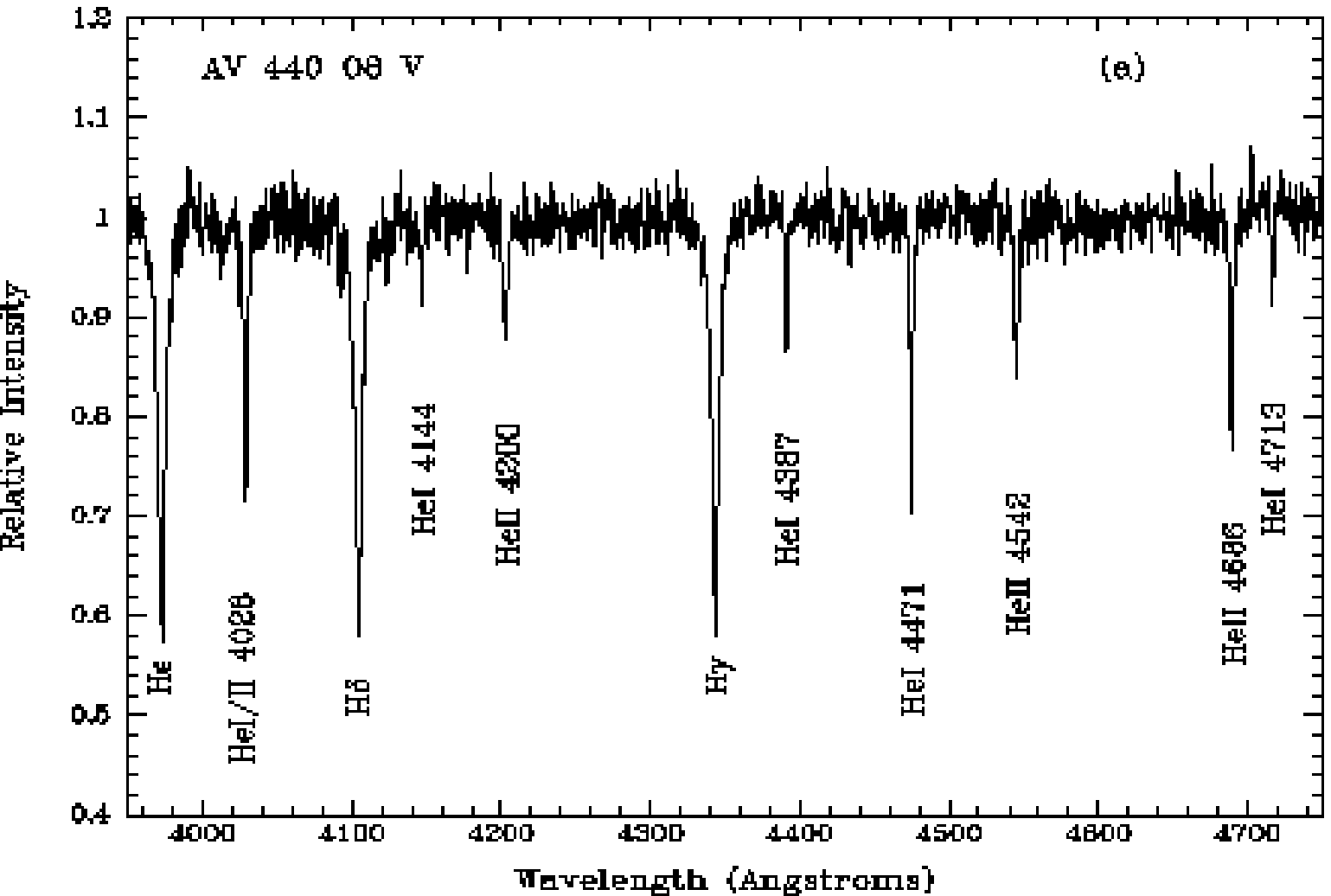}
\plotone{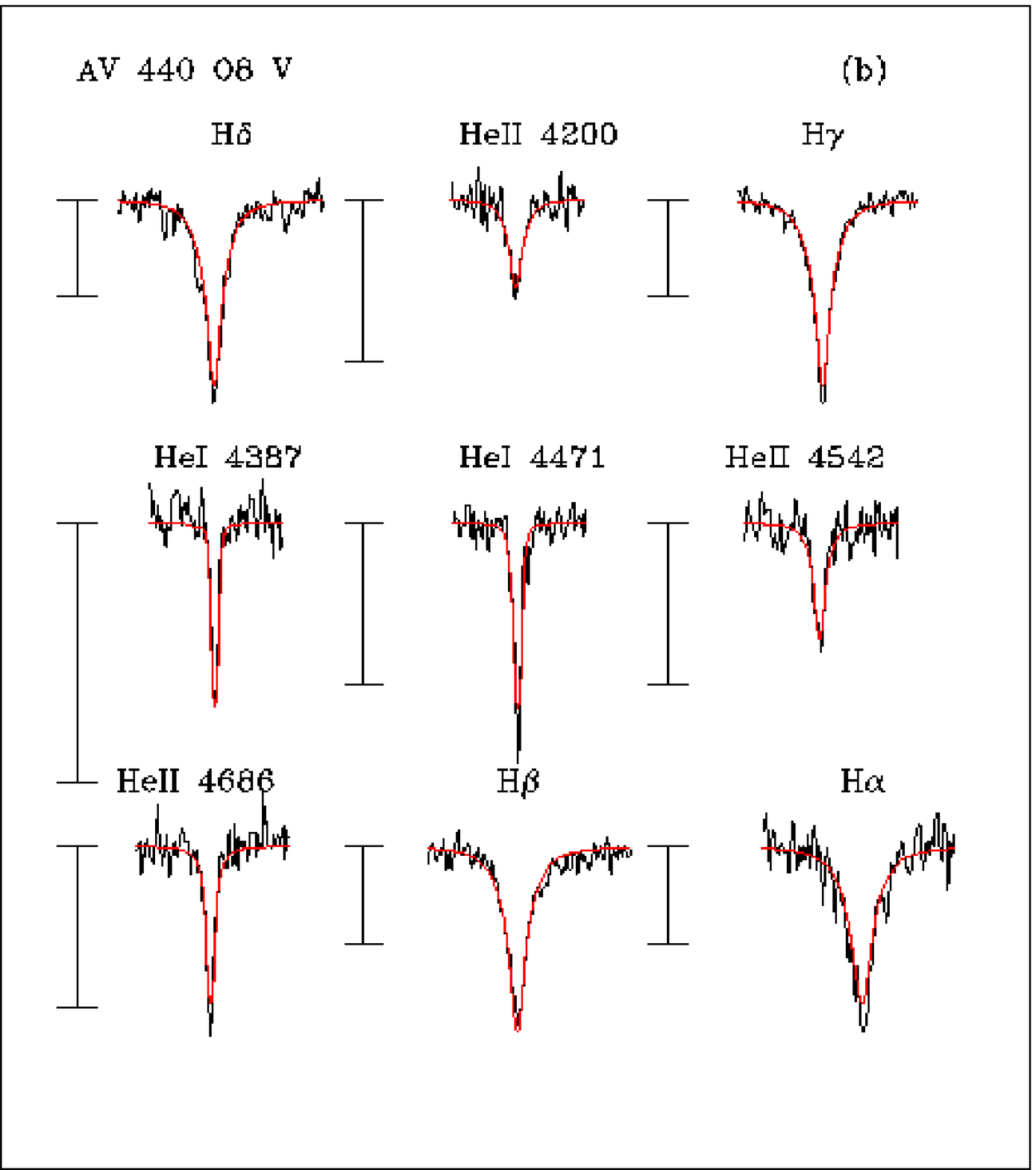}

\caption{\label{fig:av440} AV 440.  Same as Fig.~\ref{fig:av177}.  A 
radial velocity of 190 km s$^{-1}$ and a rotational
broadening $v \sin i$ of 100 km s$^{-1}$ was used in making this comparison.
}
\end{figure}

{\it AV 440.}--- Previously
the star had been classified as O7~V by Garmany et al.\ (1987), but it
is clear in our spectrum (Fig.~\ref{fig:av440}a) 
that He~I $\lambda 4471$ is somewhat stronger
than He~II $\lambda 4542$, and visually we would call this
an O8~V star.  We measure $\log W'=+0.12$, in accord with this
determination.  The absolute magnitude M$_V=-4.9$ is consistent
with the ``V" luminosity class suggested by the star's He~II $\lambda 4686$
absorption. 

In order to find a good fit, we needed to again slightly increase the 
He/H number ratio from the canonical 0.10 to 0.12.  This allowed us to then
obtain excellent matches to the He~II lines (Fig.~\ref{fig:av440}b). 
Although the central core of the the model's He~I $\lambda 4471$ line is
weaker than observed, the EWs of this line in the adopted fit and 
from the star are nearly identical (0.70\AA\ vs. 0.78\AA, respectively), and
we find excellent agreement for the He~I $\lambda 4387$ line as
well as the He~I $\lambda 4713$ line (not shown).
No ground-based H$\alpha$ was obtained, and so we rely upon the {\it HST}
observation for the mass-loss determination. Although we used a value
of $1\times10^{-7} M_\odot$ yr$^{-1}$, in fact we could only set an
upper limit on the mass-loss rate ($\dot M < 3\times10^{-7} M_\odot$ yr$^{-1}$).

\begin{figure}
\epsscale{0.6}
\plotone{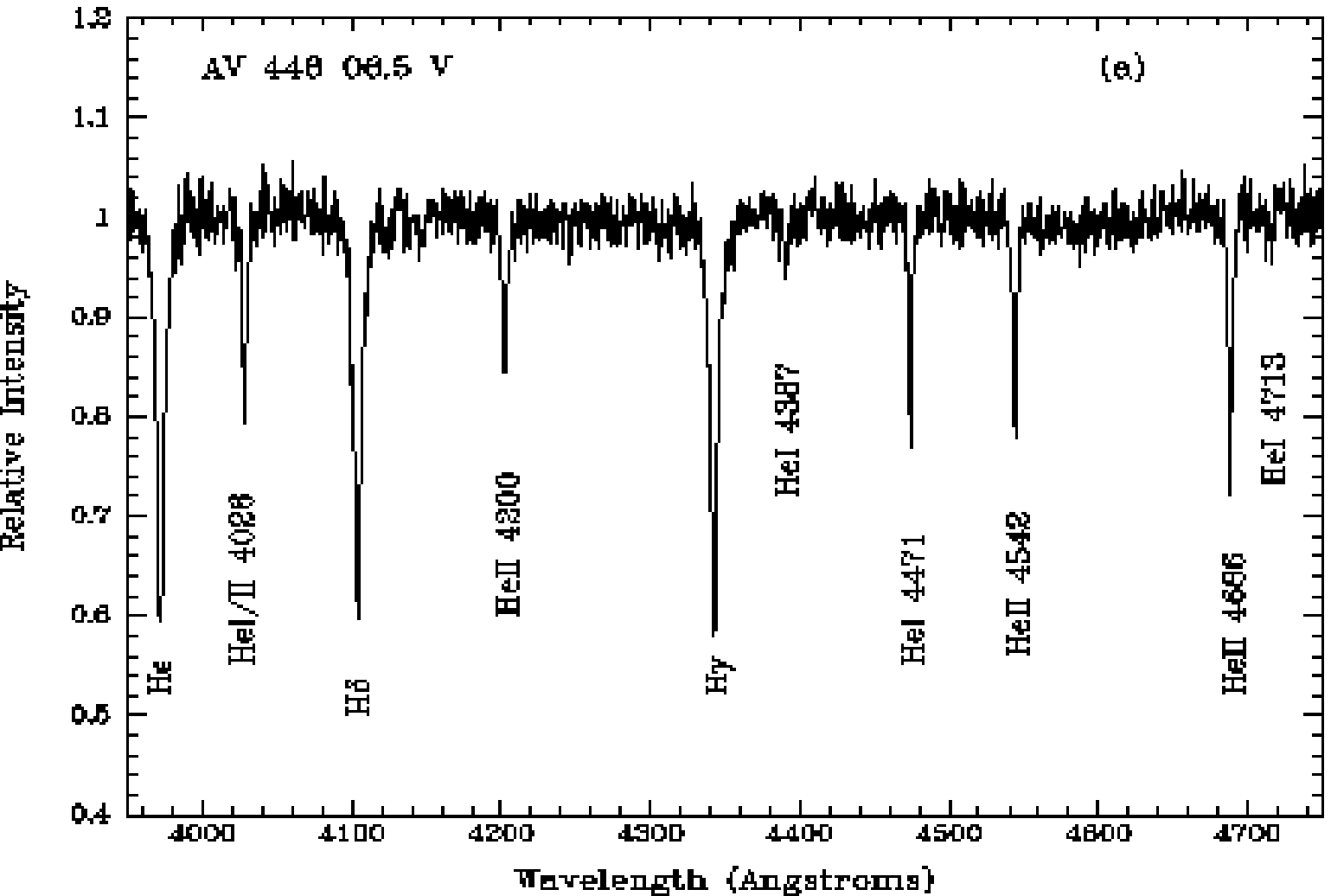}
\plotone{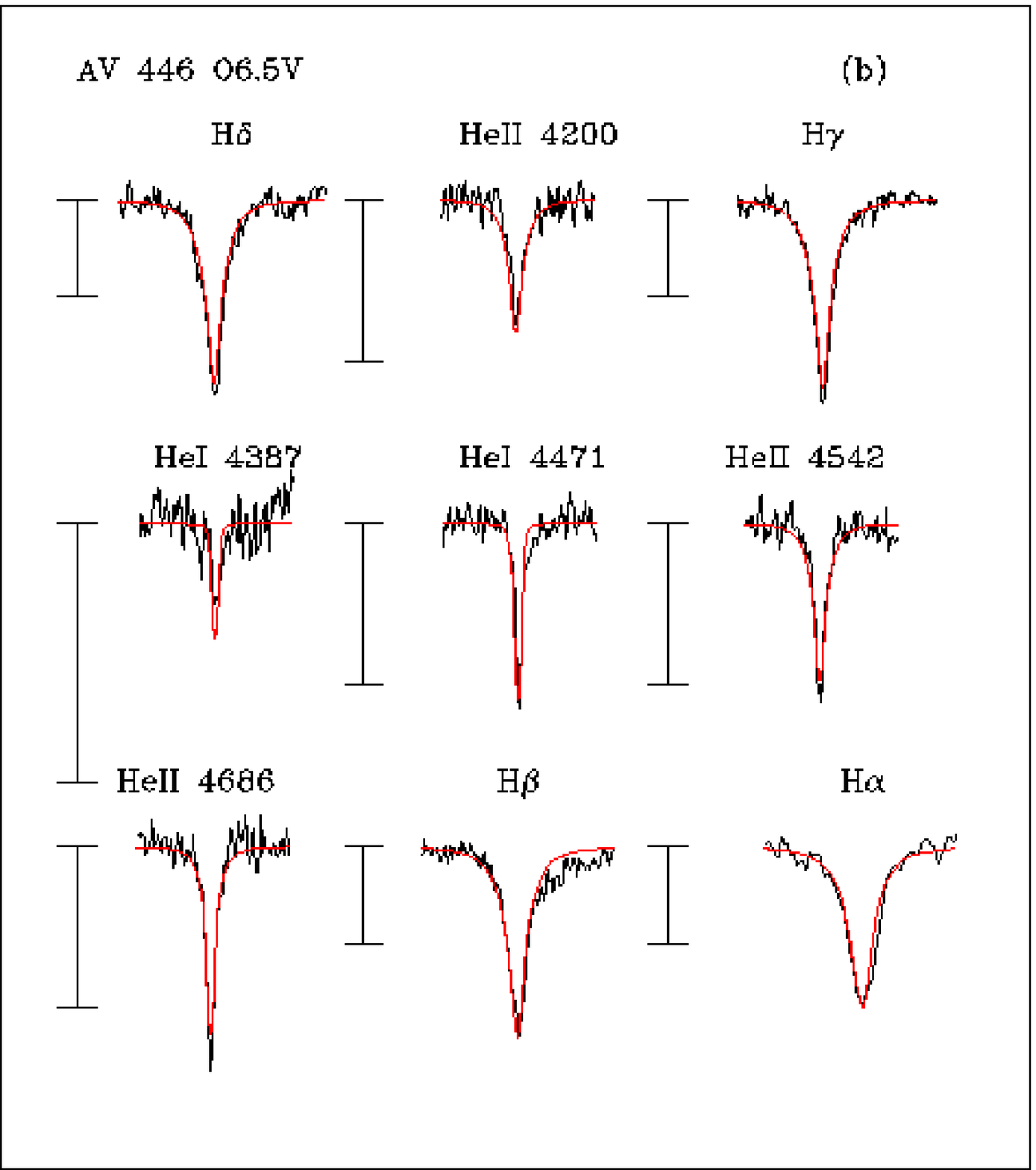}

\caption{\label{fig:av446} AV 446.  Same as Fig.~\ref{fig:av177}.  A
radial velocity of 140 km s$^{-1}$ and a rotational
broadening $v \sin i$ of 95 km s$^{-1}$ was used in making this comparison.
}
\end{figure}

{\it AV 446.}---This star was classified as O6.5~V by Garmany et al.\ (1987).
We measure $\log W'=-0.13$, consistent with this classification.  The only
spectral anomaly is that the line {\it depth} of He~I $\lambda 4471$ is
somewhat greater than that of He~II $\lambda 4542$ in our spectrum
(Fig.~\ref{fig:av446}a), even though the He~I line has a smaller equivalent
width.  This implies that the line widths differ.  This is usually
an indication of a binary (see, for example, Walborn 1973a and the follow-up
study by Massey \& Conti 1977). However, the absolute magnitude M$_V=-4.7$ is 
indicative of a single star with luminosity class ``V", consistent
with the spectral appearance.   The good agreement of the model and observed
spectra (Fig.~\ref{fig:av446}b) also supports a single-star interpretation.
A comparison of the ground-based and {\it HST} 
H$\alpha$ profile shows reasonable agreement, and so we used the higher S/N
ground-based data for the fit. Once again we could set only an upper limit
on the mass-loss rate. 

\begin{figure}
\epsscale{0.6}
\plotone{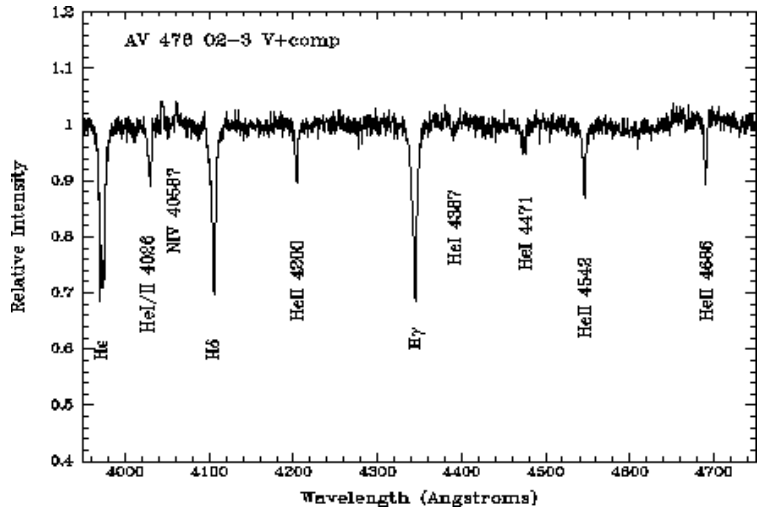}

\caption{\label{fig:av476} AV 476.  This star shows a composite spectrum.
}
\end{figure}

{\it AV 476.}---This star was classified as an O6.5~V by Massey et al.\ (1995).
Our higher S/N spectrum, shown in Fig.~\ref{fig:av476}, would suggest
a somewhat earlier type.  We measure $\log W=-0.34$, indicative of a
spectral type of O5.5.   He~II $\lambda 4686$ is comparable in strength
to He~II $\lambda 4542$, and there is no sign of N~III $\lambda \lambda 
4634,42$ emission.  Thus if this were a Galactic star, we would assign
a luminosity class of ``V".  The high absolute visual luminosity of this
star M$_V=-6.3$ is more in keeping with the star being a giant.
We argued in Paper I that these ``f" characteristics should not (and often
do not)
scale with luminosity the same way in the metal-poor SMC as they do in
the Milky Way, as the He~II $\lambda 4686$ emission that is used as luminosity
indicator will be weaker (at the same luminosity) in the SMC as the stellar
winds are weaker.  An alternative explanation could be that this star is a
binary, and the spectrum composite.  Our fitting efforts are consistent
with the latter interpretation in this case, as the hydrogen and He~II lines
have a very high radial velocity (270 km s$^{-1}$) which is not shared
by the He~I velocities.  There is a hint of N~IV $\lambda 4058$ 
emission in the spectrum (Fig.~\ref{fig:av476}), and we suggest that this
star is an O2-3~V plus a somewhat later O-type.  We note that
there is significant nebular contamination in the ground-based H$\alpha$
profile.

\subsubsection{The non-R136 LMC Stars}
\begin{figure}
\epsscale{0.6}
\plotone{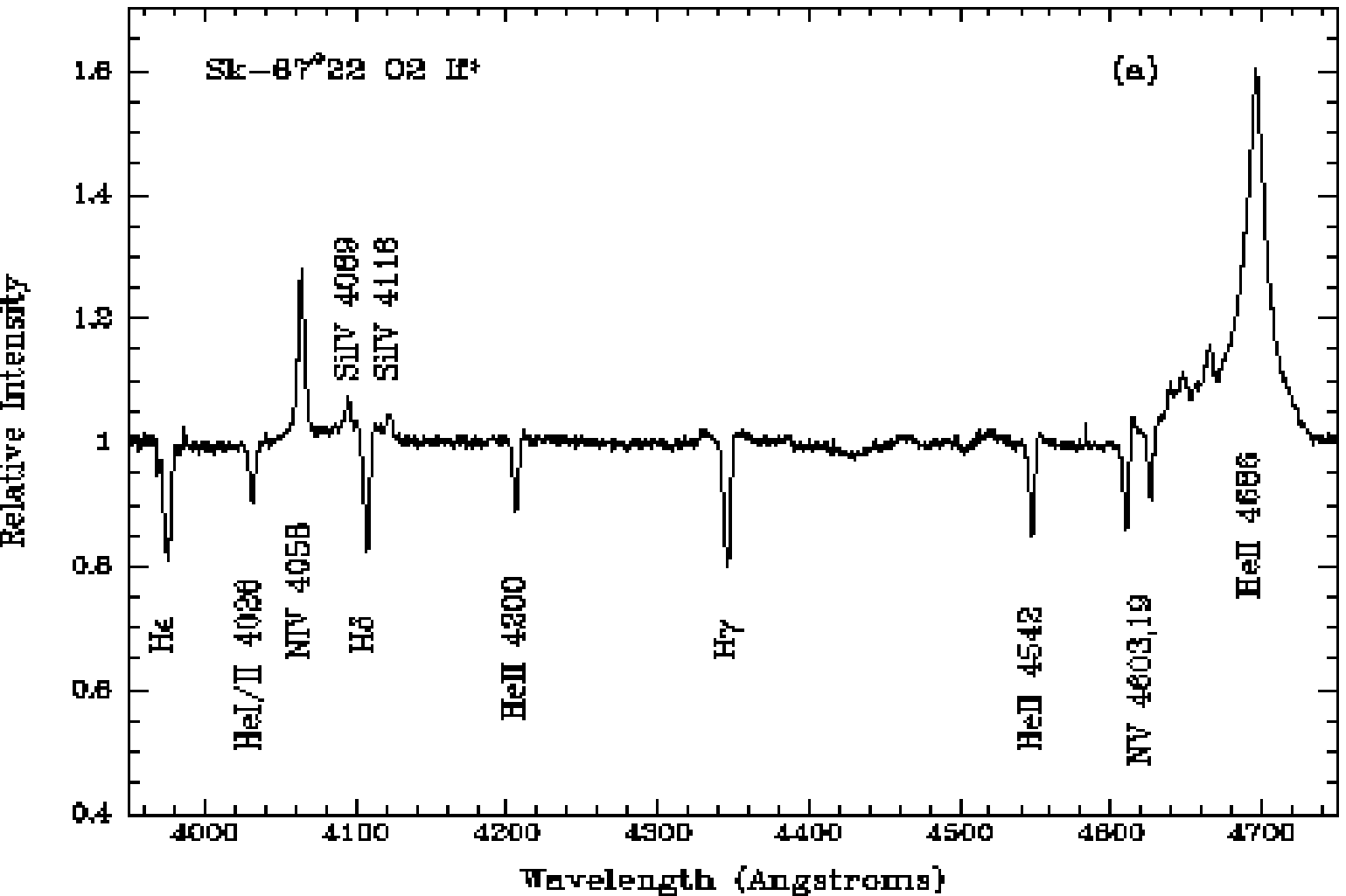}
\plotone{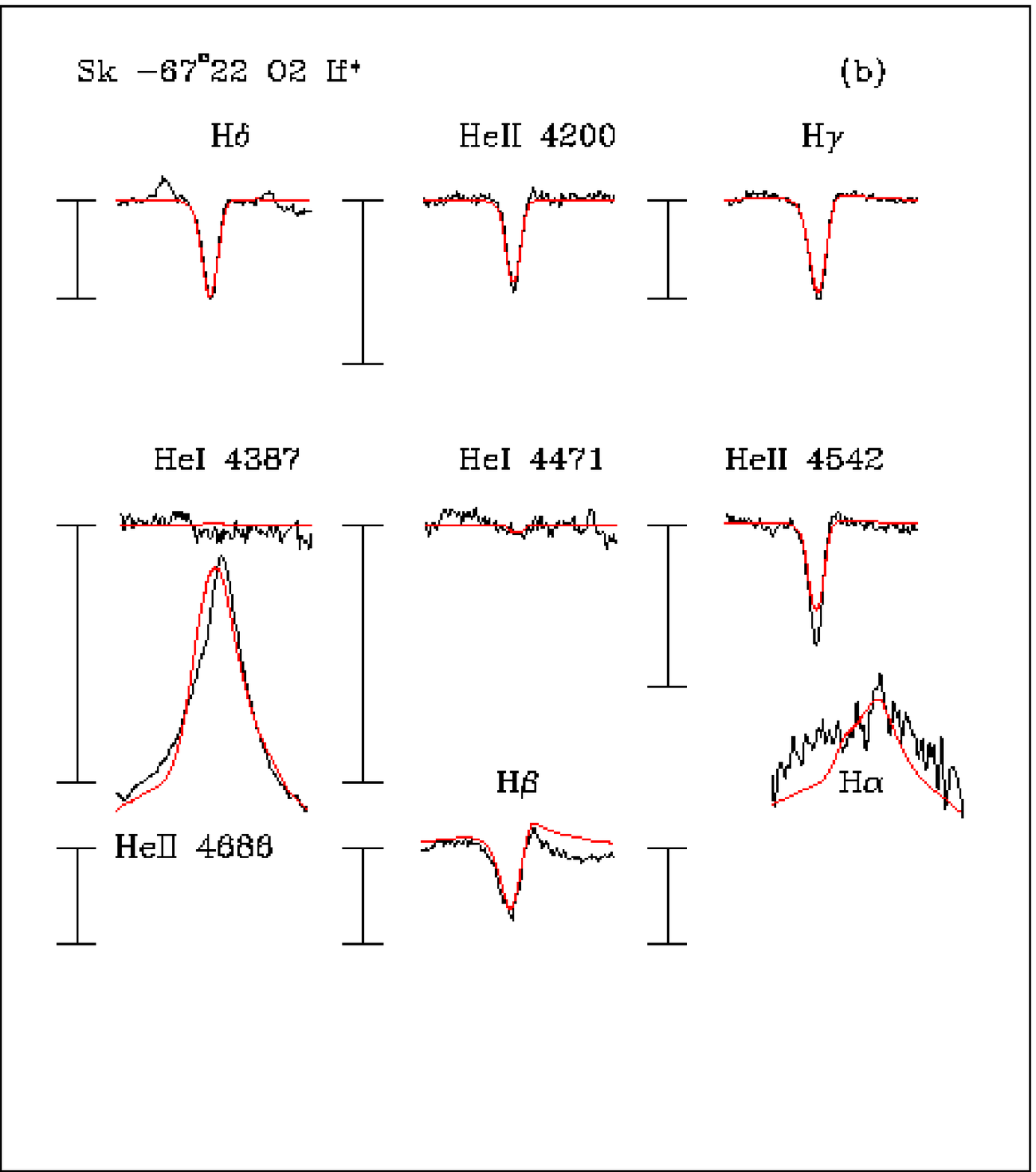}

\caption{\label{fig:sk-6722} Sk$-67^\circ22$.  Same as Fig.~\ref{fig:av177}.  A
radial velocity of 430 km s$^{-1}$ and a rotational
broadening $v \sin i$ of 200 km s$^{-1}$ was used in making this comparison.
The {\it HST} H$\alpha$ observation is shown.}
\end{figure}

{\it Sk$-67^\circ 22$.}---Classically, we would classify this star as
O3If*/WN, where the second part of the nomenclature simply denotes
that the emission features are comparable to what might be found in a
Wolf-Rayet star (see Fig.~\ref{fig:sk-6722}a). 
Indeed, this star was one of the first to be described
with this ``intermediate" designation (Walborn 1982).  Nevertheless, its
spectrum is quite unlike that of a typical Wolf-Rayet star, in that its
absorption-line spectra lacks P-Cygni components, and it is much more like
an ordinary O-type star than more extreme ``intermediate" stars, such as
Br~58 discussed below.  Nevertheless, the ``slash" designation resulted
in its being included in the recent compilation of Wolf-Rayet stars in 
the LMC (Breysacher et al.\ 1999).
Were it to be classified according to the Walborn
et al.\ (2002) criteria, it would have to be considered to be an O2If*,
as N~IV $\lambda 4058$ emission is much stronger than the slight NIII $\lambda
\lambda 4634,42$ emission visible on the blue wing
of He~II $\lambda 4686$, and there is no hint of He~I $\lambda 4471$
in our spectrum.  However, Walborn et al.\ (2002) explicitly excluded 
O3If*/WN objects from consideration when discussing the O2 spectral class.
Following in these footsteps, should we call this star an O2If*/WN? It
seems to us that {\it all} of the so-called ``O3If*/WN" stars might be
similarly reclassified as O2 If*.  Both Walborn (1982) and Conti
\& Bohannan (1989) have stressed the difficulty of classing such
``intermediate" objects, and the introduction of the O2~I classification
has added yet another wrinkle.  Furthermore, the prototype (and until
Paper I, the only member) of the O2~If* class is the star HD~93129A, which
has relatively weak He~II $\lambda 4686$.  However, 
Nelan et al.\ (2004) has now
split HD~93129A into two components ($\Delta m=0.5$ mag)
separated by 60mas, which might account for the weaker emission.
In Sk$-67^\circ 22$ the NIV $\lambda 4058$ emission is almost identical
in strength to that of R136-020 (discussed in Paper I), while 
the He~II $\lambda$
4686 line is nearly twice as strong.   Several additional examples will
be discussed below, and a comparison of their spectra made in
Sec.~\ref{Sec-O3s}.  C~IV $\lambda 4658$ is present (EW=-300m\AA) in emission.
 
Despite the strong emission, our fitting was straightforward, and good
fits were obtained (Fig.~\ref{fig:sk-6722}b).  We had
to increase the He/H ratio to 0.3 to obtain He lines of the right
strength relative to H.  Only a lower limit could be set on $T_{\rm
eff}$ since no He~I could be detected.  However, if the temperature
were 45,000~K rather than 42,000~K we would have to
significantly increase the He/H ratio even
further to maintain a good fit. We note that
this star has a large projected rotational velocity ($v \sin i=200$ km
s$^{-1}$) compared to most of the other stars in our sample, and we
offer the speculation that the chemical enrichment implied by He/H$\sim 0.3$ 
and the strong emission are both
coupled to this.  As Maeder \& Meynet (2000) demonstrate, high
rotational velocities can result in efficient mixing of processed
material from the core of a star, and it can also lead to enhanced
mass-loss, particularly from the poles (see Owocki et al.\ 1998).

The radial velocity of this star is quite large, 430 km s$^{-1}$, and is
about 150 km$^{-1}$ greater than the systemic velocity of the LMC
(Kim et al.\ 1998). A similar
radial velocity was seen in both of our blue CTIO spectrum and
{\it HST} H$\alpha$ spectrum; otherwise we would be inclined
to believe this might be a binary.  This star was cited by Massey (1998)
as an example of an isolated (field) O3 star.  Its high radial velocity
suggests that it is a runaway star, ejected from its birth place.

\begin{figure}
\epsscale{0.6}
\plotone{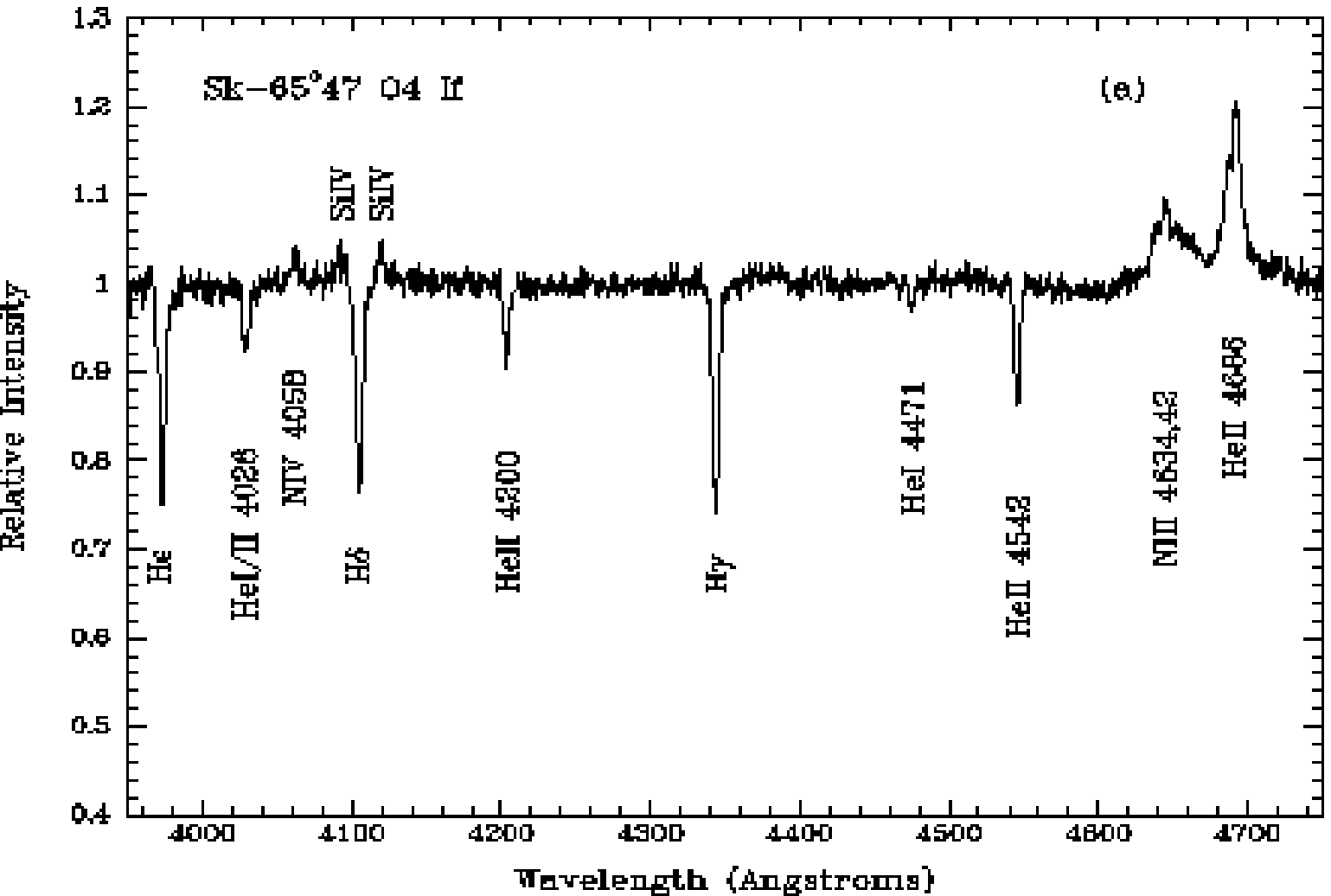}
\plotone{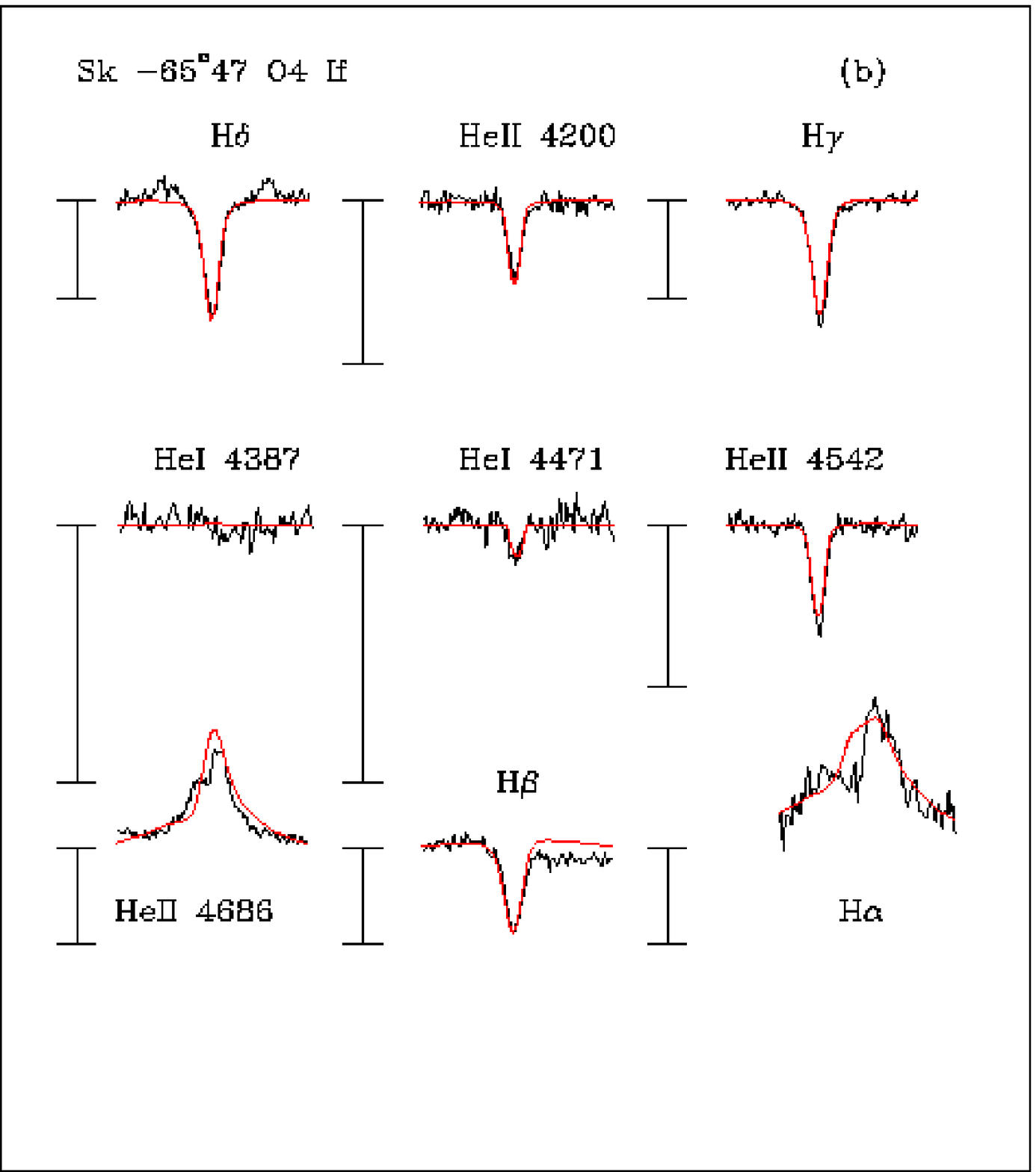}

\caption{\label{fig:sk-6547} Sk$-65^\circ47$.  Same as Fig.~\ref{fig:av177}.  A
radial velocity of 250 km s$^{-1}$ and a rotational
broadening $v \sin i$ of 160 km s$^{-1}$ was used in making this comparison.}
\end{figure}

{\it Sk$-65^\circ 47$.}---This star was classified as O4~If on the basis
of the present spectrum (Fig.~\ref{fig:sk-6547}a) by Massey
et al.\ (2000).  Here we concur with
this type. He~I $\lambda 4471$ is only weakly present, with an
equivalent width $W=115$m\AA. Given the weakness of He~I $\lambda 4471$,
we might be tempted to reclassify the star as O3, were it not for the
fact that $\log W'=-0.68$, consistent with the O4 classification.
Furthermore, Walborn et al.\ (2002) suggest that for an O3.5 N~III $\lambda
4634,42$ emission is similar in strength to N~IV $\lambda 4058$ emission,
while here we find N~III $>$ N~IV (although the latter is clearly present),
also leading us to the O4 subtype.  The strengths of the N~III $\lambda 4634,42$
and He~II $\lambda 4686$ emission lines lead to the ``If" luminosity class.
This is consistent with the absolute magnitude M$_V=-6.4$, a value which
is typical for an O4~If star (Conti 1988).
Good fits to the lines were easily found (Fig.~\ref{fig:sk-6547}b),
and the physical parameters are well constrained.

\begin{figure}
\epsscale{0.6}
\plotone{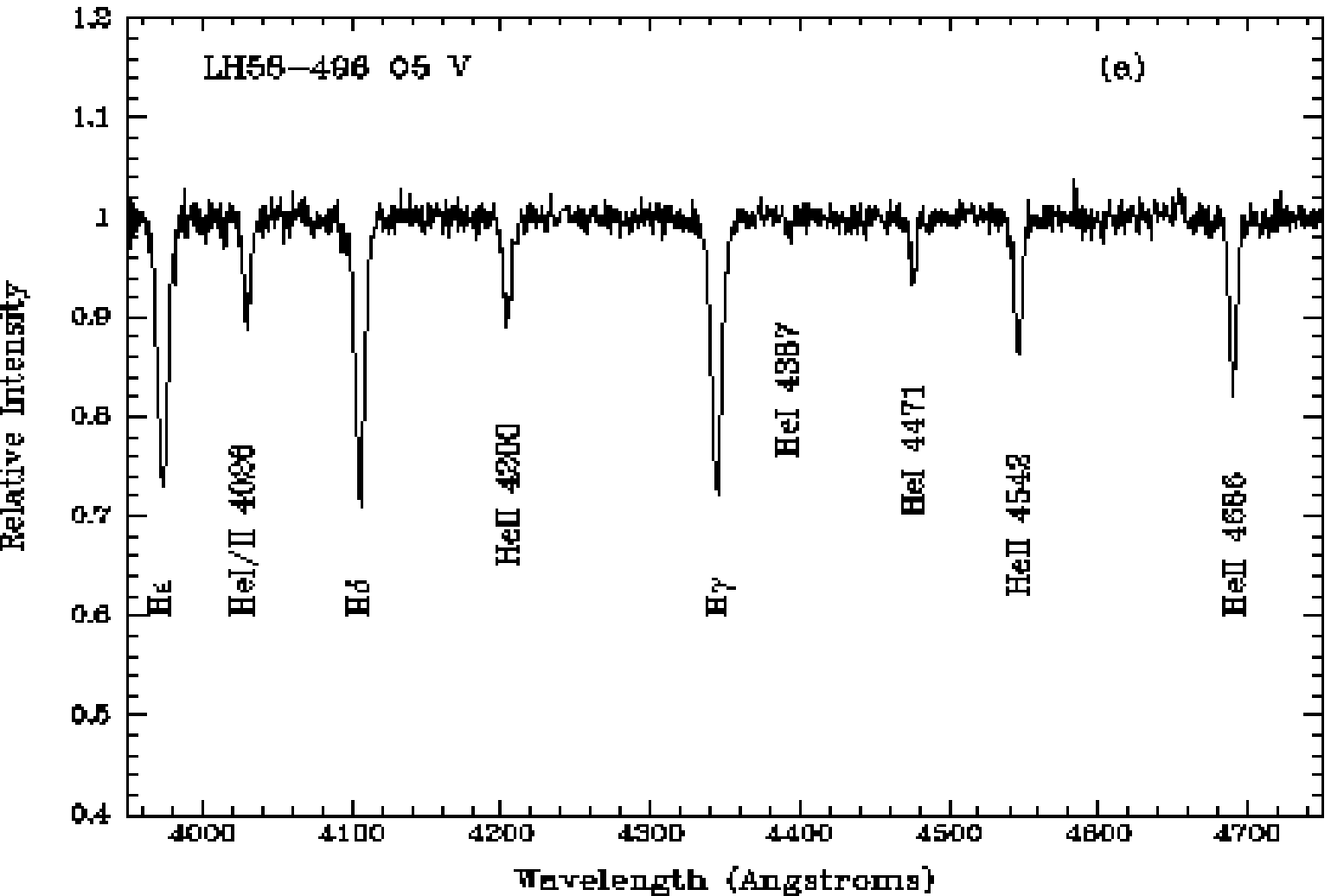}
\plotone{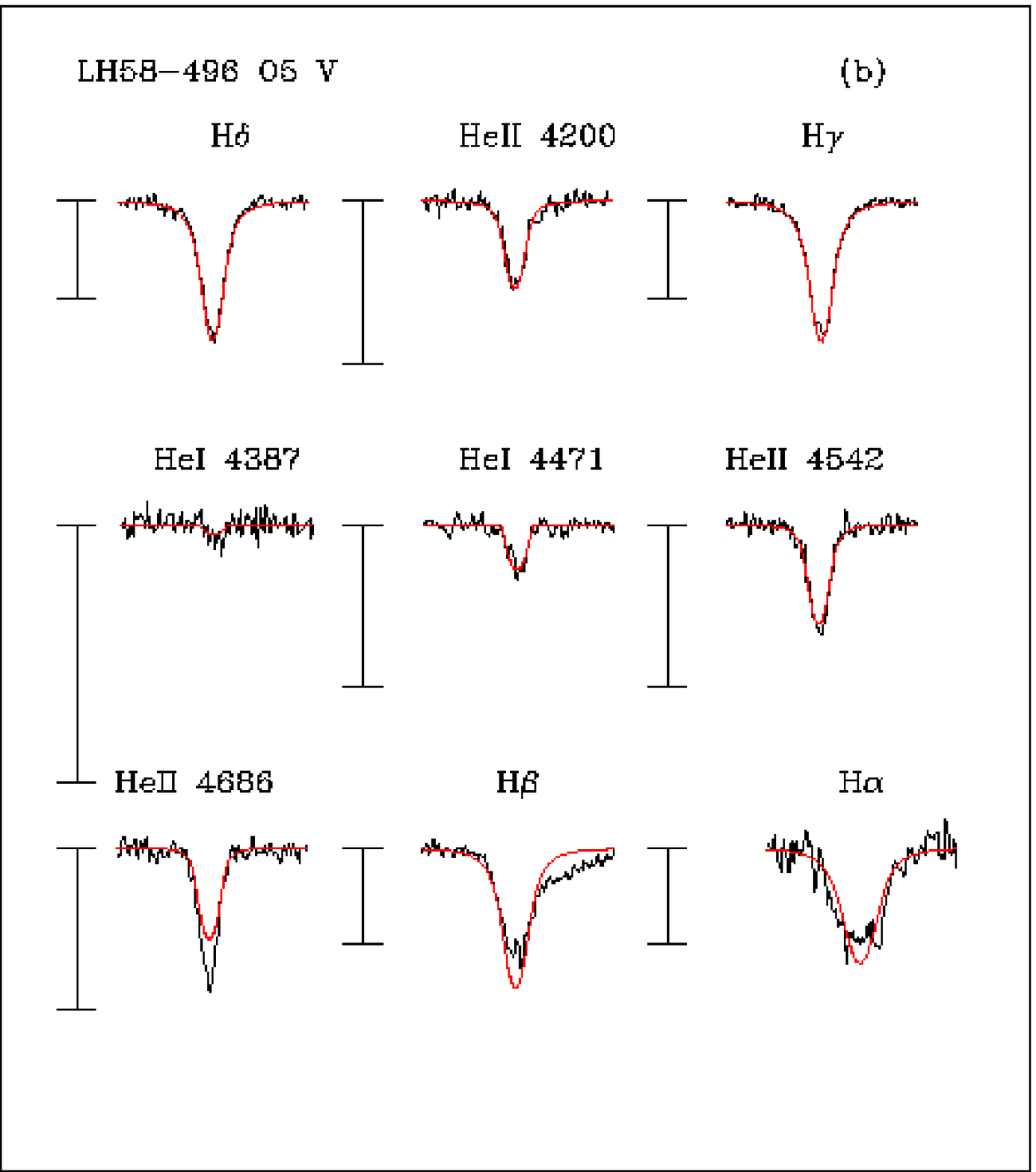}

\caption{\label{fig:lh58496} LH58-496.  Same as Fig.~\ref{fig:av177}.  A
radial velocity of 275 km s$^{-1}$ and a rotational
broadening $v \sin i$ of 250 km s$^{-1}$ was used in making this comparison.}
\end{figure}

{\it LH58-496.}---This star was classified as O3-4~V by Conti et al.\ (1986)
and Garmany et al.\ (1994).  Our higher S/N data (Fig.~\ref{fig:lh58496}a)
suggests a later type, O5~V, consistent with our measurement
of $\log W'=-0.52$.  The absolute magnitude, M$_V=-5.1$, is in good agreement
with our spectral classification of a dwarf.  
We obtained a good fit (Fig.~\ref{fig:lh58496}b), except for the H$\beta$
profile which is likely filled in by emission. The H$\alpha$ data were
obtained with {\it HST} and thus provides a relatively nebular-free 
profile.

\begin{figure}
\epsscale{0.6}
\plotone{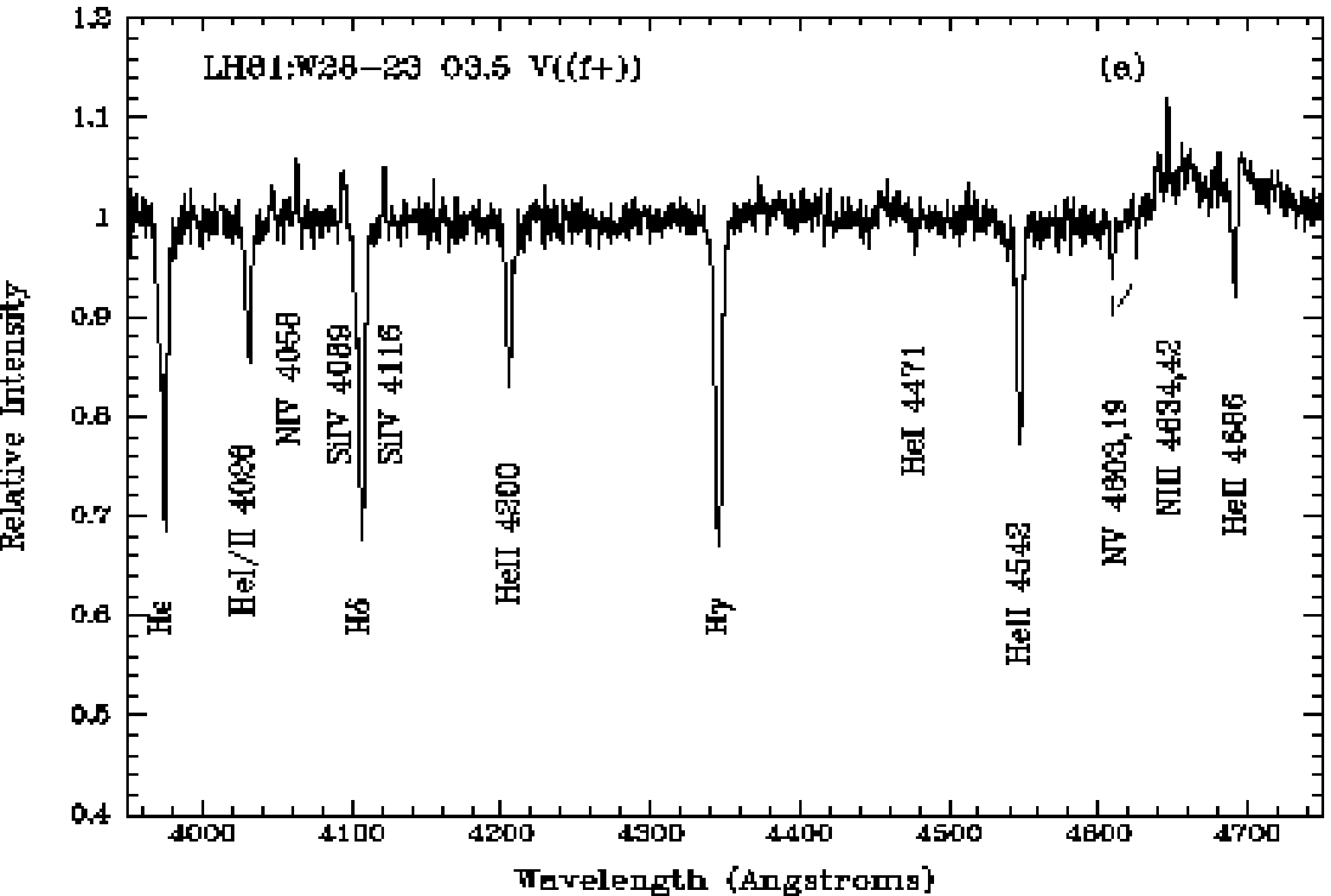}
\plotone{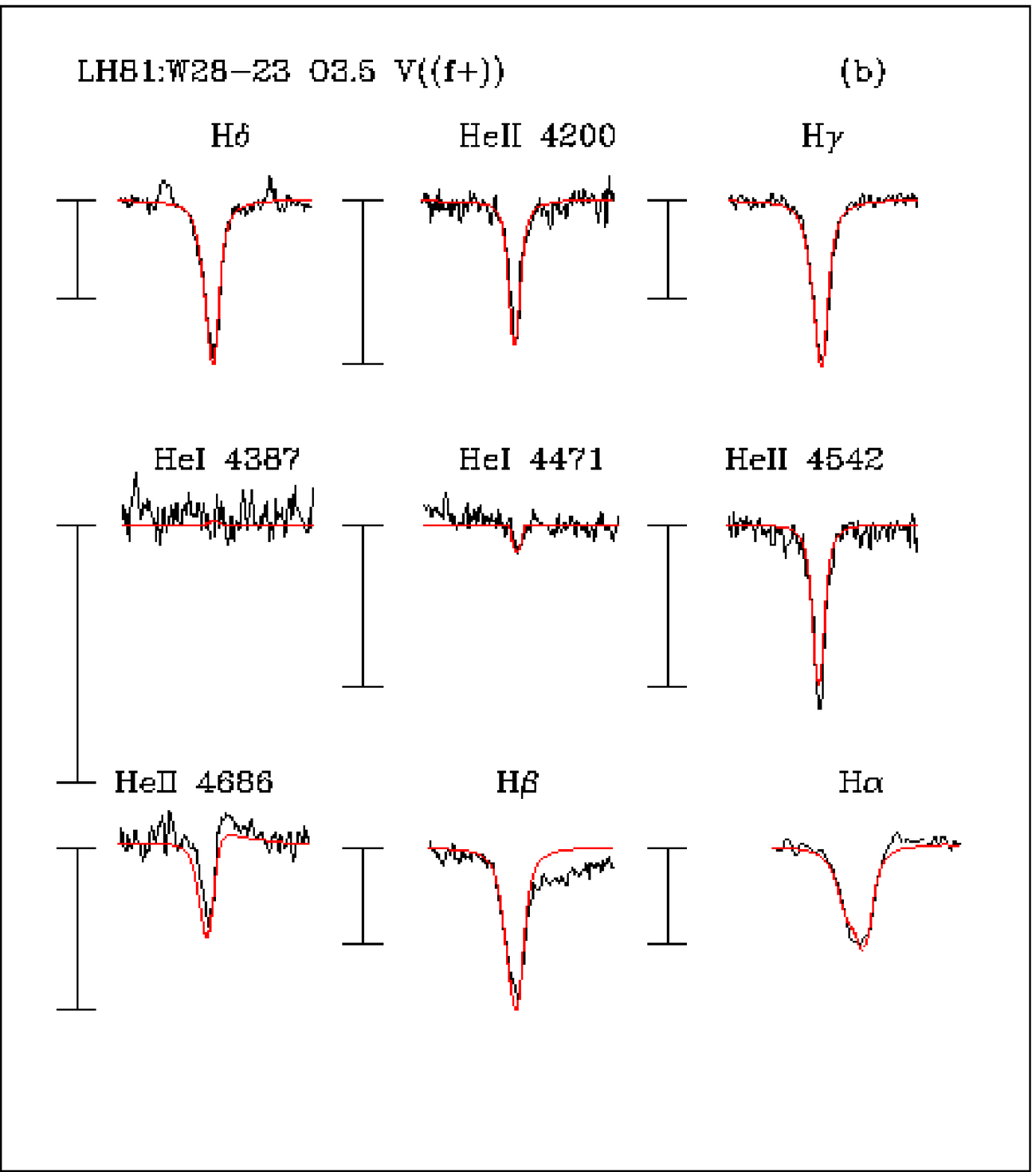}
\caption{\label{fig:lh81} LH81:W28-23.  Same as Fig.~\ref{fig:av177}.  A
radial velocity of 350 km s$^{-1}$ and a rotational
broadening $v \sin i$ of 120 km s$^{-1}$ was used in making this comparison.
A cosmic ray on the wing of the He~II $\lambda 4200$ has been removed in
this plot.}
\end{figure}

{\it LH81:W28-23.}---This star was previously classified as O3~V by 
Massey et al.\ (2000) based on the same data as used here.  
Although He~I $\lambda 4471$
is readily identified in our spectrum (Fig.~\ref{fig:lh81}a), 
its presence is due to the 
high S/N, as we measure an equivalent width of only 80m\AA. We find a value
of $\log W'=-1.0$, well below the value $-0.6$ that separates the
O4's from the O5's (Conti 1988).  N~V $\lambda\lambda 4603,19$ absorption
is clearly present, and N~IV $\lambda 4058$ emission (EW -120m\AA)
is less than
that of N~III $\lambda\lambda 4634,42$ (EW = -500m\AA).  Thus under the
Walborn et al.\ (2002) scheme, the star would be classified as 
``O3.5~V((f+))", with the ``((f+))" denoting N~III emission with
comparable Si~IV emission. 
The strong He~II $\lambda 4686$ absorption
indicates that the star is a dwarf, consistent with its M$_V=-5.1$ value.

Our modeling of this star was straightforward, although it quickly became
clear that we had to significantly increase the He/H number ratio from
its canonical 0.1 value to 0.25 in order to match the observed spectrum.
The final fits are shown in Fig.~\ref{fig:lh81}(b).
The ground-based H$\alpha$ profile agrees well with that obtained with
{\it HST}; accordingly, the former is shown, as it is considerably less 
noisy.

\begin{figure}
\epsscale{0.6}
\plotone{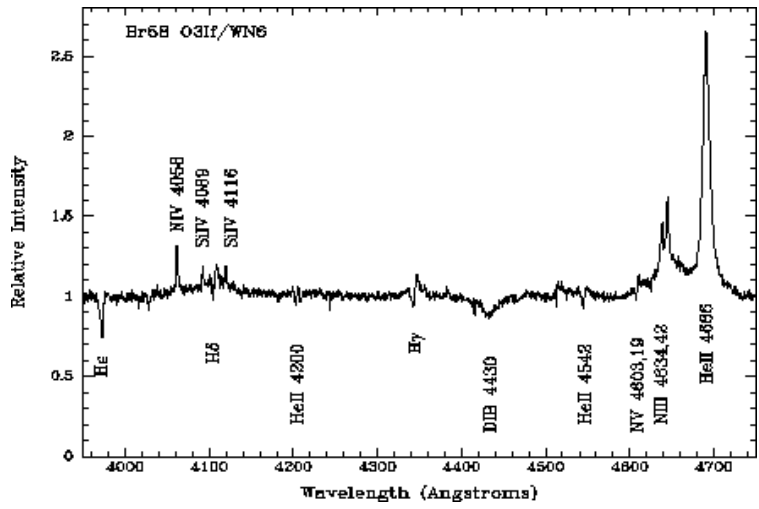}
\caption{\label{fig:br58} LH90:Br58. The strong emission in this star prevented
a satisfactory fit, although tentative parameters are given in the text.
Note the strong diffuse interstellar band at $\lambda 4430$.}
\end{figure}

{\it LH90:Br58.}---Breysacher (1981) included this star in his
catalog of LMC Wolf-Rayet
stars, where it is classifed as WN5-6. The star was reclassified
by Testor et al.\ (1993) as WN6-7.  The lower Balmer lines are in
emission (H$\alpha$) or P Cygni (H$\gamma$, H$\delta$), and He~II $\lambda
4200$ and $\lambda 4542$
 are present primarily in absorption (Fig.~\ref{fig:br58}),
leading Massey et al.\ (2000) to argue that the star is more properly
called an O3~If/WN6, although as mentioned earlier the classification of
such ``intermediate" objects is difficult.
The equivalent width of He~II $\lambda 4686$ is
$-18$\AA; usually $-10$ is taken (somewhat arbitrarily) as the dividing
line between an Of star with strong emission line and a bona-fide Wolf-Rayet
star of the WN sequence.  Clearly the emission is considerably greater
than in Sk$-67^\circ$22 (Fig.~\ref{fig:sk-6722}a).
Although Br~58 is a member of a ``tight cluster" (see 
Fig. 1a of Testor et al.\ 1993), our 1.3" slit width and position
angle excluded resolved neighbors.
We note that there is little evidence from a visual
examination of the spectrum that there is enhanced helium present at the
surface of the star, and a more detailed analysis than what we can perform
is needed to argue if nitrogen is enhanced or not.  

In any event, the strong
emission defeated our efforts to obtain an acceptable fit.  
We did establish that
a steep velocity law ($\beta=3.0$) and high mass-loss rate ($40\times 10^{-6}
M_\odot$ yr$^{-1}$) gave an excellent fit to the 
H$\alpha$ profile, with a temperature
of 40,000-42,000~K giving a reasonable match to the He~II line strength
if a $\log g=3.5$ was assumed. The high value of $\beta$ alone might
suggest that the star is {\it physically} more like a Wolf-Rayet star than 
like a normal O star, for which this large a value for $\beta$ would be
unexpected.  Our fit is not of the same quality as that we achieve
for other stars, as clearly the treatment of the stellar wind
dominates the stellar parameters.  If these parameters are even approximately
correct, the star is quite luminous (M$_{\rm bol}=-11.1$) with a 
radius of 30$R_\odot$ and hence an inferred spectroscopic mass $>100 M_\odot$.
An analysis that includes separate determinations of the elemental 
abundances (not possible with our current implementation of
FASTWIND)
is clearly warranted for this interesting object.  We do not include this
star in our analysis in Sec.~\ref{Sec-results}, and therefore it has
no effect on our derived effective temperature scale.  In addition, we exclude
this star from consideration in Sec.~\ref{Sec-O3s} since its physical
parameters are poorly determined.

\begin{figure}
\epsscale{0.6}
\plotone{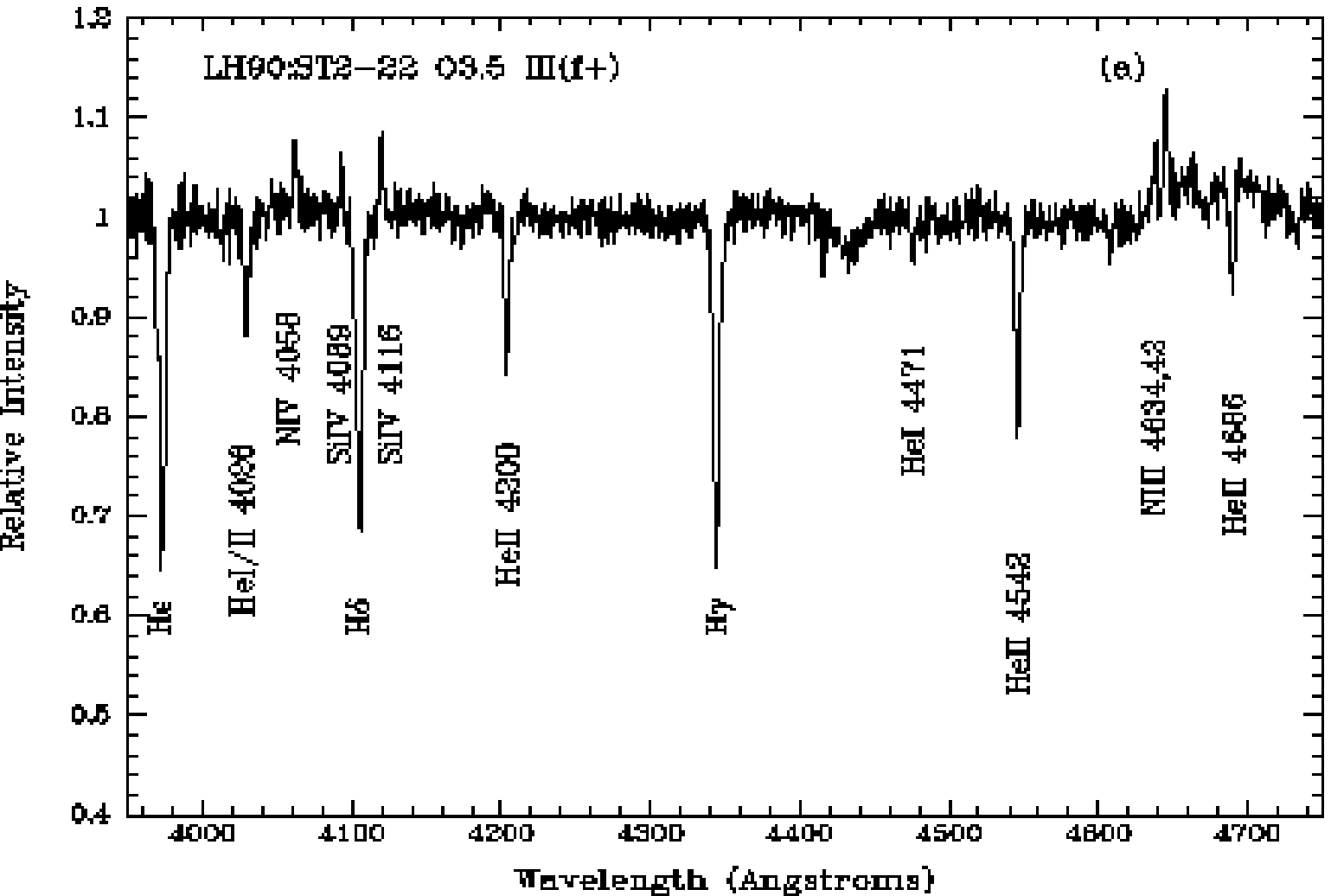}
\plotone{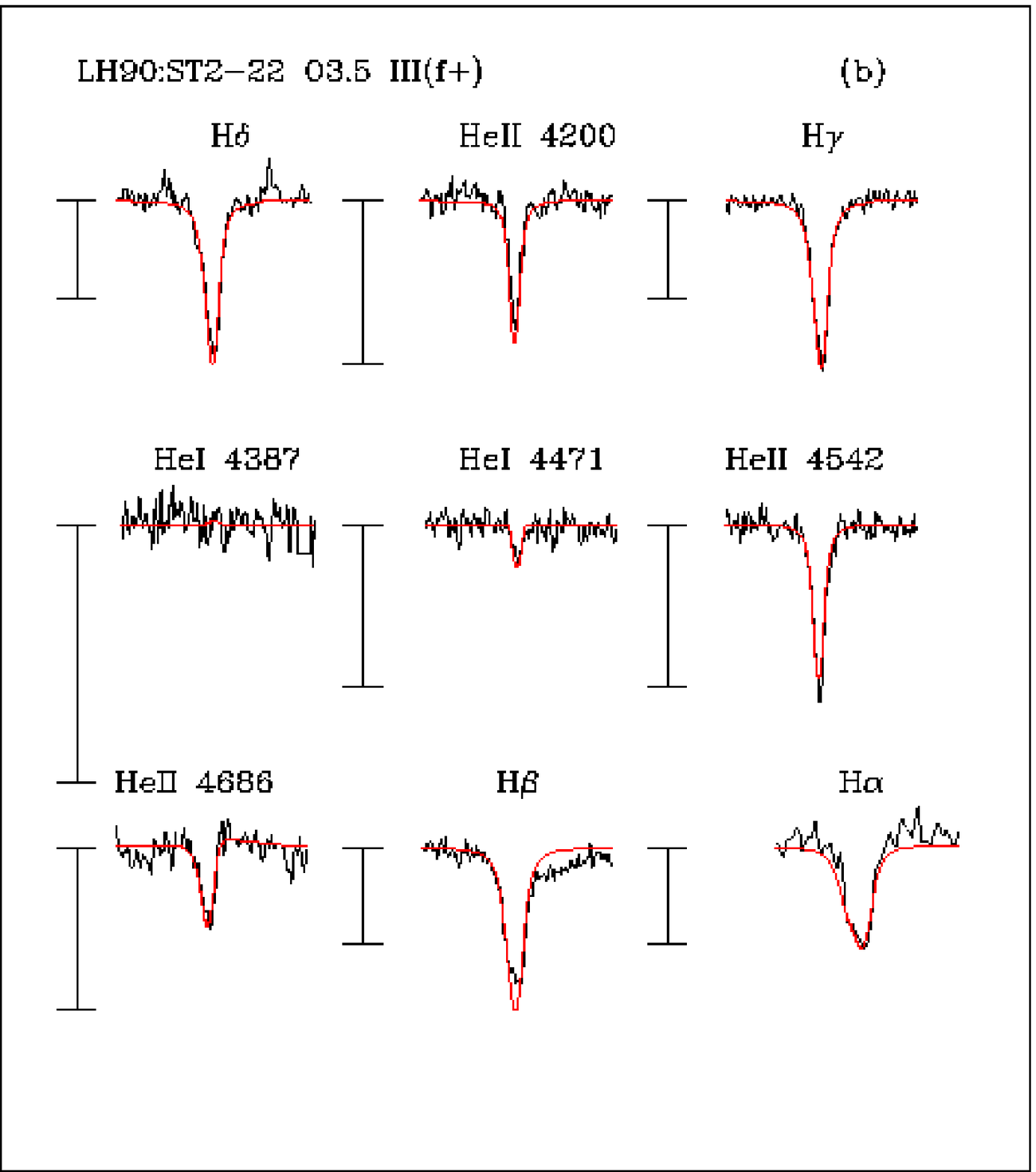}
\caption{\label{fig:st2} LH90:ST2-22. 
Same as Fig.~\ref{fig:av177}.  A
radial velocity of 250 km s$^{-1}$ and a rotational
broadening $v \sin i$ of 120 km s$^{-1}$ was used in making this comparison.
A cosmic ray on the He~II $\lambda 4200$ profile has been removed in this fit.
}
\end{figure}

{\it LH90:ST2-22.}---This star was classified as O3~III(f*) by Schild \&
Testor (1992), and reclassified as O3~V((f)) by Massey et al.\ (2000) based
upon the same spectral data as used here.  The spectrum in shown in 
Fig.\ref{fig:st2}a.  He~I $\lambda 4471$ is weakly present (EW of 140m\AA),
with $\log W'=-0.85$.  Should this star be instead classified as an O4?
We find N~IV $\lambda 4058$ emission weaker
of N~III $\lambda \lambda 4634,42$ (EW of $-200$m\AA\ versus EW=$-700$m\AA), 
with the presence of
N~V $\lambda \lambda 4603,19$ absorption uncertain.  Thus according to
the criteria given by Walborn et al.\ (2002) we would classify this
star as an O3.5 were we to rely upon the nitrogen emission lines.
As to the luminosity class, the prototype O3.5~III(f*) star Pismis 24-17
(Walborn et al.\ 2002) clearly has much stronger 
N~V $\lambda \lambda 4603, 19$
absorption (compare Walborn et al.\ 2002 Fig.~4 with our Fig.~\ref{fig:st2}
a), although the N~IV and N~III emission appears stronger in LH90:ST2-22
than in the O3.5~V(f+) star HD~93250. The appearance of He~II $\lambda 4686$
suggests a giant (III) luminosity class.
The absolute magnitude of LH90:ST2-22
is $-6.4$, which is more typical of an O3-4 {\it super}giant (O3-4~I) than
a dwarf (Table 3-1 of Conti 1988).  We adopt O3.5~III(f+) as the spectral
type, where the ``+" denotes that Si~IV $\lambda \lambda 4089,4116$
is in emission, a redundant reminder given that all stars so
classified have this feature (see Walborn et al.\ 2002).

The fitting of this star was straightforward.  The widths of the Balmer
line wings necessitated a value $\log g=3.7$.
The fits are shown in Fig.~\ref{fig:st2}b.  A slightly elevated
He/H number ratio (0.2) was needed in order to get the helium lines as
strong as what was observed.

\begin{figure}
\epsscale{0.6}
\plotone{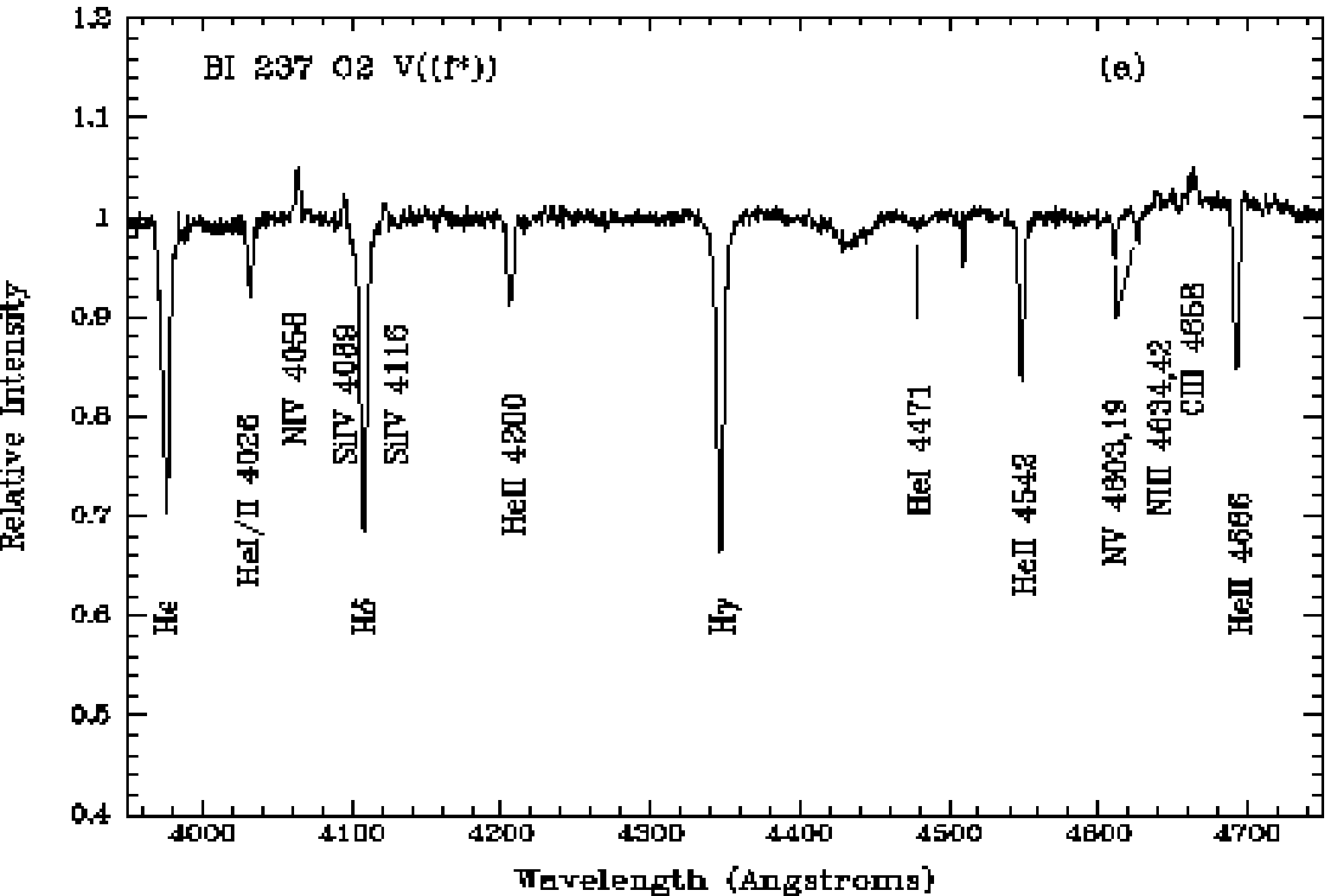}
\plotone{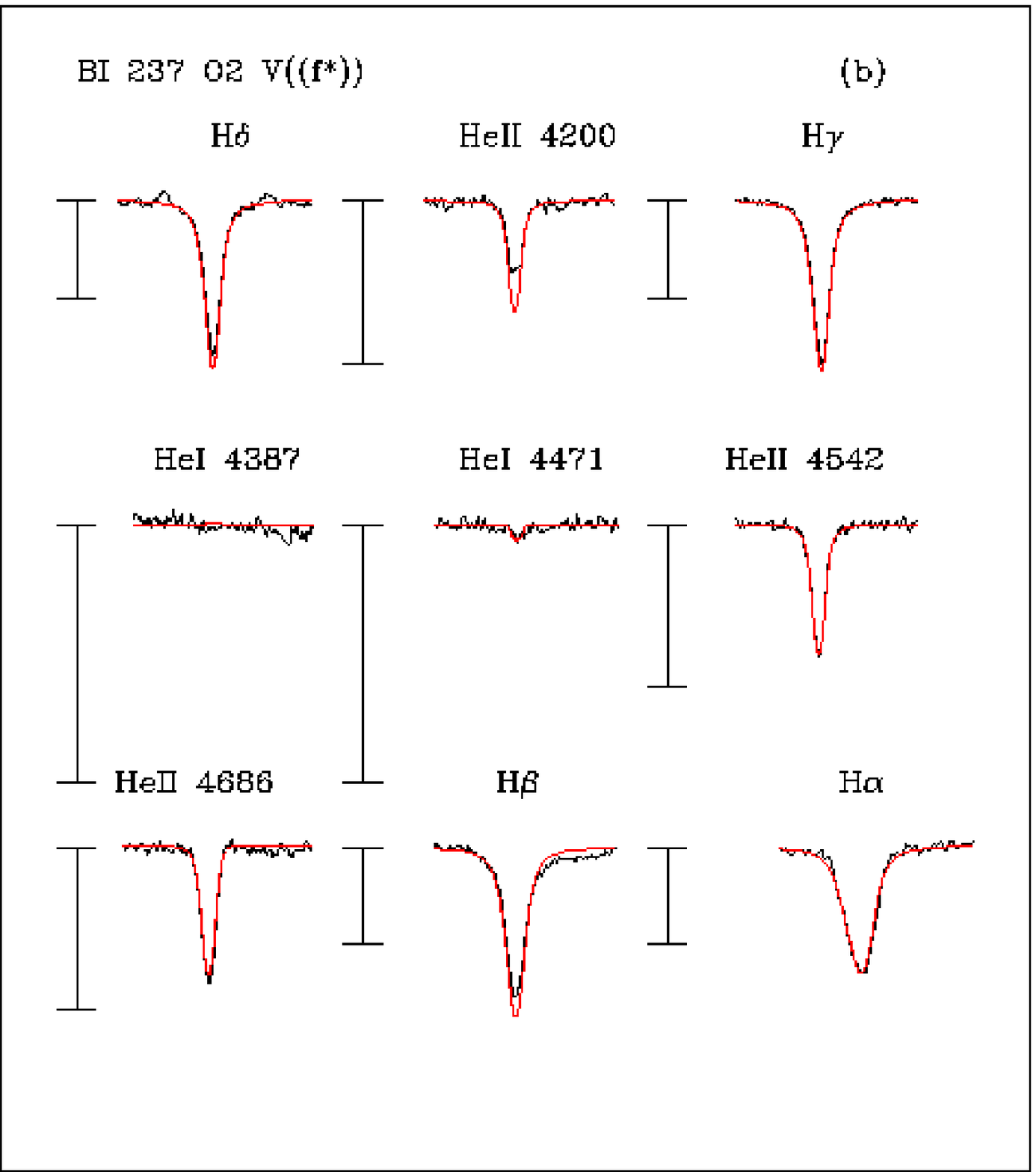}
\caption{\label{fig:bI237} BI~237.
Same as Fig.~\ref{fig:av177}.  A
radial velocity of 430 km s$^{-1}$ and a rotational
broadening $v \sin i$ of 150 km s$^{-1}$ was used in making this comparison.
The He~II $\lambda 4200$ profile was is flat-bottomed, presumably due to
a reduction problem, and not used in
making the fit.
}
\end{figure}

{\it BI~237.}---This star was classified as an O3~V by Massey et al.\ (1995)
using older, poorer S/N data than what we have obtained for the present
study.  The spectrum is shown in Fig.~\ref{fig:bI237}a.  We see that this
star is somewhat earlier than that of LH90:ST2-22 (Fig.~\ref{fig:st2}a)
discussed above, with He~I $\lambda 4471$ nearly absent. We do detect
{\it some}
He~I $\lambda 4471$, but at a very weak level ($\sim20$m\AA), a detection
possible only with our extremely high (500) S/N spectrum.  With the EW
of He~II $\lambda 4542$ $\sim$ 750m\AA\ we conclude $\log W'\sim-1.6$!
The nitrogen emission line criteria proposed by Walborn et al.\ (2002)
would make this either an O2~V((f*)) [``NIV $>>$ NIII, no He~I"]
or an O3~V((f*)) [``NIV $>$ or $\approx$ N~III; very weak He~I"],
given that N~IV $\lambda 4058$ emission
(EW=-165m\AA)
is stronger than the N~III $\lambda\lambda 4634,42$ emission (EW=-90m\AA).
  
Whether one should call the star an O2 or an O3 is of course dependent
upon the interpretation of the non-quantitative criteria enumerated by
Walborn et al.\ (2002).
However, the presence of C~IV $\lambda 4658$ {\it emission}, with an EW=-140m\AA, (also
seen in BI~253, discussed below) is taken by Walborn et al.\ (2002)
as an extreme hot O2 spectrum.  We therefore call the star an O2~V((f*)),
despite the (very) weak presence of He~I $\lambda 4471$.
The absolute magnitude (M$_V=-5.4$) is consistent with the dwarf designation
suggested by the strong He~II $\lambda 4686$ absorption feature 
(Fig.~\ref{fig:bI237}a).

Despite the extreme spectral type, the fitting was very straightforward,
with the only complication that the He~II $\lambda 4200$ profile was
corrupted.  The fit is shown in Fig.~\ref{fig:bI237}b. The
physical parameters are extremely well constrained thanks to the
very weak presence of He~I $\lambda 4471$.  The {\it HST} H$\alpha$ profile
agrees well with the higher S/N ground-based data, and thus we used the
latter.

This star is one of the ``field" early-type stars identified by
Massey et al.\ (1995), located well outside the nearest known OB
association (LH~88).  We do note that the radial velocity of this
star is quite high ($\sim 400$ km s$^{-1}$), compared to the systemic
LMC velocity of 279~km~s$^{-1}$).  Thus this field O2 star would also
qualify as
an OB runaway.  In concert with the similar results for the field O3
star Sk$-67^\circ$22, we believe this provides an explanation for the
long-standing mystery of such early-type field stars (Massey et al.\ 1995;
Massey 1998).

\begin{figure}
\epsscale{0.6}
\plotone{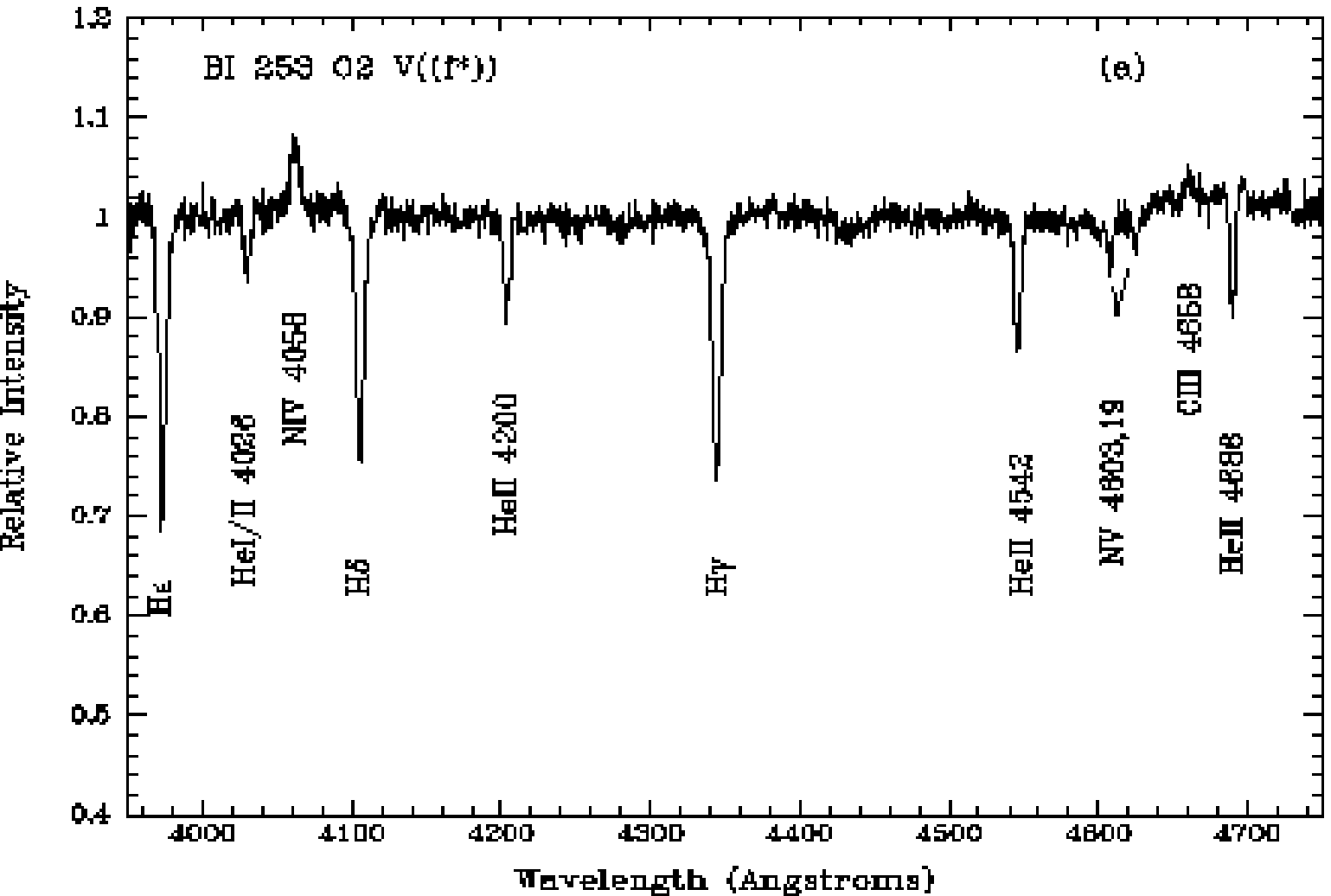}
\plotone{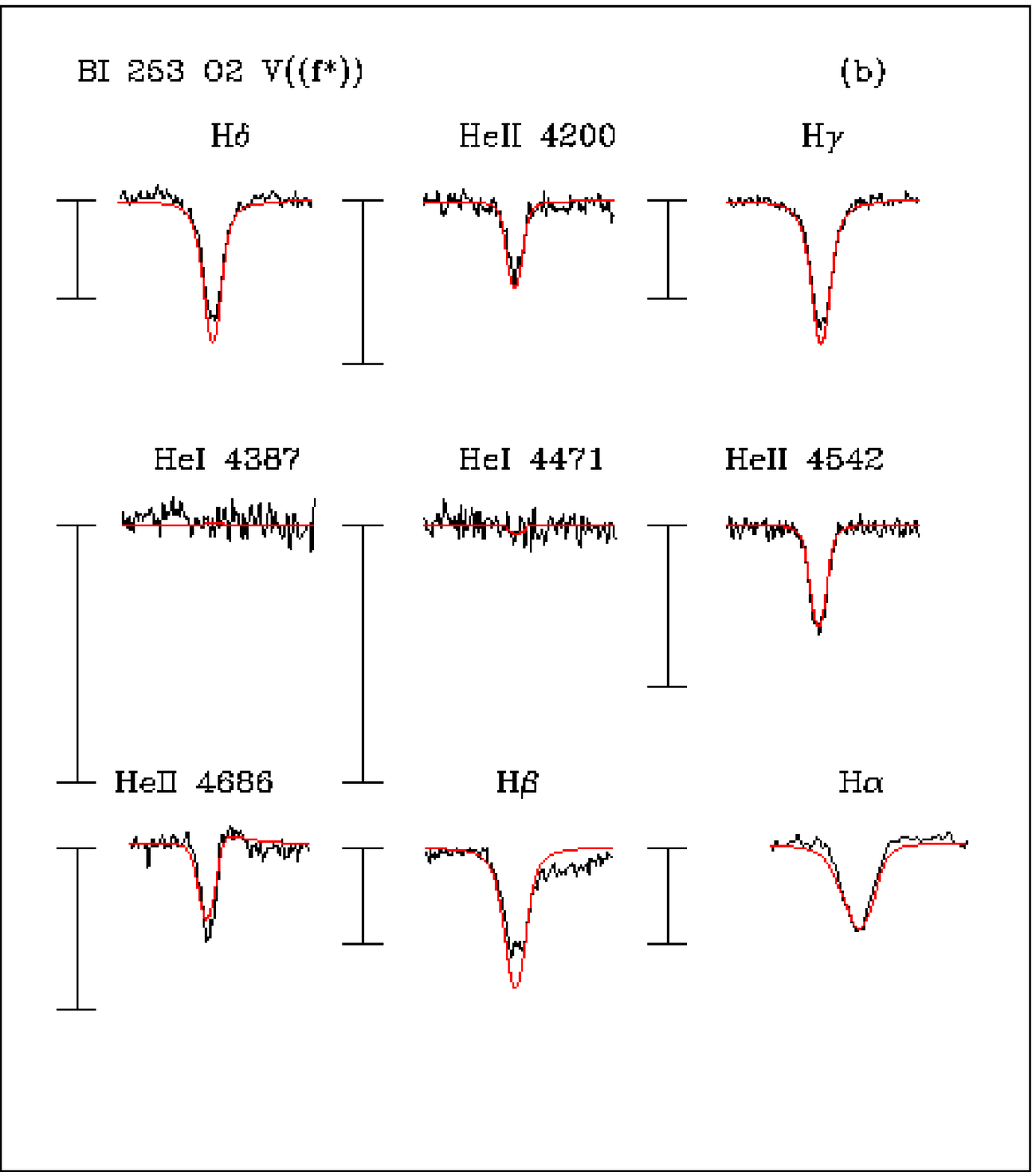}
\caption{\label{fig:bI253} BI~253.
Same as Fig.~\ref{fig:av177}.  A
radial velocity of 270 km s$^{-1}$ and a rotational
broadening $v \sin i$ of 200 km s$^{-1}$ was used in making this comparison.
}
\end{figure}

{\it BI~253.}---This star has many similarities to BI~237.  It too was
classified as O3~V by Massey et al.\ (1995), but was part of the basis
for Walborn et al.\ (2002)'s extension of the spectral sequence to O2.
Based upon the spectrum we obtained as part of the present study,
Walborn et al.\ (2002) propose BI~253 as the prototype for the
newly defined O2~V class.  It is interesting to compare its spectrum
(Fig.~\ref{fig:bI253}a)
with that of BI~237 (Fig.~\ref{fig:bI237}a),
which we argue above is also a member of the O2~V
class.  The nitrogen spectrum would suggest that BI~253 is somewhat hotter,
as N~III $\lambda\lambda 4634,42$ emission is not visible in our spectrum,
while N~IV $\lambda 4058$ emission and 
N~V $\lambda\lambda 4603,19$ absorption are both considerably stronger than
in BI~237.
The EW of NIV $\lambda 4058$ is nearly three times
stronger in BI~253, with the line having an EW of
about -450m\AA.  He~I $\lambda$ 4471 is {\it not} detectable, although
this could simply be due to the poorer S/N of this spectrum compared to
that of BI~253 (225 vs 500).
We feel confident in placing an upper limit of 25m\AA\
on its presence, requiring $\log W'<-1.5$, as the EW of He~II $\lambda 4542$
is 710m\AA.  C~IV $\lambda 4658$ emission
is present (EW=-160m\AA) in comparable intensity to that of BI~237.
The star's absolute magnitude (M$_V=-5.5$) is consistent with the dwarf
designation suggested by the appearance of He~II $\lambda 4686$.  We
describe the spectrum as O2~V((f*)), consistent with Walborn et al.\ (2002)
notation, although we note that the lack of N~III $\lambda 4634,42$ emission
makes the use of the ``f" designation arguable.

The fitting was against straightforward (Fig.~\ref{fig:bI253}b), 
although only a lower limit
could be set on the effective temperature (and hence on other stellar
parameters) due to the lack of an unambiguous detection of He~I $\lambda 4471$.
The mass-loss rate is somewhat greater in BI~253 than it is in BI~237.

\begin{figure}
\epsscale{0.6}
\plotone{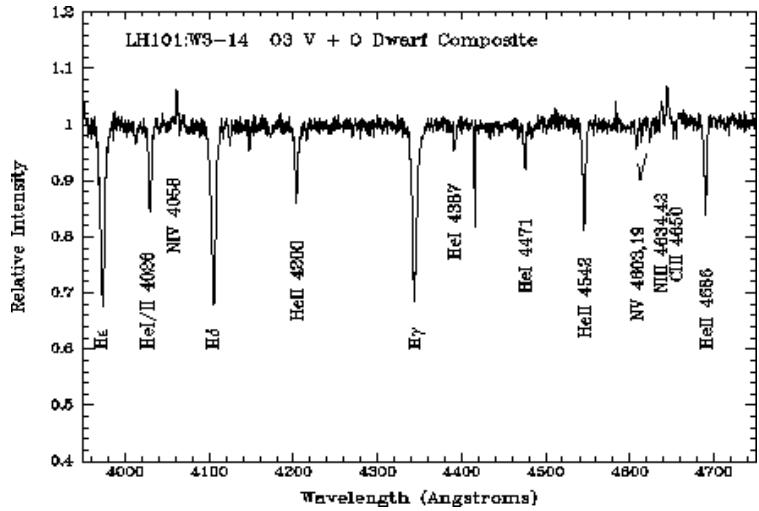}
\caption{\label{fig:lh101w14} LH101:W3-14.
The spectrum shown here is composite, with an O3 star dominating the
nitrogen and He~II spectra, and a later-type O star contributing the
He~I.  No single model was able to fit the strengths of the He~I and He~II
lines.
}
\end{figure}

{\it LH101:W3-14.}---Both LH101:W3-14 and LMC2-675 (discussed below) are
stars whose spectra are moderately early ($\sim$ O5) according to the
He~I/He~II ratio, but which show nitrogen features typical of much earlier
stars.  Are these stars spectral composites, or nitrogen-enriched objects?
We have previously shown that when a star is a spectral composite we can
seldom match the actual He~I and He~II line strengths with a single model.
The star was classified as O3~V by Testor \& Niemela (1998), and then
reclassified as ON5.5 V((f)) by Massey et al.\ (2000) based on the spectrum
used here.  The Testor \& Niemela (1998) spectrum of this star is clearly
noisy (see their Fig.~4), and it is not clear whether or not He~I absorption
is present in their spectrum or not.  It is certainly present in ours, at
a strength typical of an O5 star, with a value of $\log W'$ equal to $-0.49$.
Here we also see (Fig.~\ref{fig:lh101w14})
N~IV $\lambda 4058$ and
N~III $\lambda \lambda$ 4634,42 emission of comparable intensity, so if
we were to ignore the He~I absorption spectrum we would conclude that the star
was O3~V((f*)) according to the criteria of Walborn et al.\ (2002). N~V
$\lambda \lambda 4603,19$ absorption is also weakly present, characteristic of an
early spectral type.   On the other hand, the presence of C~III $\lambda 4650$
absorption is typical of a later spectral type (O8.5+).
The absolute
visual magnitude of the star, M$_V=-5.6$, is somewhat brighter than the
M$_V=-5.2$ value listed by Conti (1988) for an O5 dwarf, and is more consistent
with a giant classification.  The strength of He~II $\lambda 4686$ absorption
is weaker than that of He~II $\lambda 4542$, and so a giant designation is
not precluded.  Alternative, the absolute magnitude may be suggesting that
the object is the composite of two dwarfs.

The modeling quickly demonstrated to our satisfaction that this spectrum
is composite.  A 40,000~K model can match the line strengths of the He~II
lines and that of He~I $\lambda 4387$, but this model's
He~I $\lambda 4471$ line is then many times stronger than what
is actually observed.
(This goes in the
{\it opposite} sense of the ``dilution effect" described by Repolust 
et al. 2004.)
We can achieve the correct line
{\it ratio} for He~I $\lambda 4387$ and  He~I $\lambda 4471$ by lowering
the effective temperature to 38,000~K, but the He~I 
line {\it strengths} are then
much greater than what is actually observed, and the He~II lines are
too weak.  This is consistent with a composite spectrum, with an O3~V
dominating the nitrogen and He~II spectrum, and a latter O-type dominating
the He~I spectrum.
We further note that the He~I $\lambda 4471$ line appears to be broader 
($v \sin i=130$ km s$^{-1}$) than the He~II lines ($v \sin i=110$ km
s$^{-1}$), a difference that is quite noticeable with our S/N and dispersion.
This is consistent with the spectral classification of Walborn et al.\ (2002)
based upon the same data.

\begin{figure}
\epsscale{0.6}
\plotone{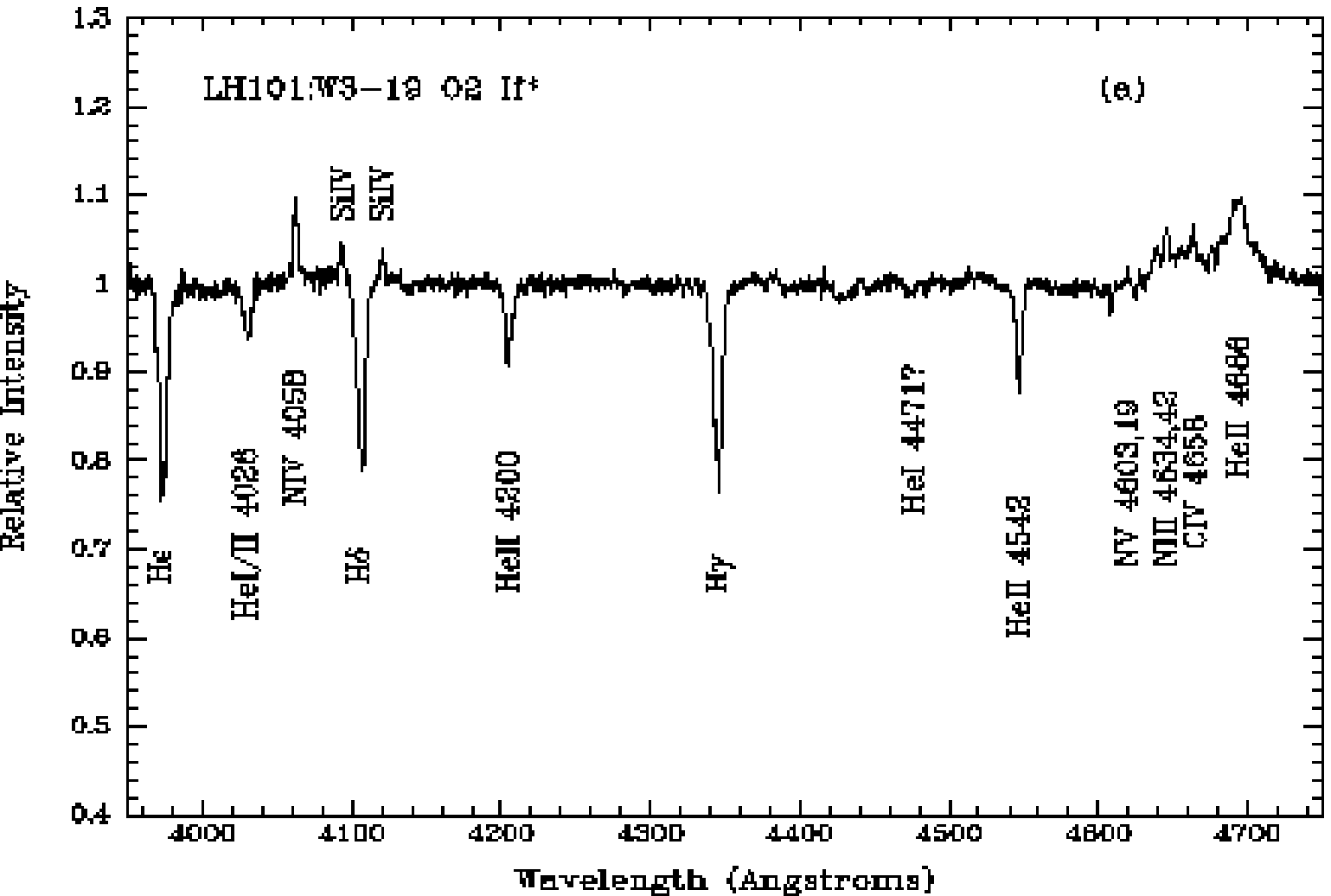}
\plotone{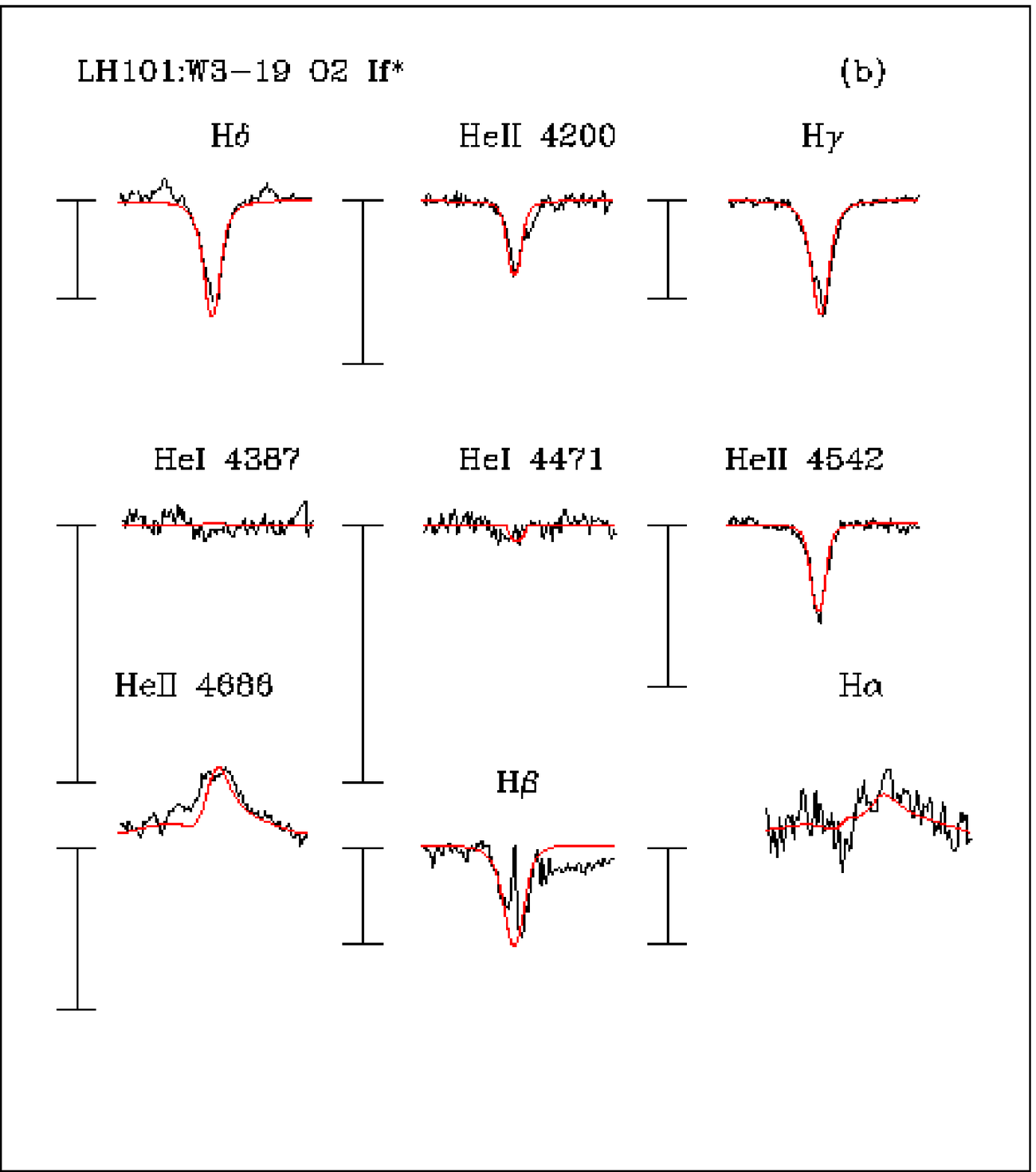}
\caption{\label{fig:lh101w19} LH101:W3-19.
Same as Fig.~\ref{fig:av177}.  A
radial velocity of 320 km s$^{-1}$ 
and a rotational
broadening $v \sin i$ of 180 km s$^{-1}$ was used in making this comparison.
}

\end{figure}

{\it LH101:W3-19.}---This star was classified as O3~If by Testor \&
Niemela (1998) and again as O3~If* by Massey et al.\ (2000); the data
for the latter are the same as what are used here.  Using the criteria
suggested by Walborn et al.\ (2002) the star would have to be
considered of spectral type O2 If*, as the N~IV $\lambda 4058$ emission
(EW=-300m\AA) is greater than the N~III $\lambda \lambda
4634 42$ emission (EW=-230m\AA).
(Fig.~\ref{fig:lh101w19}a). The presence of weak C~IV $\lambda 4658$
emission (EW=-150m\AA)  suggests it is quite hot.  N~V $\lambda \lambda 4603,19$
absorption and Si~IV $\lambda \lambda 4089, 4116$ emission are clearly visible.
He~II $\lambda 4686$ is in strong emission rather than P~Cygni
emission, attesting that it is a supergiant and not a giant.
A supergiant classification is consistent with the high absolute visual
magnitude, M$_V=-7.04$, found for this star.

We have {\it HST} and ground-based data 
for both the blue part of the spectra and H$\alpha$.  Nebulosity is
strong around this star, and the {\it HST} data proved crucial for
the H$\alpha$ region.  Although the {\it HST} data in the blue give
less contaminated Balmer lines, the S/N is so much worse (130 vs.\
420 per 1.4\AA) that we used the ground-based data for the fitting,
but checked that the {\it HST} data yielded similar results.

The fitting of this star was relatively straightforward, with
reasonably good fits.  The surface gravity of the star is somewhat
higher than what we might naively expect for a supergiant, with $\log
g=3.9$.  The inferred spectroscopic mass is very high, $>190
M_\odot$, one of the highest known.  He~I $\lambda 4471$ may be {\it
marginally} detected, with an EW of 50m\AA\ (and $\log W'\sim -1.1$); we
were unable to convince ourselves whether or not this feature was real,
and we therefore treat the effective temperature as a lower limit.

\begin{figure}
\epsscale{0.6}
\plotone{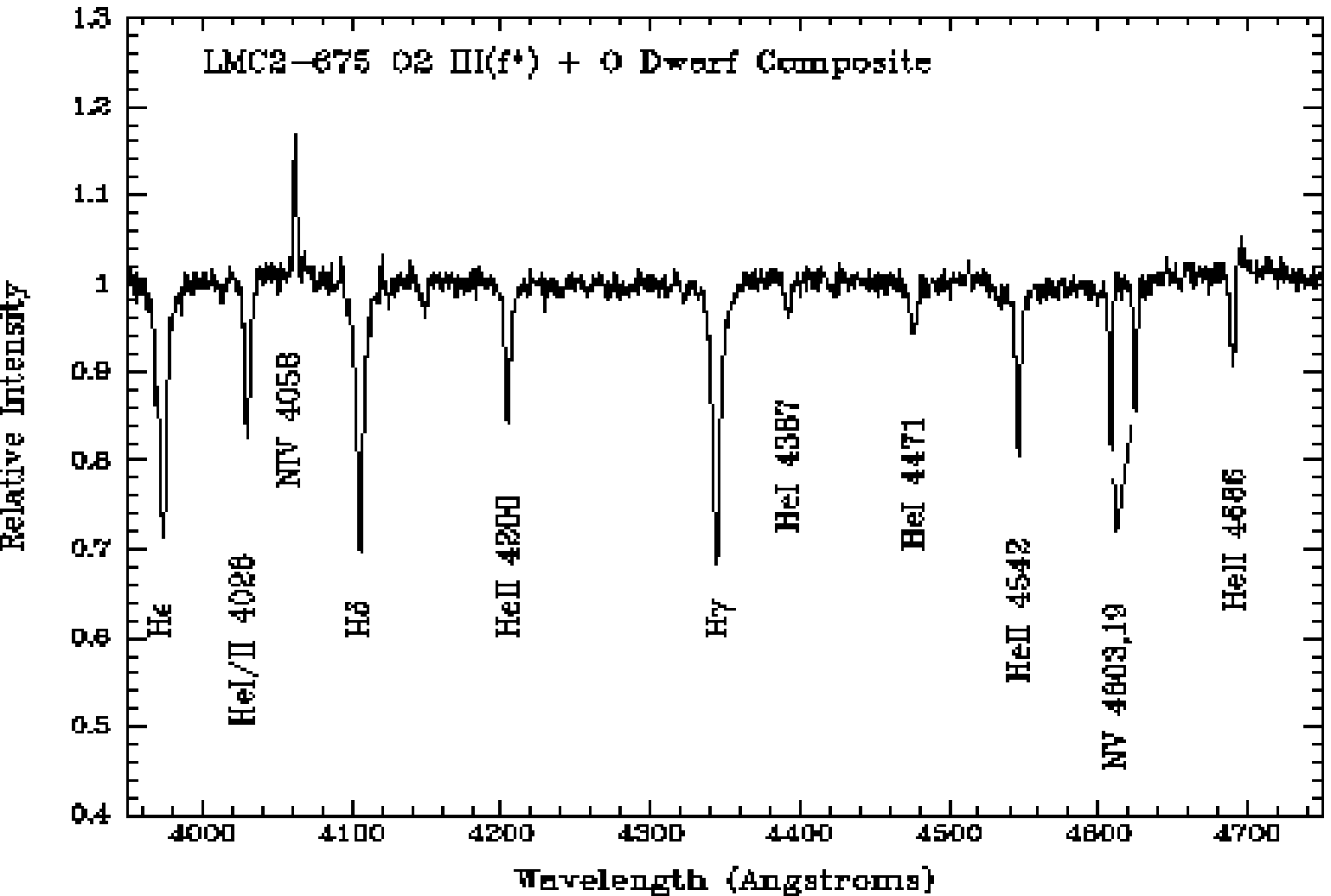}
\caption{\label{fig:lmc2675} LMC2-675.
The spectrum shown here is composite, with an O3 star dominating the
nitrogen and He~II spectra, and a later O-type star contributing the
He~I.  As in Fig.~\ref{fig:lh101w14}, 
no single model was able to fit the strengths of the He~I and He~II
lines.
}
\end{figure}

{\it LMC2-675.}---This star was classified as ``O3~If+O" 
by Massey et al.\ (1995), who recognized its composite
nature.
The spectrum of this star (Fig.~\ref{fig:lmc2675}) 
is very similar in appearance to that of LH101:W3-14 (Fig.~\ref{fig:lh101w14}),
with a He~I to He~II ratio suggesting a spectral type O5, but with strong
N features that would be characteristic of an earlier type.  Our modeling
confirms that this object is a composite.  No model simultaneously fits
the strengths of the He~I and He~II lines.  In this case we find we can
match the He~II lines and He~I $\lambda 4387$ line with a single model
of temperature 40,000~K, 
but that the He~I $\lambda 4471$ line from the model is much too strong.
Furthermore, the He~II lines are shifted in radial velocity by about 30
km s$^{-1}$ relative to those of He~I and H.  The strength of N~IV $\lambda
4058$ emission and N~V $\lambda 4603,19$ absorption, plus the absence
of any N~III $\lambda 4634,42$ emission would suggest that one of the
components is an O2 giant [i.e., O2~III(f*)]. The other component
must be an O-type dwarf.  Note that for the earliest types (O2-3.5)
the absolute
visual magnitude is not a good indicator of luminosity class, as argued
in Paper I, so this need not be in conflict with the relatively modest
M$_V=-5.0$.

\subsubsection{The R136 Stars}

The R136 cluster in the LMC contains more extreme O-type stars than
are known in total elsewhere (Massey \& Hunter 1998). For an interesting
subsample of stars, the FOS ``classification
quality" spectra of Massey \& Hunter (1998) 
have been supplemented by higher resolution STIS spectra
with considerably greater S/N for the purposes of modeling.
Since the STIS spectra cover only the wavelength region from H$\gamma$
to He~II $\lambda$4542 (plus a separate exposure at H$\alpha$; see
Sec.~\ref{Sec-data}), we make use of the FOS spectra in determining
the spectral type, but use only the STIS data for the modeling.  The
model spectra for the He~II $\lambda 4686$ line are, however, compared
to the FOS data for this line.  In presenting the blue-optical spectra.
we have spliced in the better STIS data in the wavelength region 4310\AA\ to
4590\AA. 

\begin{figure}
\epsscale{0.6}
\plotone{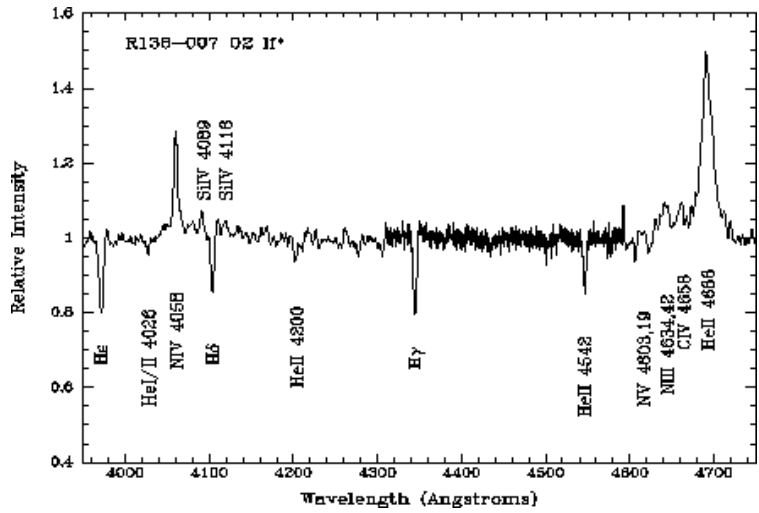}
\caption{\label{fig:r136007} R136-007.
We identify the major lines in the spectrum of R136-007.
The data are from the FOS observations of Massey \& Hunter (1998), except
for the region 4310\AA\ to 4590\AA, where we have
spliced in our higher S/N STIS spectrum.  
No combination of parameters led to a good fit, and we conclude that the
star is composite, consistent with the discovery of eclipses in the
light-curve
by Massey et al.\ (2002).
}
\end{figure}

{\it R136-007.}---This star was classified as O3 If*/WN6-A by Walborn \&
Blades (1997); Massey \& Hunter (1998) agreed with this spectral type.
The spectrum is shown in Fig.~\ref{fig:r136007}.  As noted earlier, the
``slash" designation simply denotes strong Of-like emission.  Walborn et
al.\ (2002) specifically excluded such ``slash" stars from their discussion,
but to us this division seems arbitrary and unnecessary.
The absorption
spectrum is straightforwardly interpreted as O2-O3 given the
strong He~II $\lambda 4542$ and lack of detection of He~I $\lambda 4471$.
Based upon the nitrogen emission spectrum, we would classify the star
as O2 If*, as N~IV $\lambda 4058$ emission is much stronger
than N~III $\lambda \lambda 4634,42$ emission.  C~IV $\lambda 4658$ emission
is
also present, suggesting this is particularly hot O2 star.  The presence
of N~V $\lambda \lambda 4603,19$ absorption and Si~IV $\lambda \lambda 
4089, 4116$ emission is also consistent with this designation.
The EW of He~II $\lambda 4686$ is -7\AA, somewhat 
short of the -10\AA\ (arbitrary)
boundary between WRs and Of stars.  The intrinsic emission will be slightly
stronger,
given the presence of a fainter companion (see below).
Its absolute visual magnitude M$_V=-6.9$
is similar to that of the O2 If* star LH101:W3-19 discussed above, and
is consistent with a supergiant designation.

We had difficulty matching the depth of the hydrogen and He~II lines with
our modeling, and are forced to conclude that this star is a composite.
Indeed, Massey et al.\ (2002) found light variations of this star that
are indicative of eclipses.  
Subsequent imaging has now identified a period
for the system, and a radial velocity study is in progress.

\begin{figure}
\epsscale{0.49}
\plotone{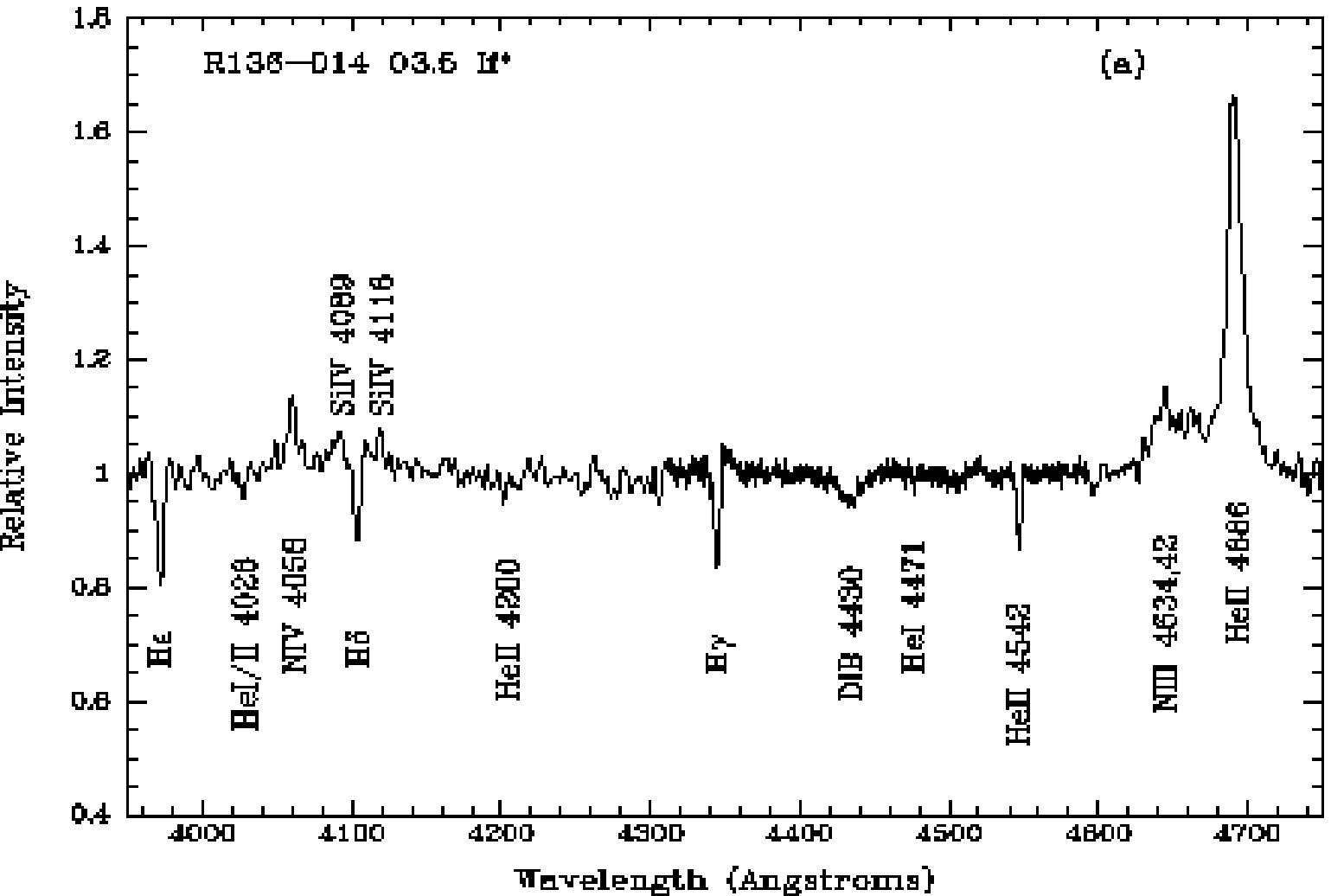}
\plotone{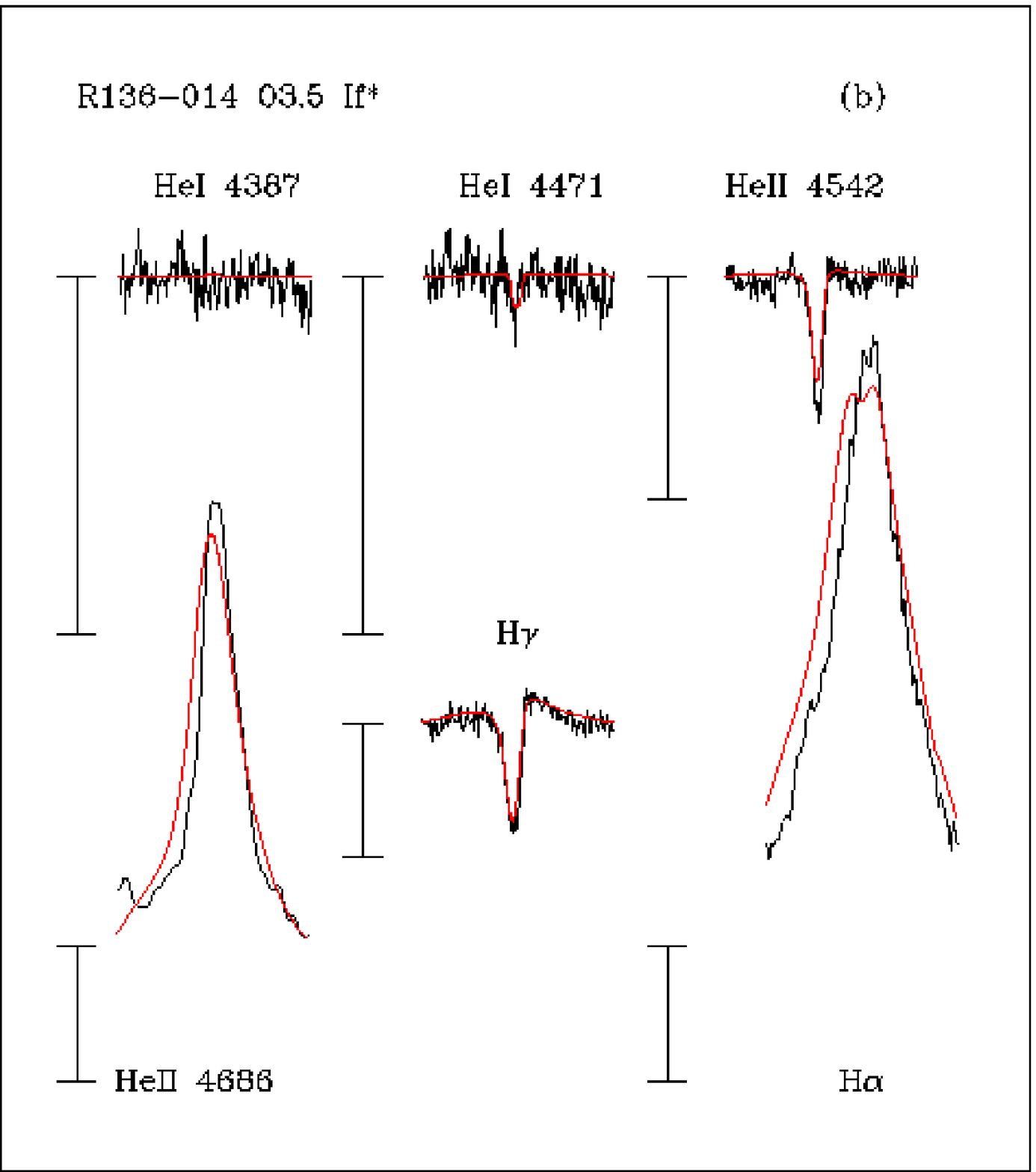}
\caption{\label{fig:r136014} R136-014.
The data in (a) come
from the FOS observations of Massey \& Hunter (1998), except
for the region 4310\AA\ to 4590\AA, where  we have spliced in
our higher S/N STIS spectrum.  In (b) the profiles are from the
STIS data, except for that of He~II $\lambda 4686$, which comes from
the FOS data.  The colors and symbols have the same meaning
as in Fig.~\ref{fig:av177}.  Problems with the STIS wavelength zero-point
preclude determining an accurate radial velocity; 
a rotational
broadening $v \sin i$ of 120 km s$^{-1}$ was used in making this comparison.
}
\end{figure}

{\it R136-014.}---This star was classified as O4~If by Melnick (1985),
under the designation Mk37) and as O4~If+ by Massey \& Hunter (1998).
Although He~I $\lambda 4471$ is {\it weakly} present in our STIS spectrum
(Fig.~\ref{fig:r136014}a, the EW is only 40m\AA, and it is difficult
to see how it could have been unambiguously detected in these earlier
studies. We derive a value of $\log W'=-1.0$, suggesting a type earlier
than O4.  Walborn et al.\ (2002) reclassify the star (under the name
``MH14")  as O3.5 If*, given the equal strengths of the N~IV $\lambda
4058$ and N~III $\lambda \lambda 4634,42$ emission lines.  The He~II
$\lambda 4686$ line has an EW of -7\AA, and the star's absolute magnitude,
M$_V=-6.4$, is consistent with it being a supergiant (Conti 1988).

The fits were relatively straightforward, and the comparison between the
model and observations are shown in Fig.~\ref{fig:r136014}b.  Note that
although the He~II $\lambda 4686$ was not used in the process, the
agreement between the model and the observation is quite good.

\begin{figure}
\epsscale{0.6}
\plotone{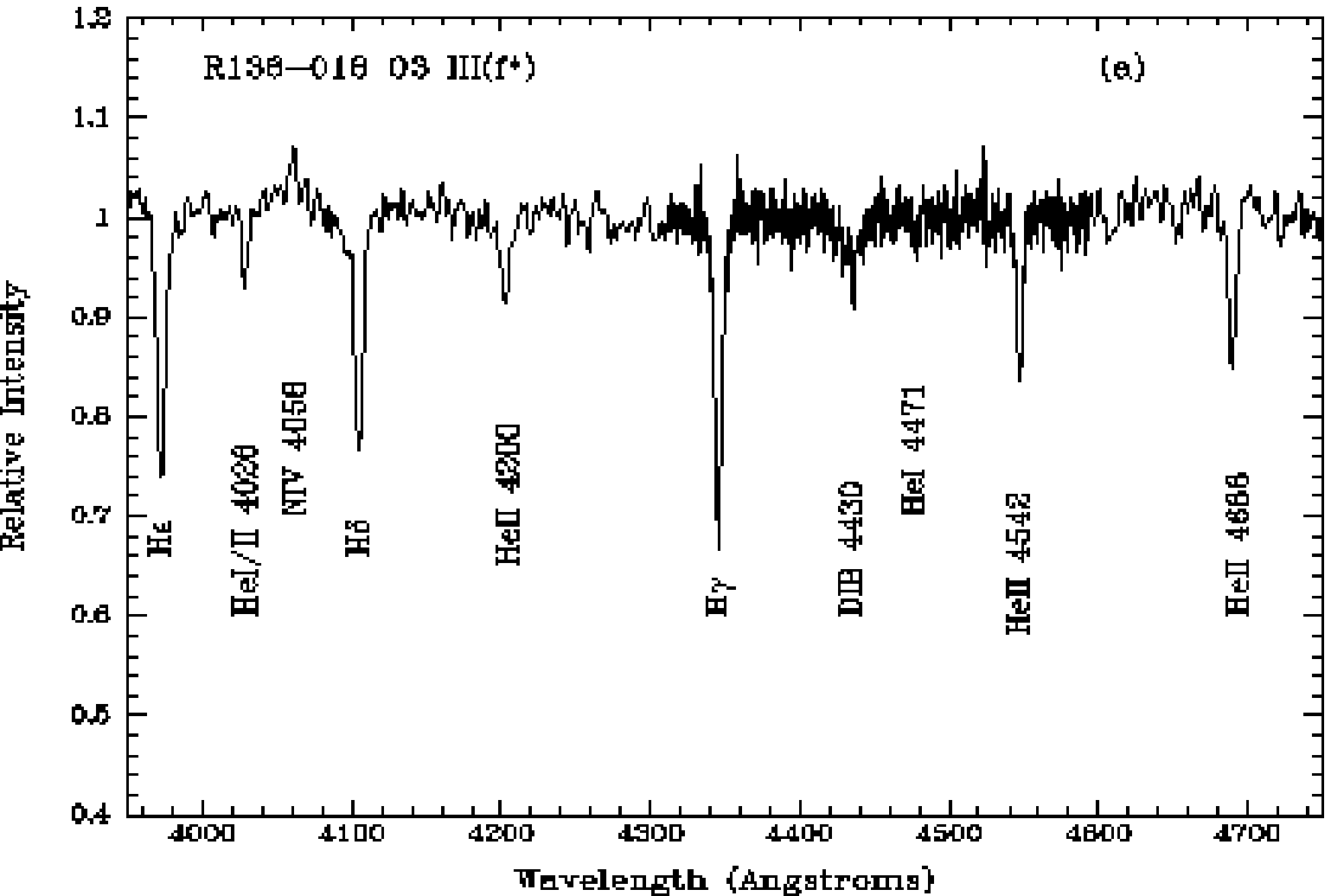}
\plotone{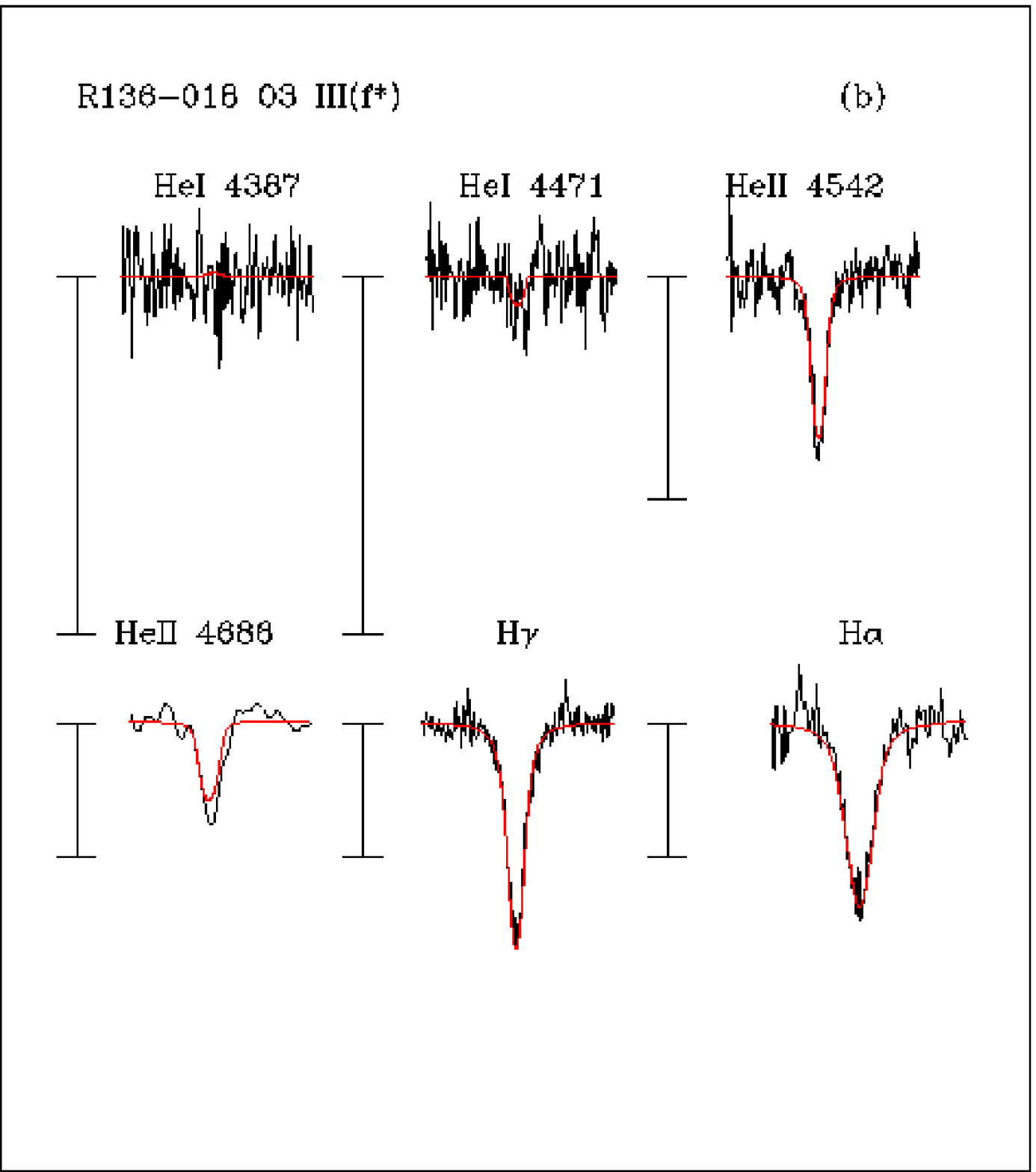}
\caption{\label{fig:r136018} R136-018.
Same as Fig.~\ref{fig:r136014}. 
A rotational
broadening $v \sin i$ of 180 km s$^{-1}$ was used in making this comparison.
}
\end{figure}

{\it R136-018.}---This star was classified as 
O3 III(f*) by Massey \& Hunter (1998), and this spectral type was
repeated by Walborn et al.\ (2002).  We show the spectrum in 
Fig.~\ref{fig:r136018}a.  He~I $\lambda 4471$ is present with an
EW of 90m\AA; we measure $\log W'=-0.87$.  NIV $\lambda 4058$ is present
in emission (EW=-750m\AA), 
although NIII $\lambda \lambda 4634,42$ is not; the N emission
lines would therefore allow either an O2 or O3 classification according
to Walborn et al.\ (2002), while the strength of He~I would argue for the
later type.  The strength of He~II $\lambda 4686$ absorption and overall
weakness of the nitrogen emission spectra argues that the star is a dwarf
or giant rather than a supergiant.  Its absolute magnitude, M$_V=-5.9$,
is intermediate between that expected for a dwarf and a supergiant (Conti 1988),
and so we retain the O3~III(f*) classification.

The fitting of this star required only two model runs to obtain an
excellent fit; we show the agreement between the observation and model
spectrum in Fig.~\ref{fig:r136018}b.  The surface gravity is found to be
consistent with a giant.

\begin{figure}
\epsscale{0.6}
\plotone{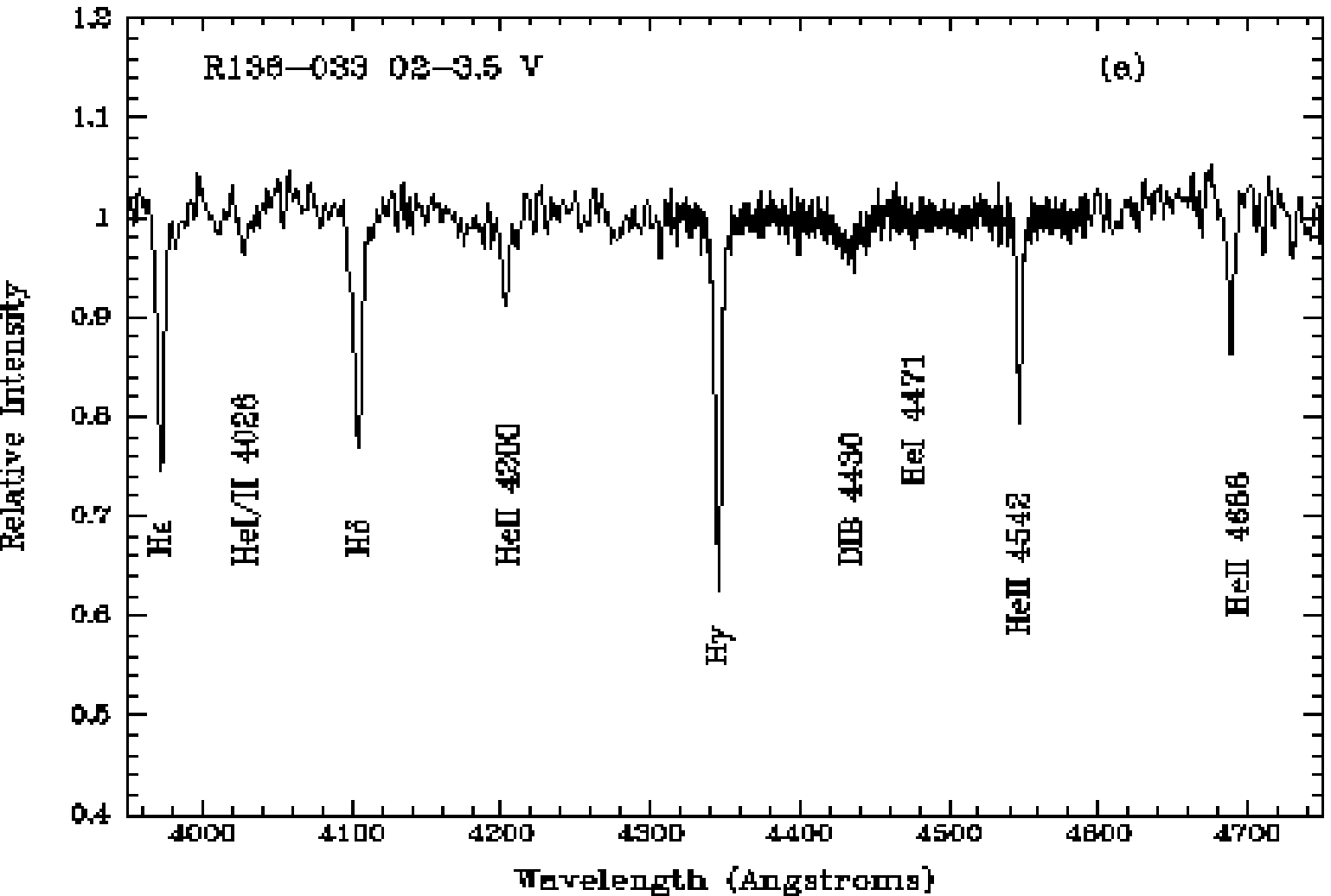}
\plotone{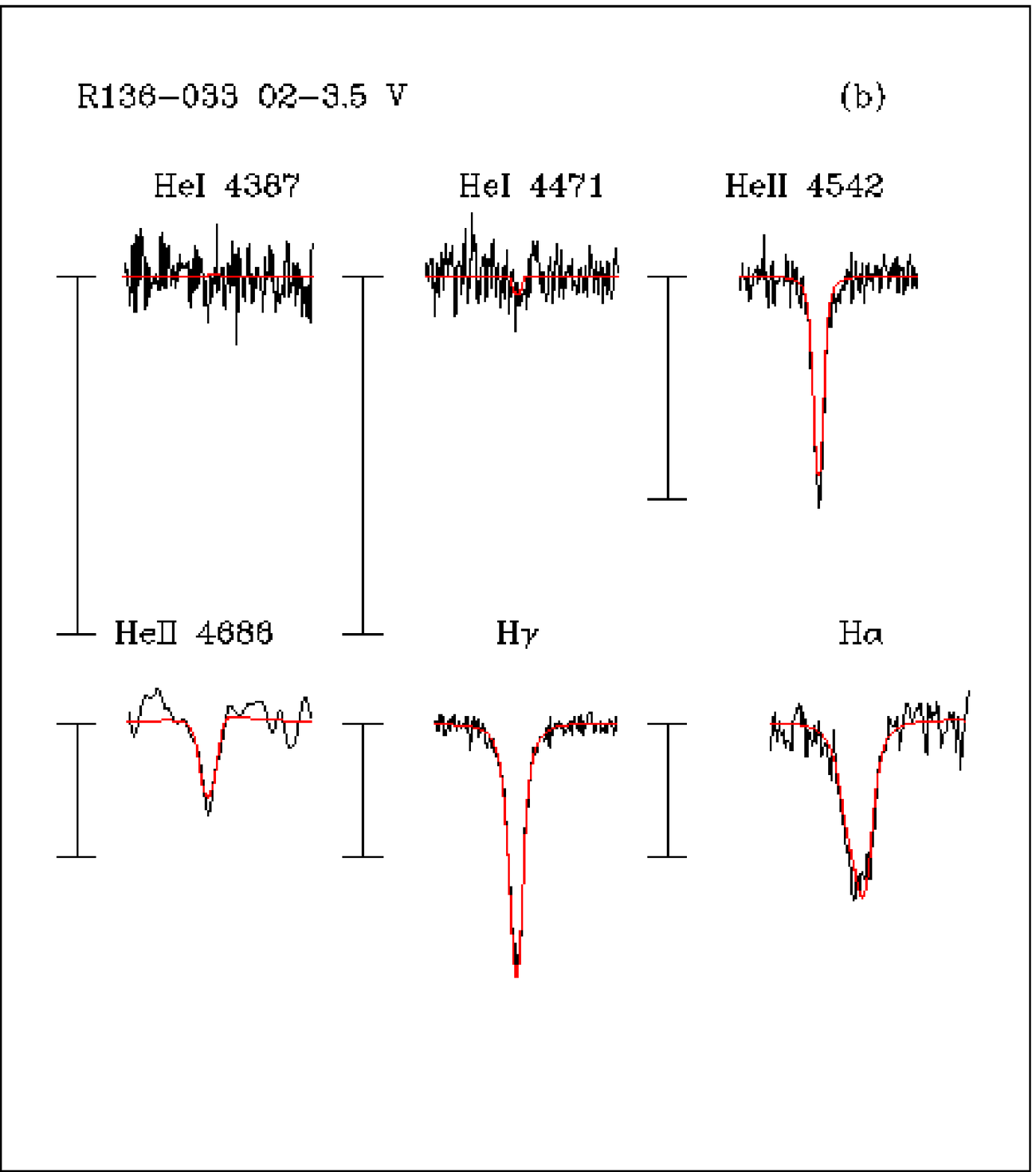}
\caption{\label{fig:r136033} R136-033.
Same as Fig.~\ref{fig:r136014}.
A rotational
broadening $v \sin i$ of 120 km s$^{-1}$ was used in making this comparison.
}
\end{figure}

{\it R136-033.}---This star was classified as O3~V by Massey \& Hunter
(1998).  He~I $\lambda 4471$ is marginally detected, with an
EW$\sim50$m\AA\ and $\log W'=-1.2$ (Fig.~\ref{fig:r136033}a).  
By the ``old" criteria, we would
call this spectral subtype O3.  Neither N~IV $\lambda 4058$ nor N~III
$\lambda \lambda 4634,42$ emission is detectable in our FOS spectrum,
and so it is difficult to know how to apply the criteria suggested by
Walborn et al.\ (2002) based upon the relative intensities of these
lines.  (We will return to issue in general in Sec.~\ref{Sec-O3s}.)
For now, we will call the star an O2-3.5~V.  The absolute magnitude
M$_V=-5.1$ is consistent with the star being a dwarf.

The fitting for this star was also straightforward, and the results are
shown in Fig.~\ref{fig:r136033}b.  The surface gravity for this star is
found to be similar that of R136-018 ($\log g=3.75$), 
despite  the fact that R136-033 is a dwarf and R136-018 is a giant.
We note in Sec.~\ref{Sec-results} that there is no clear distinction
in the surface gravities of giants and dwarfs in general among the
earlier O stars.
We list its parameters in Table~\ref{tab:results} as if the presence
of He~I $\lambda 4471$ were real, but possibly these should be viewed
as upper limits.

\subsection{Agreement with Other Wavelength Regions}

\subsubsection{Far Ultraviolet}
\label{Sec-FUV}

\begin{figure}
\epsscale{0.6}
\plotone{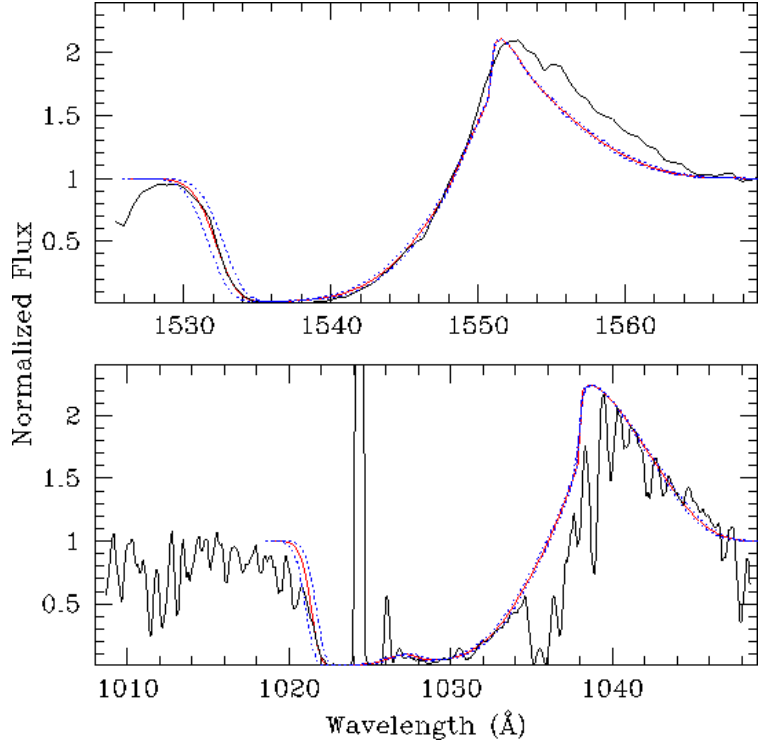}
\caption{\label{fig:fabio} 
CIV vs OVI Terminal Velocities.  For the star
Sk$-67^\circ$22 we show CIV profile (upper panel) and OVI profile
(lower panel) along with our adopted SEI fits.  In both cases we found
a terminal velocity $v_\infty=2650$ km s$^{-1}$.  The dotted curves
(blue in the on-line version)
show the sensitivity of the velocity to a change of $\pm$100~km~s$^{-1}$.
Since there has been no correction for underlying 
photospheric absorption for CIV, the poor fit on the long wavelength
side of CIV is expected; it is the short wavelength side which defines
the terminal velocity. Similarly the agreement with the
OVI could be improved by
better correction for interstellar absorption, but the short wavelength side
demonstrates that the terminal velocity is well determined.  Note the 
presence of Lyman $\beta$ emission in the OVI profile.
}
\end{figure}

There are five stars in our sample for which we have {\it FUSE} data.  Of
these, one star is composite (AV~476), and we do not consider this
further.  For the remainder, we performed SEI fitting of the OVI 
in the same manner as was carried out on the other UV data.  In
Fig.~\ref{fig:fabio} we compare the fits of the OVI line to that of CIV
for the star Sk$-67^\circ$22.  In each panel the red line shows the
adopted $v_\infty$=2650 km s$^{-1}$, with the dotted blue lines showing
$\pm$100 km s$^{-1}$.  We compare the velocities for all four stars
in Table~\ref{tab:ovi}; clearly the agreement is excellent.

We can of course go much further with both the {\it FUSE} and HST/UV
data.  Some studies of the physical properties of O-type stars have
relied solely on the ultraviolet region for modeling (see, for
example,  Martins et al.\ 2004, Garcia \& Bianchi 2004, Bianchi \&
Garcia 2002, Pauldrach et al.\ 2001), while other recent studies
have included both the optical and ultraviolet regions in their
modeling efforts (for instance, Crowther et al.\ 2002, Hillier et
al.\ 2003, Bouret et al.\ 2003).   In general the UV provides the
much-needed diagnostics of the stellar wind lines (particularly the
terminal velocities) but the various lines that have been proposed
as particularly temperature sensitive (for example, C~III $\lambda
1176$ to C~IV $\lambda$1169 by Bouret et al.\ quoting S. R. Heap
in preparation; Fe~IV to Fe~V by Hillier et al.\ 2003, Ar~VI $\lambda
1000$ to Ar~VII $\lambda 1064$ by Taresch et al.\ 1997, etc.) are
usually useful over a very limited temperature range, and, more
importantly, good matches between the model and stellar spectra
require determining the abundances of the particular element.  The
lines in the UV region are also not particularly sensitive to the
surface gravity; see discussion in Pauldrach et al.\ (2001).  We
believe that there is much to be gained by fitting the UV {\it and}
optical region, as has been shown in the past by some (see in
particular Taresch et al.\ 1997), but has generally been ignored
by others.  Although the present paper (and Paper I) has made some
{\it use} of the UV (in terms of determining the terminal velocity),
our study has been wedded to that of the optical lines, as these
lines have the greatest sensitivity to $\log g$ and $T_{\rm eff}$.
Nevertheless, the UV region contain differing ion states of various
metal lines that can be effectively employed.  In addition, wind
clumping will affect the UV and the optical in different ways.  The
most robust answers will come from fully utilizing both the optical
and UV, and we plan to make further use of our beautiful UV spectra
(both {\it HST} and {\it FUSE}) in a subsequent paper

\subsubsection{The NUV: He~I $\lambda 3187$ and He~II $\lambda 3203 $}

In Paper~I we (re)introduced the He~I $\lambda 3187$ and He~II $\lambda 3203$
lines as an interesting check on the models and effective temperatures we
adopt.  Morrison (1975) was the first to call attention to the usefulness
of these lines, noting that the He~II $\lambda 3203$ ($n$=3-5)
line was the only
accessible He~II line that did not involve transitions from $n=4$.
The He~I $\lambda 3187$ line is a triplet 2$^3$S-4$^3$P$^o$, similar to
$\lambda 4471$ (2$^3$P$^o$-4$^3$D).  

\begin{figure}
\epsscale{0.6}
\plotone{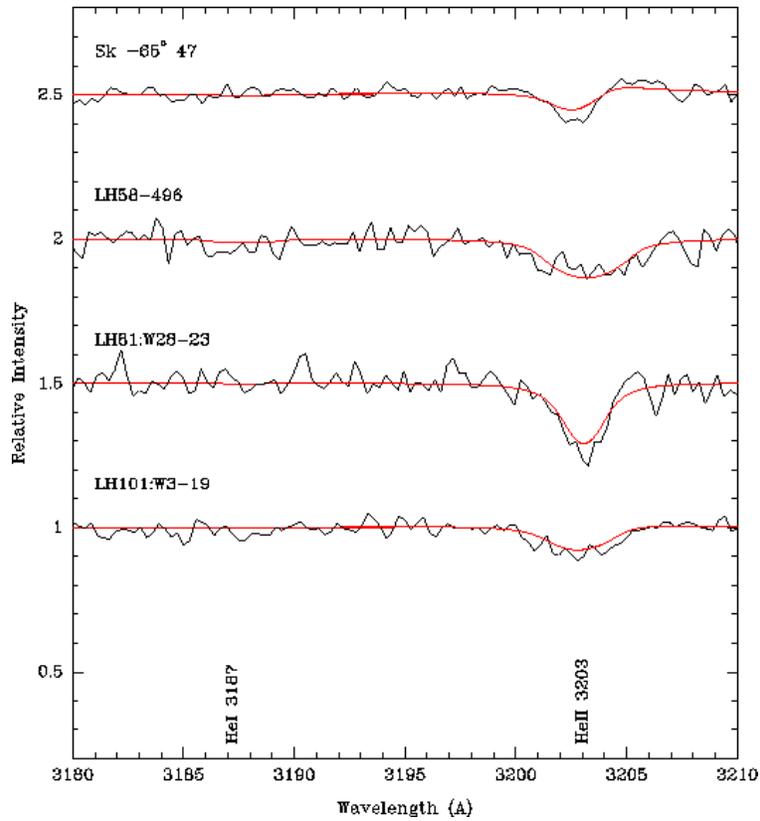}
\caption{\label{fig:3200} The NUV region.
The region containing the He~I $\lambda 3187$
and He~II $\lambda 3203$ lines are shown for the four non-composite
stars in our sample for which we have data in this region.  Although
we did not use this region in obtaining the final parameters, each star
shows good agreement with the corresponding model.}
\end{figure}

Of the six stars in this paper for which we have observations in the NUV
region, two are composites.  In Fig.~\ref{fig:3200} we show the spectra
of the other four.  Although we did not use this region in determining the
physical parameters of these stars, in all four cases there is good agreement
between the observed spectra and the models for He~II $\lambda 3203$.  In Paper
I we also demonstrated good agreement for the two stars for which we had
NUV data.   We conclude that the He~II $\lambda 3203$ line yields answers
which are consistent with that of He~II $\lambda 4200$ and $\lambda 4542$.
None of the stars were sufficiently late for He~I $\lambda 3187$ to be
detected at our S/N, consistent with the output of the stellar atmosphere
models.  Since He~I $\lambda 4471$ and, in some cases, even
He~I $\lambda 4387$ is  measurable in all of these stars (with the 
possible exception of LH101:W3-19; see Sec.~\ref{Sec-stars}), we
note that while the NUV region provides reassurance on our fitting
procedure, the additional information added is limited.

\section{The Earliest O Stars}
\label{Sec-O3s}

Although the main goal of our paper is to derive physical parameters, we have
assembled some of the highest quality optical spectra on some of the 
earliest O stars known, and we would be remiss to not use this to 
comment upon the spectral classification scheme.  Conti \& Alschuler (1971)
provided equivalent width
measurements of the primary spectral classification lines
He~I $\lambda 4471$ and He~II $\lambda 4542$ for a large sample of O stars
that had been classified in the ``traditional way" (visually
comparing the spectra of program stars to the spectra of
spectral standards) and provided a 
{\it quantitative} scale which could be used to distinguish the spectral
subtypes from one another; i.e., an O8 spectral type had a value of
$\log W'$ = EW (He~I $\lambda 4471$) / EW (He~II $\lambda 4542$) between
0.10 and 0.19, while an O8.5 star had a value of $\log W'$ between
0.20 and 0.29.   This scheme was slightly revised for the earliest
types by Conti \& Frost (1977), who used the criteria that O4 stars have
$\log W'<-0.60$ while in O3 stars He~I $\lambda 4471$ is ``absent".
(A summary of the $\log W'$ appears in Table 1-3 of Conti 1988.)  Of course,
this leads to the O3 classification as being degenerate; i.e., at a given
surface gravity, stars with any temperature 
above some value might be classified as ``O3".  Indeed, when higher
signal-to-noise spectrograms were obtained by
Kudritzki (1980) and Simon et al.\ (1983) of the prototype O3 stars
HD~93205, HD~93128, and HD~303308 (Walborn 1971a, 1973b), weak He~I $\lambda
4471$ was revealed, with EWs of 75-250 m\AA, and $\log W'$= -1.1 to -0.5.
(HD~303308, with a $\log W'=-0.5$, would today be classified as an O4 star;
see Table 3 of Walborn et al.\ 2002.  Another O3 star for which Simon et al.\
1983 detected He~I $\lambda 4471$ is HD~93129A, but today we recognize that
this is a composite spectrum; see Nelan et al.\ 2004.)
However, to date no one has actually provided a quantitative distinction 
in the He~I/He~II ratio between spectral types O3 and O4.

Detection of weak He~I $\lambda 4471$ is crucial for an accurate temperature
determination using our method, and we can ask what is our expected
detection limit?  With a S/N of 400 per 1.4\AA\ spectral resolution element,
we should be able to reliably detect (at the 3$\sigma$ level) a line
whose equivalent width is $ 3\times (1/400) \times 1.2$\AA, or about
0.009\AA\ (9m\AA).   This is consistent with our measurements of the 20m\AA\ in
the spectra of some stars (i.e., BI~237).  For the R136 stars, with a
S/N of 100 per 0.4\AA\ spectral resolution element, a limit of about 20m\AA\ is
expected.
In a few cases, the STIS spectra had worse S/N, and only an upper limit to
He~I $\lambda 4471$ can be detected,  and we note such cases.  What about the
effects of normalization on such weak lines?  In normalizing our spectra, we
used a low order cubic spline fit through line-free regions.  Care was taken
with this procedure.  Comparing the continuum levels in our normalized spectra
in the region around He~I $\lambda 4471$ we find that this procedure worked 
very well, with some very early stars (BI~237) the normalization proved
good to 0.05\%; in general, we the agreement was good to a few tenths of
a percent.   What effect will that have on the measured equivalent widths?
If our continuum placement was in error by 0.5\% (which, we believe, would
be an extreme case), then we would measure a 20m\AA\ line as having an EW
of 28m\AA.  Thus we expect that any error introduced by our normalization
will have an effect comparable to that of the large but
finite S/N of the spectra.

Walborn et al.\ (2002) 
attempted to reduce the O3 degeneracy problem by using the 
N~IV $\lambda$ 4058 and N~III $\lambda 4634, 42$ emission lines to
extend the spectral classification to spectral type O2 (``N~IV $>>>$
N~III") with the introduction of an intermediate type (O3.5) for which
N~IV $\sim$ N~III.  (The O2 class would then be the degenerate subtype.)
The classification criteria are not quantitative,
and the system 
still relies upon statements such as ``very weak He~I" or ``no He~I"
as secondary criteria.  The criticism has been made (both here and in Paper~I)
that relying upon the relative strengths of the optical nitrogen lines lacks
a solid theoretical underpinning; i.e., although a unique spectral subclass
may be defined, it is not clear that a star's effective temperature
is the primary thing that distinguishes O2's from O3's in this scheme.
The mechanism for N~III emission in O stars has been well described
by Mihalas \& Hummer (1973), who demonstrated the N~III $\lambda 4634,42$
lines will come into emission even in a static, non-extended atmosphere
if the effective temperature is sufficiently high, due to a complicated
NLTE effect.  Thus the presence of N~III $\lambda
4634,42$ in dwarfs and giants should be related to effective temperature.
However, for stars with significant mass-loss rates, an additional process
known as the Swings mechanism (Swings 1948) will come into play, 
enhancing
the N~III emission (see discussion in Mihalas \& Hummer 1973).  
No such detailed analysis exists for the N~IV $\lambda 4058$ line, although 
Taresch et al.\ (1997) show its behavior as a function
of effective temperature holding the surface gravity constant and maintaining
the same H$\alpha$ profile by slightly varying the mass-loss rate.  Indeed,
the Taresch et al.\ (1997) analysis of HD 93129A showed the power of
using the N~III and N~IV emission lines as a constraint
on the effective temperature, {\it if} other parameters (such as $\log g$
and $\dot{M}$) were constrained by other observations.

In order to provide a more quantitative assessment of the classification criteria, in 
Table~\ref{tab:o3s} we provide for the first time equivalent width measurements
and ratios for the He~I $\lambda 4471$/He~II $\lambda 4542$ absorption lines,
and N~III $\lambda 4634,42$ / N~IV $\lambda 4058$ emission lines, for a
sample of the earliest O-type stars.  We have classified the stars using
the Walborn et al.\ (2002) criteria; otherwise, all of the stars earlier
than O4 would simply be described as ``O3".  Some of the dwarfs in Paper~I
had no detected N~III {\it or} N~IV emission and were simply called ``O3~V"; 
here we ``revised" the type to ``O2-3.5~V".  (The need to detect these lines,
which are weak or non-existent in dwarfs, particularly at low metallicities,
is an obvious drawback to the new classification scheme.)

First, let us note that the
dividing line between O3's and O4's occurs at roughly a $\log W'=-0.8$, and
we would propose this as an improvement over the old description that
He~I be ``absent".  However, even a casual inspection of Table~\ref{tab:o3s}
shows the effect that mass-loss has on the derived effective temperature
for a given He~I/He~II ratio.  LH101:W3-24, an O3~V((f+)) star,
and BI~237, an O2~V((f*)) star, both have similar $T_{\rm eff}$ and
$\log g$, although their He~I/He~II ratios differ by a factor of 6!
This can be readily attributed to the significantly greater mass-loss rate
of BI~237 due to its larger radius, mass, and luminosity. (Recall from the
discussion in Paper~I and here that the effect of stellar winds will be
to {\it decrease} the derived effective temperature for a given HeI/HeII
ratio, in part because of the filling in of HeI due to emission produced in
the wind.)  

\begin{figure}
\epsscale{0.6}
\plotone{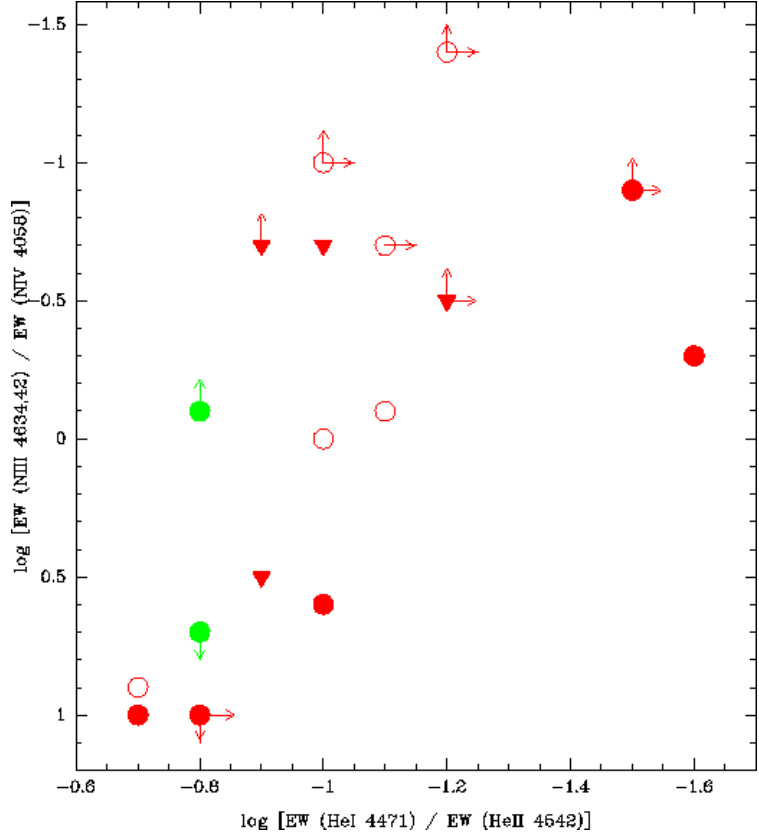}
\caption{\label{fig:o3lines} NIII/NIV vs HeI/HeII for O2-3.5 stars.  
The log of the ratio
of the equivalent widths (EWs) of N~III $\lambda 4634,42$ to N~IV $\lambda 4058$ emission is shown as a function of the log of the ratio of the EWs
of He~I $\lambda 4471$ to He~II $\lambda 4542$.  Typical errors (as given
in Table 7) are 0.2 dex on each axis, but are not shown to prevent confusion
with lower and upper limits.
Dwarfs are shown as filled
circles, giants are shown as filled triangles, and supergiants are shown as
open circles.  
Red symbols are for the LMC stars; green for the SMC.
}
\end{figure}

In Fig.~\ref{fig:o3lines}a, we show the relationship between the HeI/HeII and
NIII/NIV ratios.  
There is a reasonable correlation in this figure, in that
stars of a given luminosity class which
have a small HeI/HeII ratio (right side of the diagram) also tend to show
a smaller NIII/NIV emission line ratio (top of the diagram).  The
correlation is quite good for the dwarfs (filled circles),  with
giants (triangles) and supergiants (open circles) 
exhibiting more scatter.  To us this is as expected.
Although HeI absorption is affected by stellar winds (by the filling in of
emission), the strength of the 
nitrogen {\it emission} lines are likely to be more
affected by mass loss (due to, for instance, the Swings mechanism,
as discussed above).

\begin{figure}
\epsscale{0.49}
\plotone{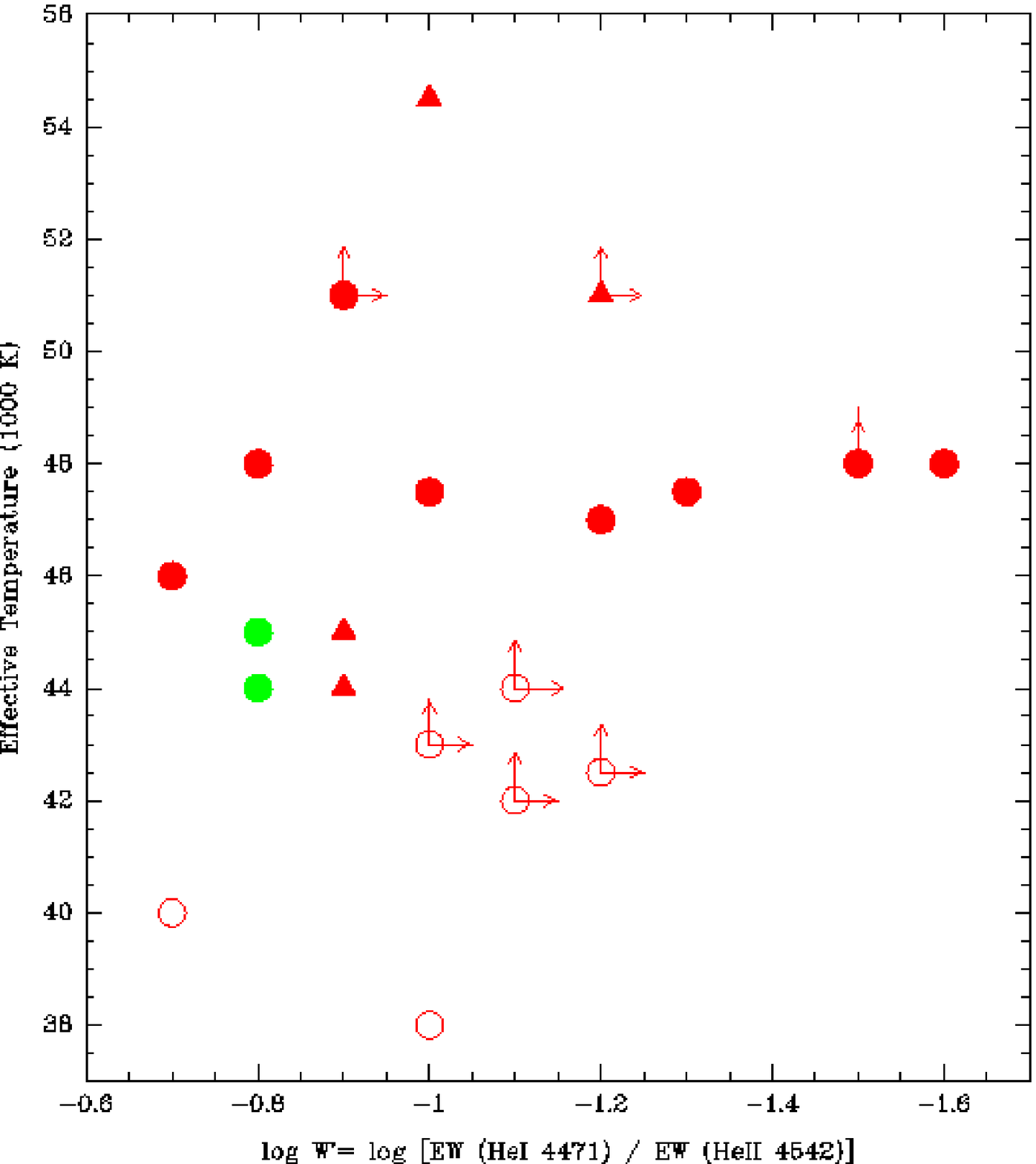}
\plotone{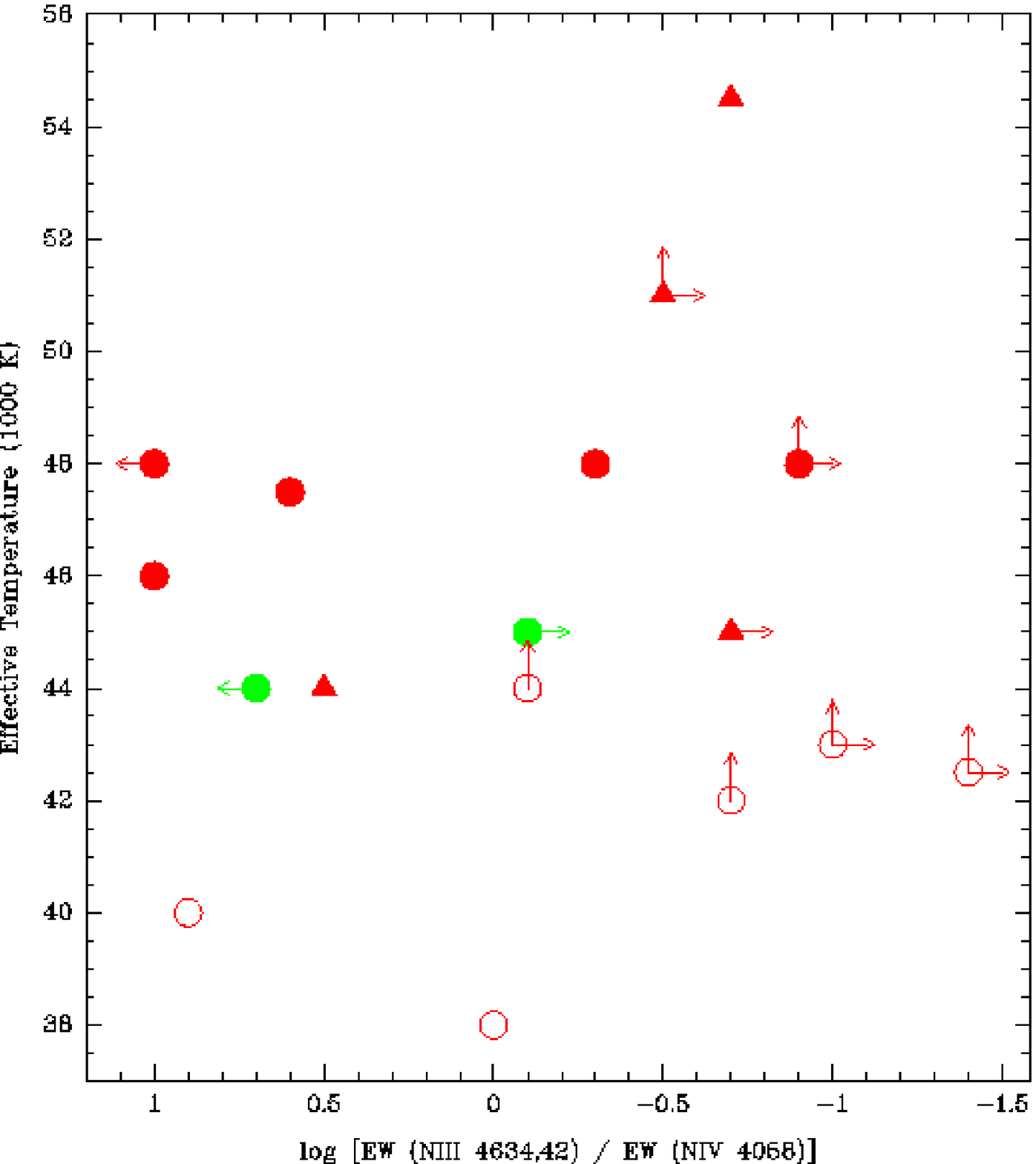}
\caption{\label{fig:o3teff} Effective temperatures of the earliest-type
O stars (O2-O3.5) is shown as a function of 
(a) the HeI/HeII line ratio, and (b) the NIII/NIV emission
line ratio.  Again, the uncertainty in the line ratios are typically
0.2 dex; see Table~7.
The symbols have the same meaning as in Fig.~\ref{fig:o3lines}.
}
\end{figure}

We address this in a more quantitative way in Fig.~\ref{fig:o3teff}
where we show the correlation of effective temperature with
(a) the HeI/HeII ratio and
(b) the NIII/NIV ratios.  Again, this considers only the earliest-type
O stars.  For the dwarfs (filled circles) there are very large changes
in both HeI/HeII and NIII/NIV with little change in effective temperature.
In other words, the extension of the spectral classification
through O2-O3.5 for the dwarfs is unwarranted in terms of effective
temperature.  For the giants (denoted by triangles) and supergiants 
(open circles) there
is more of a correlation of effective temperature with either spectroscopic
criteria, but neither is very good. Consider, for example, the star
with the {\it hottest} temperature in our sample, LH64-16.  An independent
analysis (using the same optical data, however) yielded a similar
effective temperature (Walborn et al.\ 2004) as what we derived in Paper~I.
However, the giant R136-018 has a NIII to NIV emission ratio which is
at least as small, 
but has nearly the {\it coolest} effective temperature for any of
the giants with a measured NIII/NIV ratio in our sample!  (Recall from
the discussion above that the lack of NIII emission in the spectrum of
R136-018 would allow either an O2 or O3 designation; the argument for
the O3 classification came about from the strength of He~I absorption.)
Of course, we argue in Paper~I that the star LH64-16 is likely the
result of binary evolution, and should not be considered a prototype
of a new type (cf.\ Walborn et al.\ 2002).

\begin{figure}
\epsscale{0.50}
\plotone{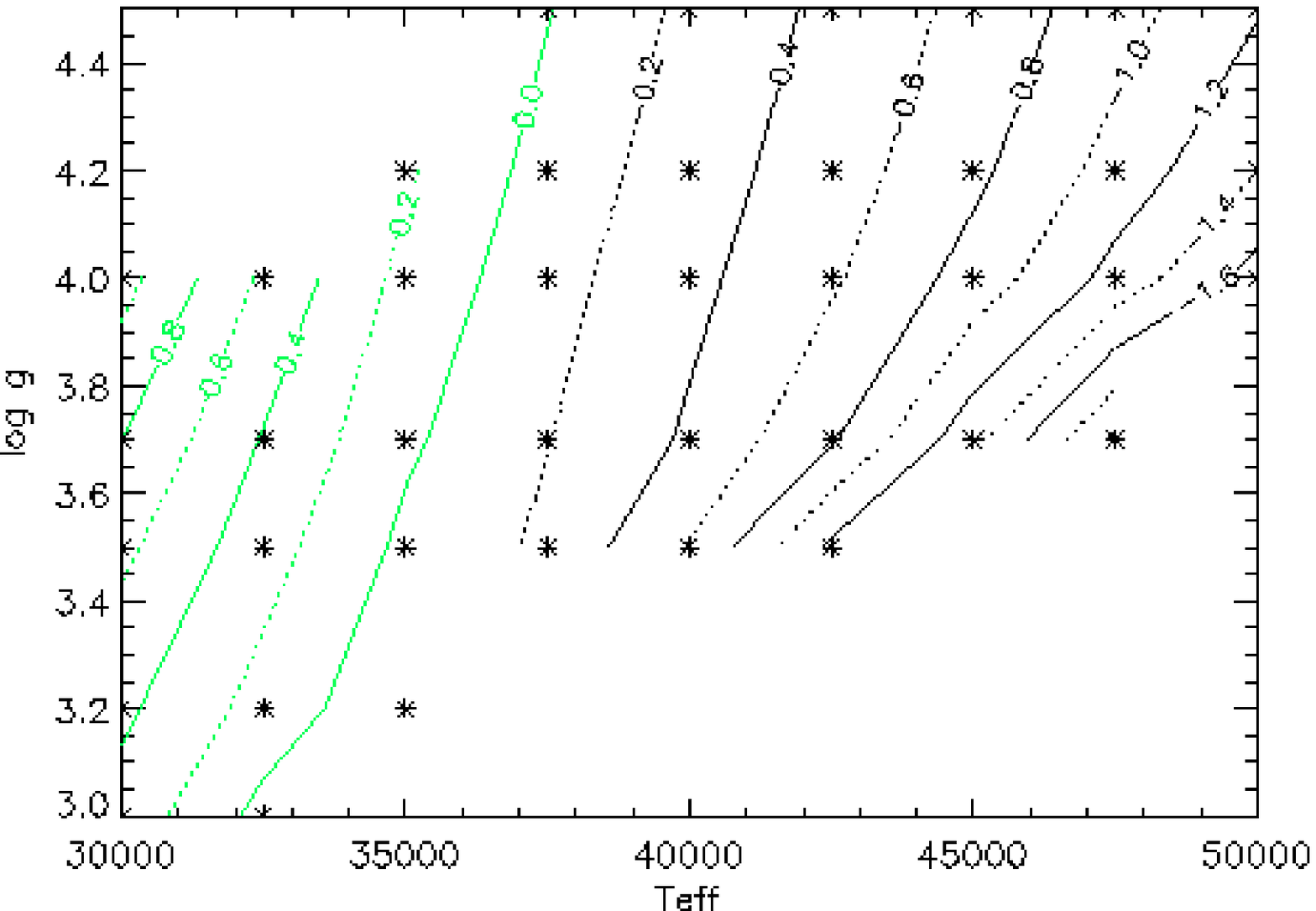}
\plotone{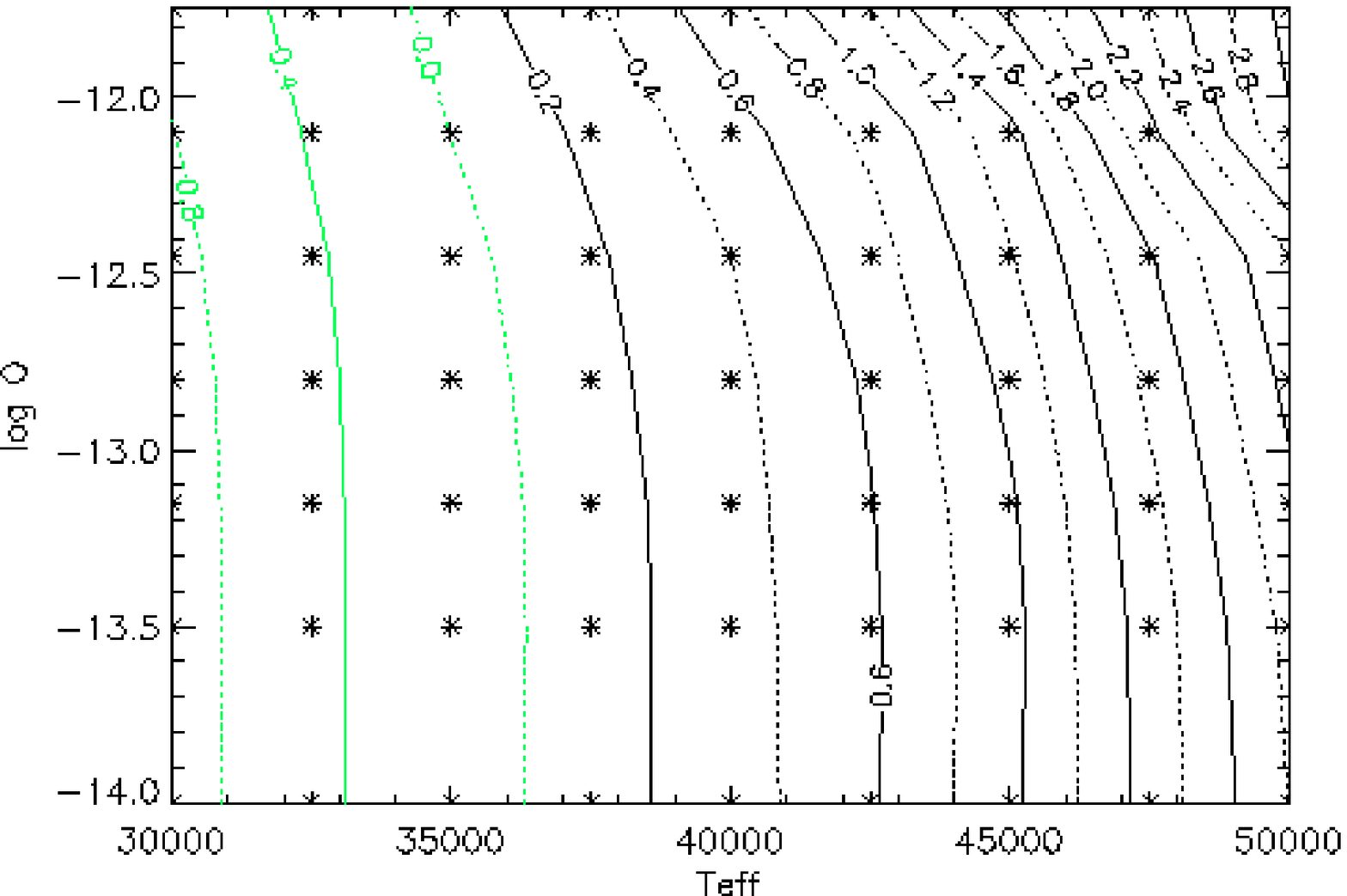}
\caption{\label{fig:jo}The dependence of $\log W'=\log$[EW(He~I$\lambda 4471$)/(He~II$\lambda 4542$) on effective temperature and mass-loss rates according to FASTWIND.  The contours show values of constant
$\log W'$ (green for positive, black for negative, with the values indicated).
In (a) we show the dependence of $\log W'$ on $\log g$ and $T_{\rm eff}$
with the value of $\log Q$ (as defined below) held constant
(-12.45, correspond to a
mass-loss rate of $10^{-6} M_\odot$ yr$^{-1}$ for $v_\infty=2000$ km s$^{-1}$
for a star with radius 10$R_\odot$).  Although 
$\log$ W' is primarily a temperature indicator over {\it most} of parameter
space, it becomes a sensitive function of surface gravity for the hottest 
stars.
(b) There is also a strong
dependence of $\log W'$ 
on the mass-loss rates for hot stars with high rates. The value $Q$
is defined as $\dot{M}/(v_\infty * R)^{1.5}$.   The data here all have
$\log g=3.8$.}
\end{figure}

The degeneracy of $T_{\rm eff}$ with $\log W'$ is somewhat surprising,
since afterall this is the basis for the original MK types (although not
the extension to O2-3.5 described by Walborn et al.\ 2000).  By examining
the large grid of models computed as part of the Puls et al.\ (2005) study,
we came to understand that the effect is due primarily to the dependence
of $\log W'$ on {\it surface gravity} for stars of the hottest temperatures.
This is demonstrated in Fig.~\ref{fig:jo}(a).  There is also a significant
dependence of $\log W'$ 
upon the mass-loss rate for the hottest stars with the highest
rates, and highest temperatures (Fig.~\ref{fig:jo}b).

It is also worth noting that the spectroscopic luminosity criteria proposed
for the earliest types (based upon the strength of
N~III $\lambda\lambda 4634,42$ and
He~II $\lambda 4686$) do not prove to be
a reliable indicator of the effective surface gravities. Although 
the supergiants in Table~\ref{tab:o3s} do have lower surface gravities
than the dwarfs, one cannot differentiate the dwarfs and giants on the
basis of $\log g$.  We return to this point in Sec.~\ref{Sec-results}.

Thus in terms of the {\it physical properties} of early-type O stars, and
in particular the effective temperatures, we are forced to conclude that
there is little benefit to the extension to spectral type O2 proposed by
Walborn et al.\ (2002).  For a star of the same effective temperature
and similar surface gravity either the HeI/HeII ratio or the
NIII/NIV ratio can vary by essentially the full range found between O2
and O3.5 (for example, BI~237 and LH101:W3-24).  We believe that careful
{\it modeling} of the N~III and N~IV emission lines, in concert with other
lines will provide useful diagnostics (following Taresch et al.\ 1997),
and modifications to FASTWIND to allow this are planned.  
However, our work here suggests that it is naive to expect that any
one line ratio (N~III/N~IV, or even He~I/He~II)
gives insight into the effective temperature of the star.
{\it Although the earliest O-type stars (what would have
classically been called ``O3") contain stars with a significant range of
effective temperatures, no one spectroscopic diagnostic (such as the 
He~I/He~II
or N~III/N~IV line ratios) 
provides a good clue as to the effective temperature without
knowledge of the other physical parameters 
(mass-loss rate, surface gravity).}  This general principle has been
previously emphasized by others, notably Sellmaier et al.\ (1993).

\section{Results}
\label{Sec-results}

With the work described here and Paper~I, we have attempted to model
the spectra of 40 O-type stars in the Magellanic Clouds.  We succeeded
for 13 SMC stars (8 dwarfs, 2 giants, and 3 supergiants) and 20 LMC
stars (9 dwarfs, 4 giants, and 7 supergiants), with the other 7 stars
showing signs of composite spectra\footnote{For one of these stars,
AV469, Evans et al.\ (2004), obtained a fit that they deemed satisfactory.
Possibly we were too fussy in being dissatisified with our fit, and
declaring this a composite, but a radial
velocity study is underway.}.  This sample is now large enough
for us to determine preliminary effective temperature scales for O
stars in the SMC and LMC, and compare these to that of the Milky Way,
using the sample of stars analyzed by by Repolust et al.\ (2004) using
the same methods and models atmospheres. We also wish to make good on
our promise in Paper~I to examine the Wind-Momentum 
Luminosity relation
from
these new data, and, finally, to compare the spectroscopically derived
stellar masses with those derived from stellar evolutionary tracks.

In Table~\ref{tab:finalresults} we summarize the derived parameters for
our final sample of 33 stars.  In what follows we assume the fitting
errors as quoted in Paper~I, namely an uncertainty of about 1000K
(2-3\%) in $T_{\rm eff}$, 0.1~dex in $\log g$, and 20\% in $\dot{M}$.
(The later assumes that $\beta$ is the same for all stars and is precisely
known; the actual error on $\dot{M}$
 may be significantly higher, depending upon the
validity of this assumption.\footnote{A range of 0.7 to 1.3 is
reasonable for O-type stars.   Our inability to derive values for
$\beta$ when H$\alpha$ is in absorption produce a typical error
of $\pm 0.1$ dex (25\%), with a maximum error of $\pm 0.2$~dex, which
is considerable (65\%).  We can safely say that at worse the values
of $\dot{M}$ are uncertain by a factor of two if H$\alpha$ is in absorption;
however a 20-30\% error is more typical of what we expect.  This error
is comparable to the uncertainity we derive in the fits by varying
$\dot{M}$ while holding $\beta$
constant.})
In general our values for $v_\infty$ are good to 100 km s$^{-1}$
(5\%).  Using the propagation of errors analysis by Repolust et
al.\ (2004) [see their equation 8] we would then expect the uncertainty
in the derived stellar radius R to correspond to $\Delta \log R \sim
0.03$, or about 7\%, where we have allowed for a 0.1~mag uncertainty in
M$_V$.  (The uncertainty in the radii of Galactic stars is about twice
as great, given the much greater uncertainty in the distances and hence
a larger uncertainty in M$_V$.)  The total luminosity of the star is
uncertain by about 0.12~mag in M$_{\rm bol}$, or 0.05 in $\log
L/L_\odot$.

For the stars with only lower limits on $T_{\rm eff}$ we expect that
the values for the stellar radii, $D_{\rm mom}$, and $M_{\rm spect}$ are
all {\it approximately} correct; 
nevertheless, we do not use them in
the analysis, in order not to mix good values with bad.

In Fig.~\ref{fig:hrds} we show the location of our stars in the H-R diagram,
where we have over-plotted the evolutionary tracks of the Geneva
group (i.e., Charbonnel et al.\ 1993 for the SMC, and Schaerer et al.\ 1993
for the LMC).  For simplicity we show only the H-burning part of the
tracks.  Since the time of these calculations various improvements have
been made in the evolutionary code, such as the inclusion of the
Vink et al.\ (2000, 2001) prescription for mass-loss, improved opacities,
and, most importantly, the inclusion of the effects of rotation.  However,
a full set of these tracks are not available for low metallicities.
We do show the new tracks (for an initial rotation velocity of
300 km s$^{-1}$) 
for initial masses of $60 M_\odot$ and 
$40 M_\odot$ from Meynet \& Maeder (2004); we are grateful to Georges 
Meynet for making these tracks available to us.

\subsection{Effective Temperatures,  Surface
Gravities, and Bolometric Corrections}
\label{Sec-Teff}

In Fig.~\ref{fig:teffs} we show the effective temperatures as a function
of spectral subtype for the complete data set: SMC stars are shown
in green, the LMC in red, and the Milky Way in black, with
different symbols representing the
various luminosity classes.  Due to the constraints of limited
observing time,
and our emphasis on trying to understand the effect that metallicity has
on the physical parameters of stars, our LMC sample (intermediate
metallicity) is incomplete,
but biased towards the earliest spectral types, while the SMC
stars (low metallicity) was chosen to cover the full range of spectral
subtypes for direct comparison with the Milky Way 
sample of Repolust et al.\
(2004). 

First, we can see from Fig.~\ref{fig:teffs} that both in the Milky Way
and in the SMC that the effective temperatures of supergiants are 3000
to 4000 K cooler than dwarfs of the same spectral types for early O
stars (O4-O6). For stars of spectral type O8 and later, there is
essentially no difference in the effective temperatures of dwarfs and
supergiants.  In both galaxies the giants (denoted by triangles) appear
to follow the same sequence as the dwarfs (filled circles).  For the
early O stars (O4-O6) SMC stars are roughly 4000 K hotter than their
Galactic counterparts.  By spectral type O8 the difference is about
2000 K, and by O9.5 the data are consistent with no difference.
Thus, by coincidence the effective temperature scale of SMC {\it
supergiants} is very similar to that of Milky Way {\it dwarfs}.

This is consistent with what we expect by way of
the effects that stellar winds will have on the effective temperature scale
of stars: at higher temperatures, stars with high luminosities (supergiants)
will have higher mass-loss rates than stars of lower luminosity (dwarfs).
At higher metallicities (Milky Way) the mass-loss rates will be higher than
at lower metallicities (SMC).  By the later spectral types (O8-B0) the
differences become minor.  While this is expected
for dwarfs, as these stars have progressively weaker stellar winds with
spectral types (as the luminosity decreases with later spectral type
as well), it is somewhat
surprising for the supergiants, as even B0-2~I stars can still have very
high 
mass-loss rates. We can offer a few possibilities.
Perhaps the effect is caused by 
the fact that the 
continuum edges of H and He~I are becoming increasingly important
at cooler temperatures.  These edges would do their own ``blanketing", and
block out significant amounts of the flux.  Also at cooler temperatures
the peak flux occurs at longer wavelengths, which we expect to then
result in a decrease in the effects of wind blanketing.  But, in all
this we should keep in mind that we have included only one SMC late O-type
supergiant in our analysis, and that possibly additional data will
invalidate this effect.

The LMC data on the earliest spectral types (O2-O3.5) emphasizes again
how diverse a group of stars these are.  Nevertheless, some clear
patters emerge, and in particular the fact that the supergiants are
significantly ($\approx 5000$ K) cooler than dwarfs of similar spectral types.

With these trends in mind, we provide in Table~\ref{tab:teff} a
provisional effective temperature scale for the three galaxies.  Owing
to the lack of  data points for intermediate and late LMC O stars, we
have adopted values for the LMC which are intermediate for that of the
SMC and Milky Way; what limited data we do have (i.e., spectral type O5
in Fig.~\ref{fig:teffs}) supports this.  We emphasize that this scale
is not the final word on the subject; indeed, we were in a quandary as
to what effective temperatures to assign for the O3-O4 class where the
SMC dwarfs in our sample have a lower effective temperature scale
than the O5~Vs; we chose to assign a higher temperature than our
analysis of AV 177 and AV 378 would indicate.  
The effective
temperature scales can be made more trustworthy by observations and
analysis of LMC stars of all luminosity classes at intermediate and
late O-type, and SMC stars of the earliest types, as well as
SMC late-O supergiants.
Another priority but
much needed would be studies of stars of additional O and early-B type
in the Milky Way to complement the efforts of Repolust et
al.\ (2004).

We present the adopted temperature scale in Fig.~\ref{fig:scale}(a).
We include for comparison the Vacca et al.\ (1996) scale for Galactic
stars, which is much hotter; the scales are shown in comparison to the
data in Figs.~\ref{fig:scale}(b) and (c).

We were initially surprised to find that the effective temperature
scales for giants and dwarfs were indistinguishable, but an inspection
of the surface gravities in Table~\ref{tab:finalresults} reveals the
reason: in general, there is no difference in the surface gravities
derived for these stars, except for the two latest types.  In our sample
this might be due to the fact that all but these two giants are of
early type (O2-3.5) where, as we have previously emphasized, the
physical properties are not obviously correlated with the details of
the spectral properties.  However, even for the Galactic stars studied
by Repolust et al.\ (2004) there is only a modest difference found in the
surface gravities of dwarfs and giants, with averages of 3.7 (dwarfs)
and 3.6 (giants) found from their Table~1.  Conti \& Alschuler (1971)
note that in the original MK system luminosity classes were defined for
O stars only for the O9 subclass, and this was based upon the strength
of ratio of Si IV $\lambda 4089, 19$ to He~I $\lambda 4143$ (see
Morgan et al.\ 1943).  Subsequently, Walborn (1971b) used
the N~III $\lambda 4634, 42$ emission lines and the He~II $\lambda
4686$ line (either absorption or emission) to establish luminosity
classification criteria.  As we have previously noted, these lines are
not gravity-sensitive {\it per se} but rather may reflect the effects of
surface abundances, effective temperatures, and what is happening in
the stellar winds (both mass-loss rate and density/temperature
profiles) in a complicated manner.  It is clear that the supergiants
(both in the Milky Way and in the Magellanic Clouds) have lower surface
gravities than do the stars spectroscopically classified as dwarfs, but
there is so far no evidence to support that the early- or mid-type
giants (defined primarily as having ``weakened "He~II $\lambda 4686$
absorption) as having lower surface gravity than the dwarfs.
Theoretical modeling that includes the N~III $\lambda 4634,42$ line
will help address this important issue.

However, for now let us note a particularly egregious example,
namely the star LH64-16. Walborn et al.\ (2002, 2004) make a point of
declaring this the prototype of the O2 giant class.  Yet, compare its
physical properties to the O2 dwarfs BI~253 and BI~237.
LH64-16 may be hotter, but the three stars have the same surface gravities,
similar stellar radii, mass-loss rates, and bolometric luminosities.  
The spectral
appearance of LH64-16 differs from that of BI~253 and BI~237 primarily
in the fact that the He~II $\lambda 4686$ shows more of a P Cygni profile
in LH64-16, plus the nitrogen lines (N~IV and N~III emission, and N~V
absorption) are significantly stronger than in BI~237 and BI~253.
The only
physical difference we can discern between the two stars is one of 
chemical abundances (Paper~I and Walborn et al.\ 2004).  In Paper~I we
speculated that perhaps this star is the product of binary evolution.
In any event, the luminosity criteria used by Walborn et al.\ (2002) in
this case do not seem to be tied into the properties one usually
uses to distinguish amongst the second MK dimension, such as actual
luminosity.

How does our effective temperature scale compare to that from other
recent studies using line-blanketed models?  We give the comparison
in Fig.~\ref{fig:others}.  First, let us consider the supergiants
(Fig.~\ref{fig:others}a). The overall agreement is fairly modest.
In particular, the results of
WM-BASIC modeling of the UV spectra of
Galactic supergiants by Bianchi 
\& Garcia (2002) and Garcia \& Bianchi (2004) [shown by black squares]
give effective temperatures considerably lower than those indicated
by the Repolust et al.\ (2003) data on which our scale is based.
The FASTWIND study of Cyg~OB2 stars by
Herrero et al.\ (2002) [black open circles] 
is in better agreement, unsurprising since the
same models and methodology was used. Even so, their
data would suggest our Galactic scale should be hotter
for the earliest supergiants and cooler for the latest  than our scale.
As for the Magellanic Cloud data, the results are quite mixed, and 
there is not good internal agreement of the various CMFGEN studies.
The Crowther et al.\ (2002) and Hillier et al.\ (2003) studies of
several LMC and SMC supergiants would indicated lower temperatures than
what we find (in general), while
the
Evans et al.\ (2004) analysis of an
SMC O8.5 supergiant suggests a somewhat higher
effective temperature than that of the O7 supergiants in the other studies.
As discussed earlier,
Puls et al.\ (2005) have found significant difference for the He~I singlets 
produced by FASTWIND and CMFGEN in certain temperature regimes.
Until this matter is resolved and the
same data re-analyzed, it is
hard to know what to make of these differences.

For the dwarfs and giants (Fig.~\ref{fig:others}b)  we again see that the
FASTWIND modeling indicates higher effective
temperatures than that obtained by the WM-BASIC modeling of the
UV spectra by Bianchi \& Garcia (2002) and Garcia \& Bianchi
(2004), who  find temperatures that are about 4000 K cooler than the mean
relationship we derive from the Repolust et al.\ (2004) 
data. (The single giant modeled with FASTWIND by Herrero et
al.\ 2000 agrees with this mean Galactic relationship.)
The WM-BASIC modeling by Bianchi \& Garcia (2002) and Garcia
\& Bianchi (2004) involves careful fitting of the absolute
strengths of such lines as C~IV $\lambda 1169$, C~III
$\lambda 1176$, P~V $\lambda \lambda 1118,1128$, 
Si~IV $\lambda \lambda 1123, 1128$,
etc.  In Paper~I we note that the {\it flux distributions} of FASTWIND
and WM-BASIC models agree well despite the former's use of approximate
line-blanketing and blocking; this is now demonstrated at length by
Puls et al.\ (2005).  We do not have a ready explanation for
the systematic difference apparent between the parameters derived by
the WM-BASIC and FASTWIND modeling.   However, we believe that the first
step in investigating this is to use FASTWIND modeling on the optical
spectra of the same stars studied by Bianchi \& Garcia (2002) and
Garcia \& Bianchi (2004), and, similarly, to use WM-BASIC to model
the UV spectra of the same stars 
studied 
by Repolust et al.\ (2004) using WM-BASIC and 
in order to understand the specific differences.
We have earlier alluded to the need for such a complete modeling effort.

In our comparison of the dwarfs and giants 
we include the results
of the CMFGEN modeling of Martins et al.\ (2002), which, for the Galactic
stars, indicate a {\it higher} effective temperature scale than our own,
in contrast to the WM-BASIC results.
The Martins et al.\ (2002) study is
theoretical in the sense that no actual stars were
used; instead, the synthetic 
He~I $\lambda 4471$ to He~II $\lambda 4542$ equivalent width ratios
were used to assign ``spectral types" to the models, which were run
with reasonable mass-loss rates. We note that this procedure avoids the
potential modeling problem involving the He~I singlets (Puls et al.\ 2005)
discussed earlier. 

For the SMC dwarfs modeled by Martins et al.\ (2004) with CMFGEN
there is very good agreement with our scale, despite the fact that their
observations were all obtained in the UV.   Of course, this again means
that the CMFGEN modeling was unaffected by any potential problem with
the He~I singlets.  We do note in passing an intrinsic uncertainty of the
placement of these stars in our diagram; Martins et al.\ (2004) estimated
the spectral types based upon the optical spectra produced by the models,
since they had no optical observations.

However, there are other studies of SMC dwarfs and giants
which are at variance
with our results. First, there is AV~69, an
OC7.5~III((f)) star 
whose UV and 
optical spectra were analyzed by Hillier et al.\ (2003) [denoted by
the green star in Fig.~\ref{fig:others}b], find a low effective
temperature compared to our scale. 
More disturbing, at first blush, is the discrepancy between our effective
temperature scale for SMC dwarfs and the analysis of stars in the NGC~346
H~II region using
TLUSTY and CMFGEN modeling with apparently similar results from the two
codes (Bouret et al.\ 2003)\footnote{In a recent preprint Heap et al.\ (2005)
use the same optical data and derive identical answers from their
modeling. The resulting effective temperature scale is considerably
lower than ours (6000 K) for the early SMC O stars.}.  
However, we will note that NGC~346 is the strongest H$\alpha$
source in the SMC, and that the data used by Bouret et
al.\ (2003) (and Heap et al.\ 2005)
was obtained with short-slit echelle data, which limited their
ability to do sky (nebular) subtraction\footnote{The version of
the data shown by Walborn et al.\ (2000) lacked nebular
subtraction, and made no correction for the moonlight continuum
contamination (although the solar {\it line} spectrum was removed in
an ad hoc manner).
This was partially corrected by a re-reduction of the data for
the versions used by Hillier et al.\ (2003), Bouret et al.\ (2003),
and Heap et al.\ (2005), although no correction could be made
for NGC~346-368 or NGC346-487 since there weren't enough sky pixels.
We are indepted to Chris Evans for correspondance
on this subject. We have recently obtained our own high S/N optical
spectra of the NGC 346 stars, and an analysis is underway.}.

However, since the Bournet et al.\ (2003) study does involve stars in 
a single cluster, it does give us a chance to allow stellar evolutionary
theory to weigh in on the issue of the effective temperature scales.
Using what became the Chlebowski \& Garmany (1991) effective
temperature scale, Massey et al.\ (1989) derived an H-R diagram for
this cluster which was highly coeval, with an age of 2-4~Myr and a typical
age spread  of $1$~Myr (see also Massey 1998, 2003).  
The H-R diagram shown by
Bouret et al.\ (2003) [their Fig.~12] implies an age spread of $>7$~Myr
rather than 1~Myr\footnote{Note that although Bouret et al.\ (2003) incorrectly
attribute the source of the photometry used in constructing their H-R
diagram to Walborn et al.\ (2000), it was actually that of Massey et
al.\ (1989); i.e., both studies started with the same {\it UBV} values,
and thus the differences in the H-R diagrams due purely to the different
effective temperature scales and slight differences in the treatment of
reddenings. See below for more on the latter point.}.  
Even excluding the oldest star (NGC346-MP12, a
nitrogen rich O9.5-B0~V star which Walborn et al.\ 2000 suggest is not
a member of the cluster), the age range is 5~Myr.  Furthermore, although
Bouret et al.\ (2003) fail to comment on this point, the O2~III
star would have the youngest age (0.5~Myr); its location in the H-R
diagram would be consistent with it being a {\it dwarf} and not a giant.

We compare the H-R diagram derived from their values to one which we
construct using our new effective temperature calibration in 
Fig.~\ref{fig:N346}.
For consistency
we have used the reddenings and spectral types adopted by Bouret et al.\ (2003)
in making this 
comparison\footnote{It should be noted that the 
intrinsic colors used by Bouret et al.\ (2003) lead to low values
for the reddenings of the earliest stars, $E(B-V)=0.05$ to 0.12.
Massey et al.\ (1989) find a somewhat greater color excess (0.09 to 0.15).
By comparison to other early-type stars in the SMC, the reddening of
the early O stars in NGC~346 is average or slightly high, 
as might be expected; 
Massey et al.\ (1995) find an average reddening of 0.09 for the SMC.
Were the Bouret et al.\ (2003) intrinsic colors correct, then many
of the early-type stars in the SMC would have reddenings less than the
expected foreground reddening to the SMC (Schwering \& Israel 1991,
Larsen et al.\ 2000).  The Bouret et al.\
(2003) values also lead to a progression in reddenings with spectral
type.}.  The effective temperature scale we adopt here lead to more 
consistent ages for the stars in the cluster.  While this doesn't prove
our scale is right, and the Bouret et al.\ (2003) values wrong, it does
emphasize an often overlooked point, namely that changes in the
effective temperature scale of O stars do have implications in the
interpretation of H-R diagrams and star-formation in clusters; see discussion
in Hanson (2003) and Massey (2003).  We plan a re-analysis of the
NGC~346 stars using FASTWIND using high S/N spectra with good sky (nebular)
subtraction.

In Fig.~\ref{fig:bcs} we show the BCs as a function of effective
temperatures.  As was the case with unblanketed models, there is no
apparent difference with surface gravity, as is shown by the small
scatter.   There {\it is} a slight shift
with metallicities, with stars of low metallicity (SMC) having a BC which
is perhaps 3-4\% more negative (i.e., more significant) than the
higher metallicity (Milky Way) stars. We find a relationship
\begin{equation}
\label{equ:bc}
{\rm BC} = -6.90 \times \log T_{\rm eff} + 27.99
\end{equation}
fit the data with an RMS of 0.03~mag.  This is very similar to that found
by Vacca et al.\ (1996) considering unblanketed models.

\subsection{The Wind Momentum-Luminosity Relationship}
\label{Sec-WLR}

Kudritzki et al.\ (1995) introduced the {\it Wind-Momentum
Luminosity Relation (WLR)}, where
\begin{equation}
\label{equ:WLR}
D_{\rm mom}\equiv \dot{M} v_\infty R_\star^{0.5} \propto L^x,
\end{equation}
with $x = 1/\alpha_{\rm eff} = 1/(\alpha - \delta)$, (Puls et al.\ 1996).
The force multipliers $\alpha$ and $\delta$ have been described by
Kudritzki et al.\ (1989): $\alpha$ would equal 1 in the case that
only optically thick  
lines contributed to the line acceleration force, and would equal
0 if only optically thin lines contributed.  Typical values are 0.5 to 0.7 
(Kudritzki 2002). The parameter $\delta$ describes
the ionization balance of the wind, and typically has a value between
0.0 and 0.2 (Kudritzki 2002).   
Since the value of $\alpha_{\rm eff}$ 
is expected to have only a weak
dependence upon the effective temperature and metallicity (and in a way
that can be theoretically predicted from radiative-driven wind theory;
see Puls et al.\ 2000, Kudritzki 2002, as well as Vink et al.\ 2000, 2001, but
see also Martins et al.\ 2004 for counter-examples)
one can in fact use Eq.~\ref{equ:WLR} to find the distances to galaxies
using basic physics combined with quantitative spectroscopy of the bright
supergiants.

In Fig.~\ref{fig:wlr}(a) 
we show the WLR for the stars in our sample in comparison
to the Galactic sample studied by Repolust et al.\ (2004). Although there is
considerable scatter, it is clear that the SMC stars show a lower value for
$D_{\rm mom}$ than do the Galactic stars, with the LMC stars being somewhat
intermediate.  A more subtle effect is that the supergiants of a given
galaxy seem to lie somewhat higher than do the dwarfs and giants.

We can reduce the scatter in this diagram dramatically by correcting the
mass-loss rates given in Table~\ref{tab:finalresults} for the ``clumpiness"
of the stellar wind.  Puls et al.\ (2003) and 
Repolust et al.\ (2004) argue that any kind of instability
requires some time to grow, and thus the lower part of the wind should be
minimally affected by clumping.  Thus, the only stars whose H$\alpha$
profile should be affected are those which for which the profiles
are formed in a major volume 
of the wind, i.e., the stars with H$\alpha$ in emission. If
H$\alpha$ is in absorption, this immediately suggests that the only wind
contribution is from layers very close to the sonic point, and thus
absorption type profiles should be little affected by wind clumping.
If radiatively-driven
wind theory is correct, then the WLR should be independent of luminosity
class, and Repolust et al.\ (2004) derived a numerical correction
factor of 0.44 by forcing
Galactic supergiants to follow the same WLR as do the dwarfs and giants.
We have applied the same correction factor
to our data for the stars with H$\alpha$ in emission; only three 
LMC supergiants are affected, if we discount the stars for which only
upper limits have  been determined. It is not clear what metallicity 
dependence, if any, there might
be in this correction factor, but since only the LMC data 
are affected (other than
upper limits), this is probably safe.
We show the revised plot in Fig.~\ref{fig:wlr}(b).  Indeed the supergiants
in the LMC are in better accord with the dwarfs, 
just as Repolust et al.\ (2004)
found for the Galactic stars.

In general, the SMC stars have a considerably smaller $D_{\rm mom}$ than do
the Galactic stars. What would we expect on theoretical grounds?
Vink et al.\ (2001) calculates that for O-type stars, 
$\dot{M}\propto Z^{0.69},$
if the Leitherer et al.\ (1992) relation $v_\infty \propto Z^{0.13}$ is
taken into account.  We can therefore expect that
$$D_{\rm mom}(Z) \propto Z^{0.82}.$$  If $\alpha$ remained constant, we
would then expect that the {\it slopes} in Fig.~\ref{fig:wlr}(b) would
be the same for all three galaxies, but that the {\it intercepts} would be
-0.25 dex lower for the LMC than for the Milky Way, and -0.57 dex lower
for the SMC than for the Milky Way.  We show these relationships in 
Fig.~\ref{fig:wlr}(b), where we have adopted the Vink et al.\ (2001) 
slope and intercept as given by Repolust et al.\ (2004) for the
Milky Way stars. We consider the agreement excellent in this diagram: 
the plot shows a clear effect between the Galaxy and the LMC along the
lines prediced by theory, and if the upper limits for the SMC stars
are close to the true values, then the SMC also seems to agree well,
with the exception of two objects.  The analysis of the UV spectra we have
planned should provide better constrains in the cases where H$\alpha$
provides upper limits only.
Kudritzki (2002) has performed
careful calculations of the effects of metallicity and effective temperature
on the force multipliers, and his calculations show that at an SMC-like
metallicity that $\alpha$ will be 3\% ($T_{\rm eff} \approx 40000$ K) and
15\% ($T_{\rm eff} \approx 
50000$ K) larger at an SMC-like metallicity than in the 
Milky Way.  We would thus expect $x$ to be somewhat smaller for the SMC,
and the relationship slightly more shallow, which is certainly not
excluded by our data.

In evaluating Fig.~\ref{fig:wlr} one should keep in mind the typical errors
discussed above, i.e., an uncertainty of 0.05~dex in $\log L/L_\odot$ for
the Magellanic Cloud objects, and of 0.15 for the Galactic objects.
The error in $\log D_{\rm mom}$ is about 0.15~dex for both the Magellanic
Cloud and Galactic objects.  We show the typical error bars in the
figure.  Most of the stars follow the relationship fine; the notable
exception is the star AV~14, which shows an upper limit of $D_{\rm mom}=28.15$
at $\log L/L_\odot=5.85.$

\subsection{Comparison of Spectroscopic and Evolutionary Masses}
\label{Sec-Mass}

The analyses of our stars have yielded values for the ``spectroscopic
mass", $M_{\rm spect} = (g_{\rm true}/g_\odot)R^2$ 
given in Table~\ref{tab:finalresults}.  We remind the reader that
these $g_{\rm true}$ values have resulted from a modest correction
for centrifugal acceleration to the measured $g_{\rm eff}$ values
obtained from the model fits.
It is of interest to compare these values with the mass $M_{\rm evol}$
which we derive from
stellar evolutionary models based upon a star's $\log L$ and $T_{\rm eff}$.
This comparison is shown in Fig.~\ref{fig:massdis}.  We have included 
error bars, with the uncertainty in the evolutionary mass assumed to be
due purely to the uncertainty in $\log L/L_\odot$ (i.e., $\Delta \log
{M_{\rm evol}} = -0.2 \Delta$ M$_{\rm bol}$; see equation 4 of Massey
1998).  These evolutionary masses were derived from the ``standard"
(non-rotating) models using the older opacities; we return to this
point below.

We see that most of the stars in our sample cluster around the
mean relationship
$M_{\rm spect} \sim M_{\rm evol}$, but that both samples contain a fair
number of objects in which the
evolutionary mass which is considerably greater than the spectroscopic
mass.  This {\it mass discrepancy} was first found by Groenewegen et al.\ 
(1989) and studied extensively by
Herrero et al.\ (1992) for Galactic stars.
This problem was investigated by many authors; comparison with 
the masses determined from binaries have tended to support the evolutionary
masses (Burkholder et al.\ 1997; Massey et al.\ 2002).  The lowering of
the effective temperatures for Galactic supergiants has resulted in
decreasing, if not eliminating, the discrepancy for Galactic stars
(Herrero 2003, Repolust et al.\ 2004).  The expected
reason is two-fold: first,
lower effective temperatures imply a smaller luminosity of a star, and hence
the deduced evolutionary mass will be less.  Secondly, the new line-blanketed
models resulted in larger photospheric 
radiation pressure, so a higher surface gravity
is needed to reproduce the Stark-broadened wings of the Balmer lines.

Although the Galactic mass discrepancy was considerably worse for
Galactic supergiants than dwarfs, we see that the discrepant stars in
Fig.~\ref{fig:massdis} include both dwarfs and giants, and an
inspection of Table~\ref{tab:finalresults} suggests that the problem
stars all have $T_{\rm eff}>45000$K, and indeed all stars this hot show
a significant mass discrepancy.  We indicate these stars with filled
symbols in the figure.  (We note that the two stars with $T_{\rm
eff}>45000$K in the Galactic sample of Repolust et al.\ 2004 did not
show a similar problem.)

\begin{figure}
\epsscale{0.65}
\plotone{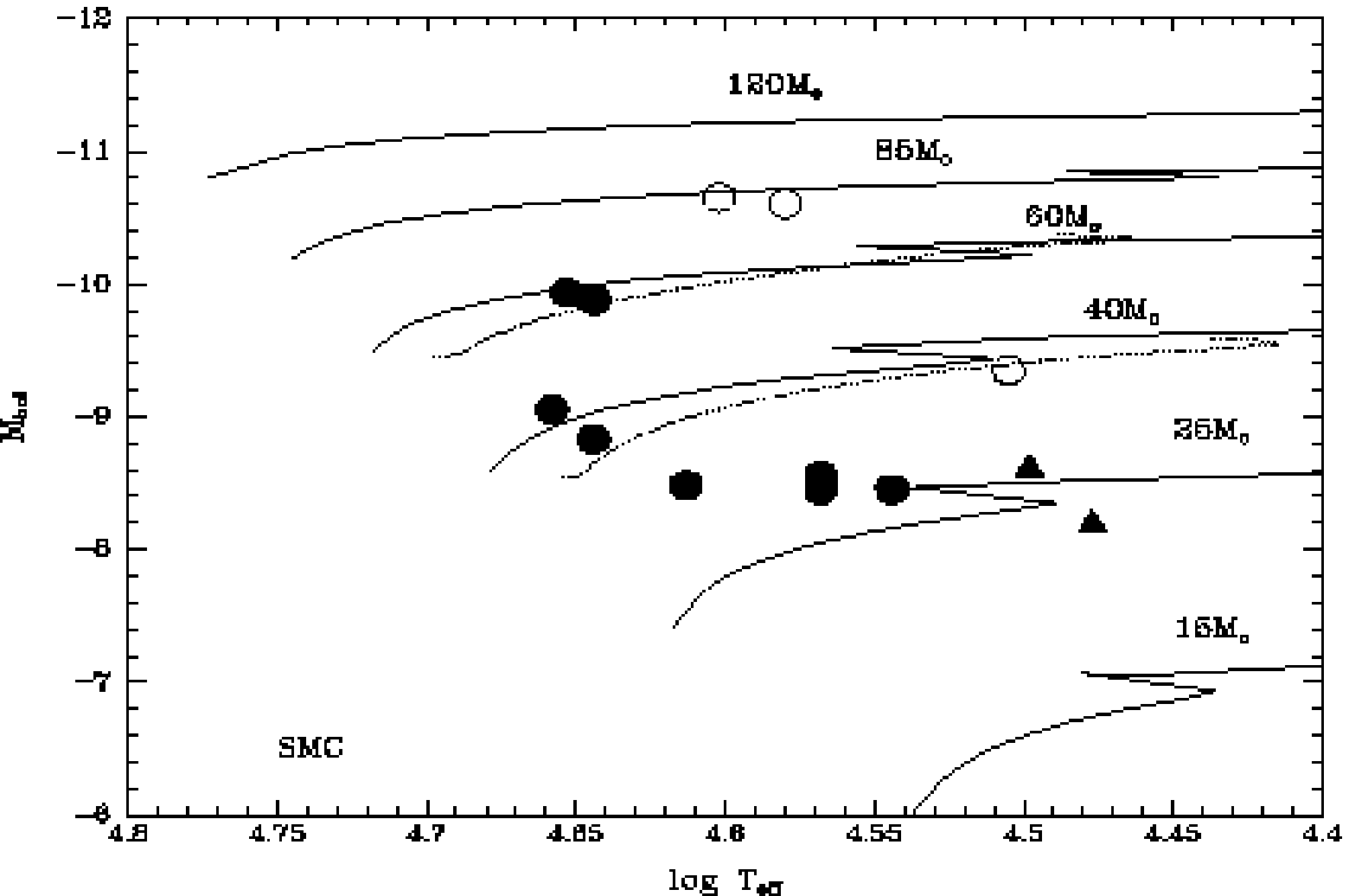}
\plotone{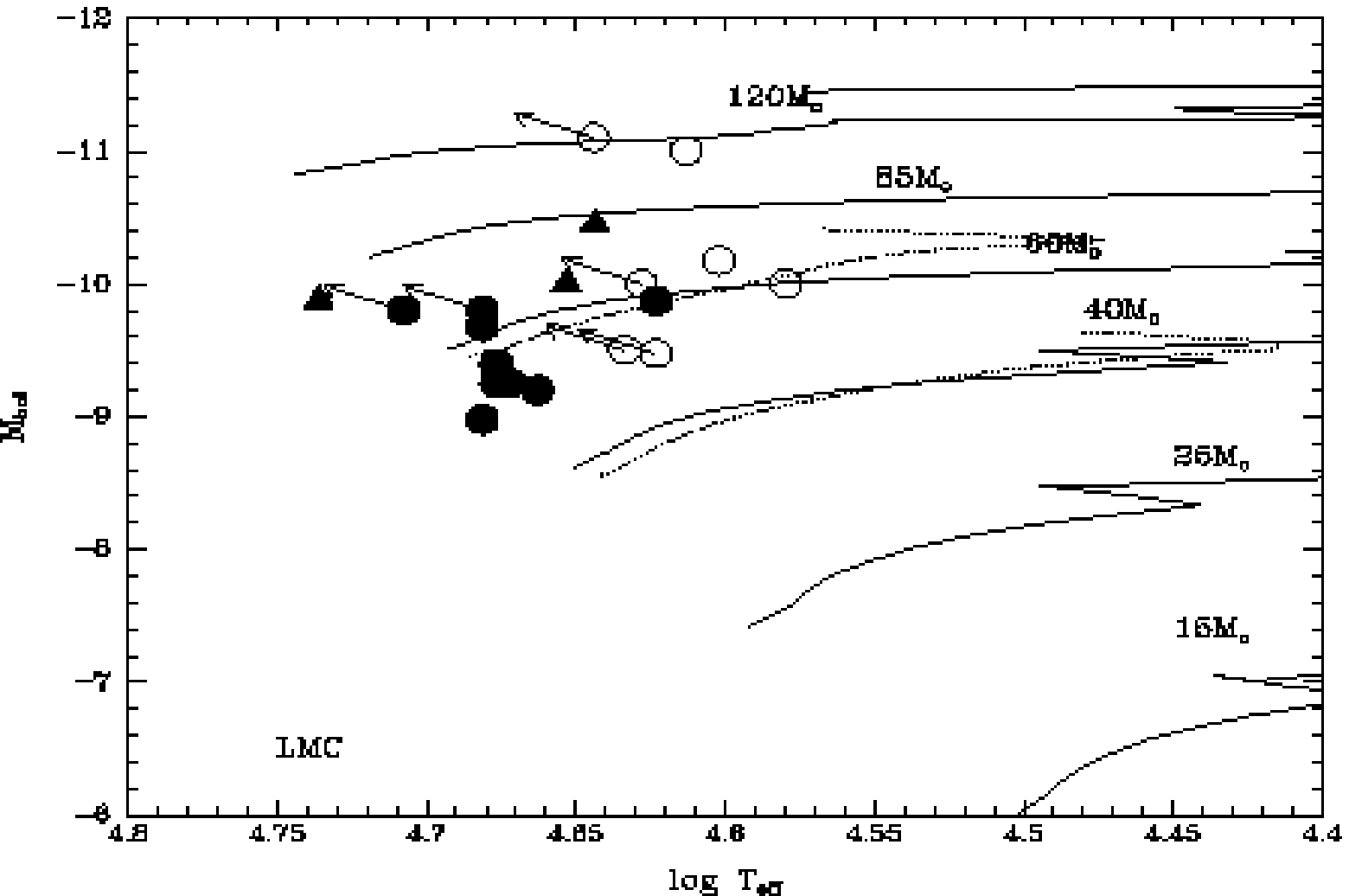}
\caption{\label{fig:hrds} The location of our stars in the H-R diagram.
Dwarfs are indicated by filled circles, giants by triangles, and supergiants
by open circles.  (a) The non-rotating stellar evolutionary models of
Charbonnel et al.\ (1993) are shown for the SMC, and (b) those of
Schaerer et al.\ (1993) are shown for the LMC.  Only the H-burning
part of the tracks are shown.  For comparison, we include the newer
rotation models (dotted lines) for initial masses of $60 M_\odot$ and
$40 M_\odot$.
}
\end{figure}

In determining the evolutionary masses we have relied upon the older
tracks of Charbonnel et al.\ (1993) for the SMC and Schaerer et
al.\ (1993) for the LMC.  The effects of newer tracks, including the
effects of an initial rotation speed of 300 km s$^{-1}$, is shown in
Fig.~\ref{fig:hrds} by the dotted lines for the $60 M_\odot$ and $40
M_\odot$ tracks.  We see that the inclusion of rotation will have a
modest effect on the deduced masses compared to the non-rotating
models.  For the hotter stars the effect will be to make the
evolutionary masses lower; at cooler temperatures (on the
main-sequence) the effect is in the other direction.

For some of the
discrepant stars in Fig.~\ref{fig:massdis} use of the newer evolutionary
models could potentially bring the evolutionary masses into closer agreement
with the spectroscopic masses.
However, the effect is small compared
to the high temperature mass discrepancy we note: for stars near the
ZAMS, the difference is about 0.25~mag, which is equivalent to
0.05~dex in $\log M$, or
about 10\%

It is possible that the objects with large mass discrepancies
in Fig.~\ref{fig:massdis} represent the results of binary
evolution.  A good candidate is LH64-16, discussed in Paper~I.  This
``ON2~III" star shows highly processed material at the surface (Paper~I and
Walborn et al.\ 2004), and in Paper~I we argue that this star might be
the result of binary evolution.  We note that this star is found to
the left of the ZAMS in Fig.~\ref{fig:hrds}, a nonphysical location.
A similar problem was found
for some of the relatively close binaries in the R136 cluster by 
Massey et al.\ (2002), suggesting that the stars had suffered from
some binary interactions.
However, we are also left with the possibility that at high $T_{\rm eff}$
the models may be underestimating either $\log g$ or $R$; given that
$R$ is derived from $M_V$ and $T_{\rm eff}$, a problem with $\log g$ appears
to be the most likely.

\section{Summary, Discussion, and Future Work}
\label{Sec-sum}

\begin{figure}
\epsscale{0.6}
\plotone{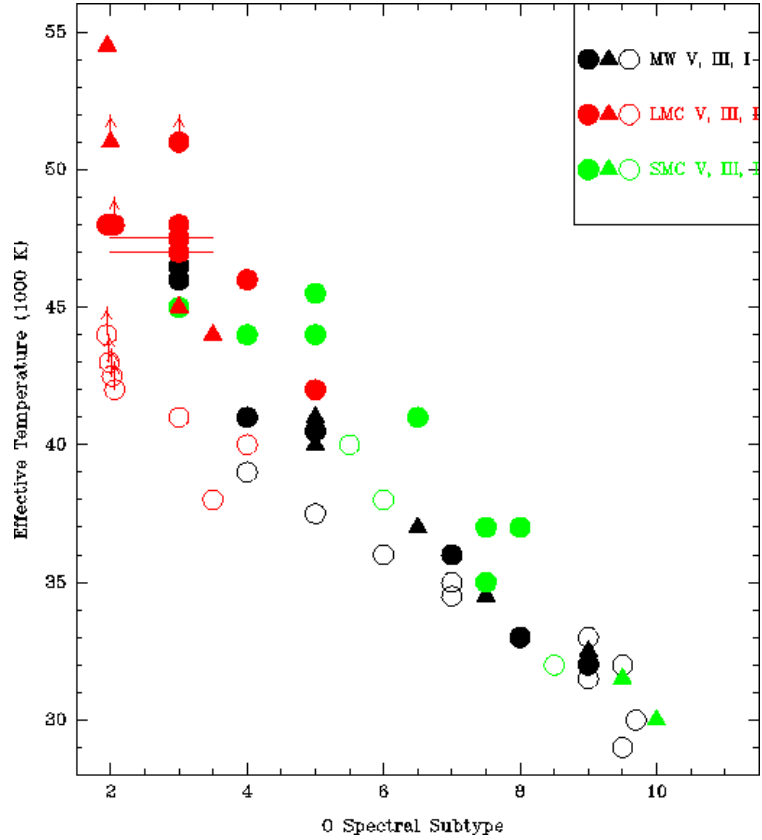}
\caption{\label{fig:teffs} The effective temperatures as a function of
spectral subtype, with a value of 2 denoting spectral type O2, 
5.5 denoting O5.5,
and ``10" denoting B0.  The filled circles for dwarfs; the filled
triangles are giants, and the open circles are supergiants.  Black
symbols correspond to the Milky Way, red to the LMC, and green to the
SMC.   The data for the Milky Way is taken from Repolust et al.\ (2004);
the data for the SMC and LMC are taken from Paper I and the present study.
}
\end{figure}

\begin{figure}
\epsscale{0.4}
\plotone{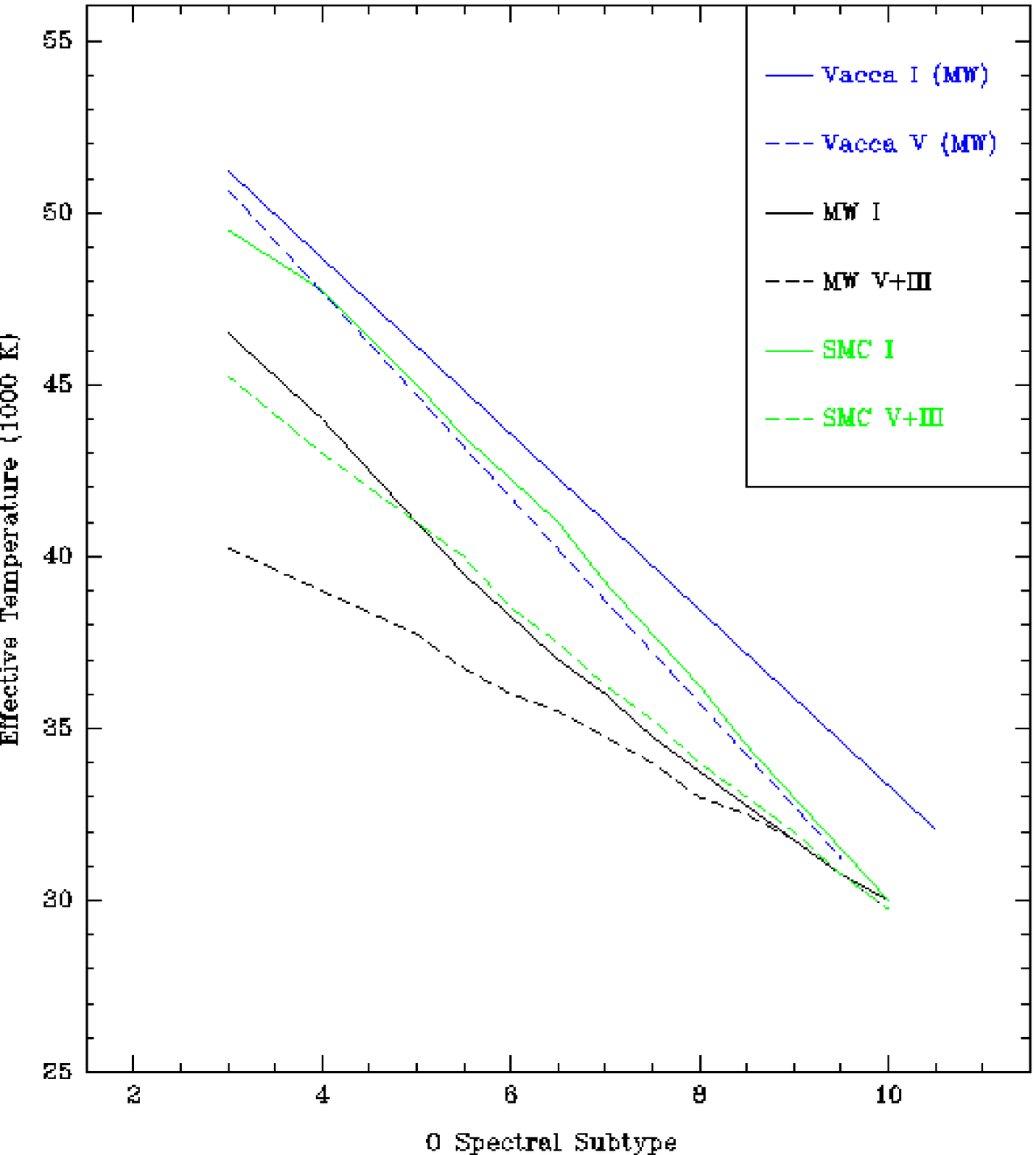}
\plotone{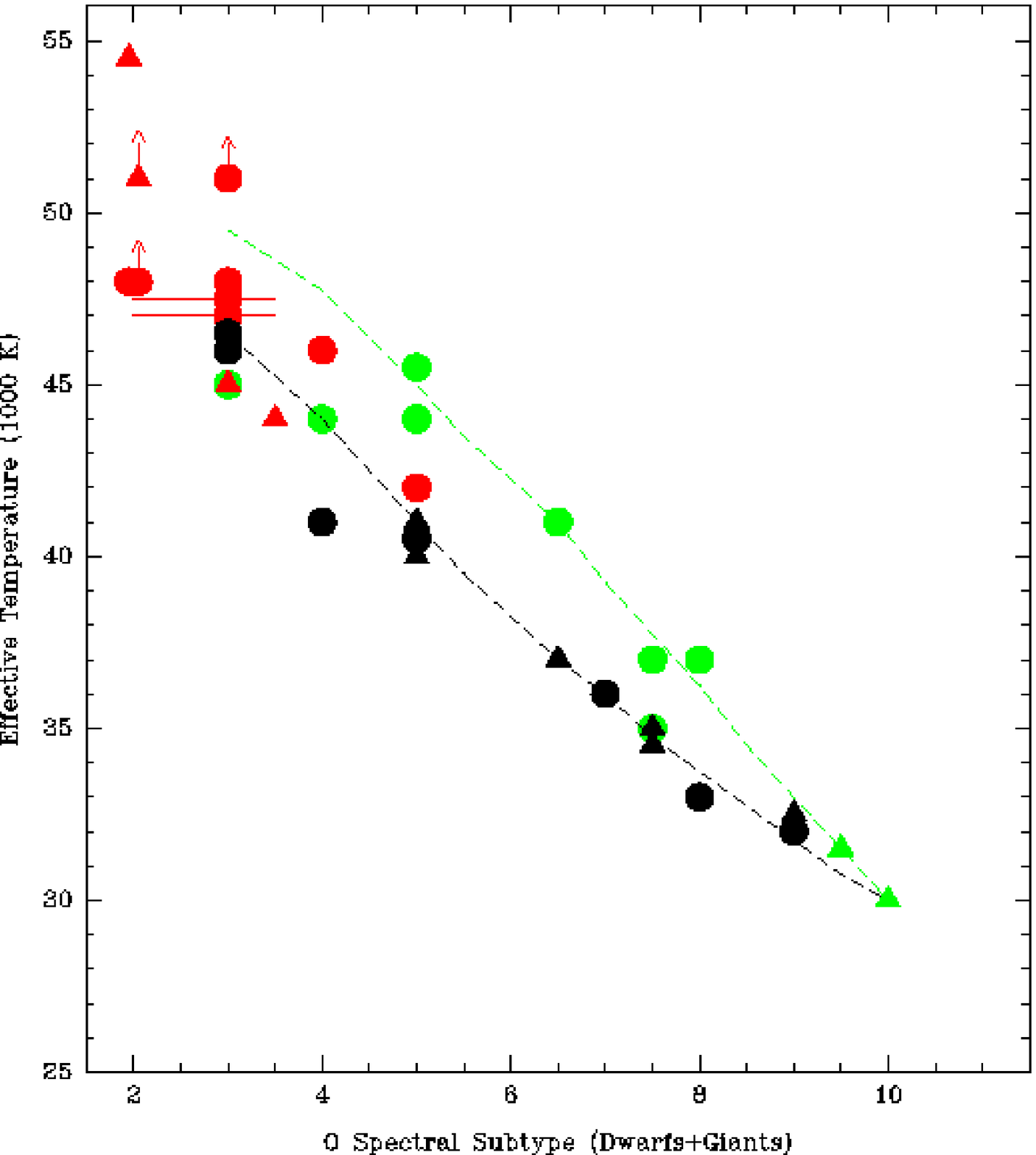}
\plotone{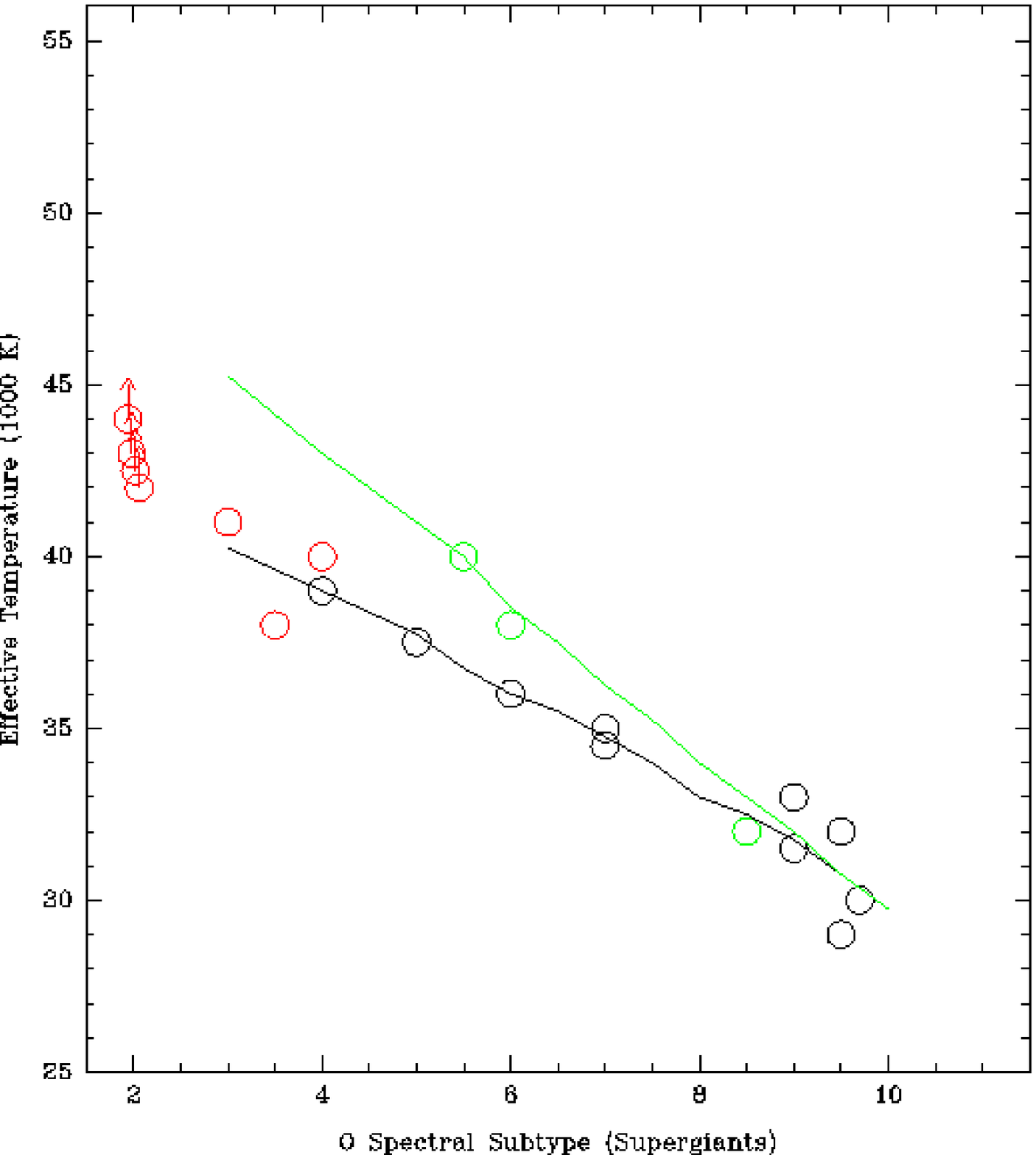}
\caption{\label{fig:scale} The effective temperature scale adopted here
for the SMC (green) and the Milky Way (black).  Solid lines denote the
scale for supergiants; dashed lines are for dwarfs and giants.  
(a) The supergiant and dwarf effective temperature scales from
Vacca et al.\ (1996) are shown for comparison in blue.
(b) The effective temperature scale is shown with comparison for the
data for the
dwarfs and giants.  (c) The supergiant 
effective temperature scale is shown in comparison with the data.
The LMC scale is assumed to be intermediate between that of the SMC and
the Milky Way.
}
\end{figure}

\begin{figure}
\epsscale{0.4}
\plotone{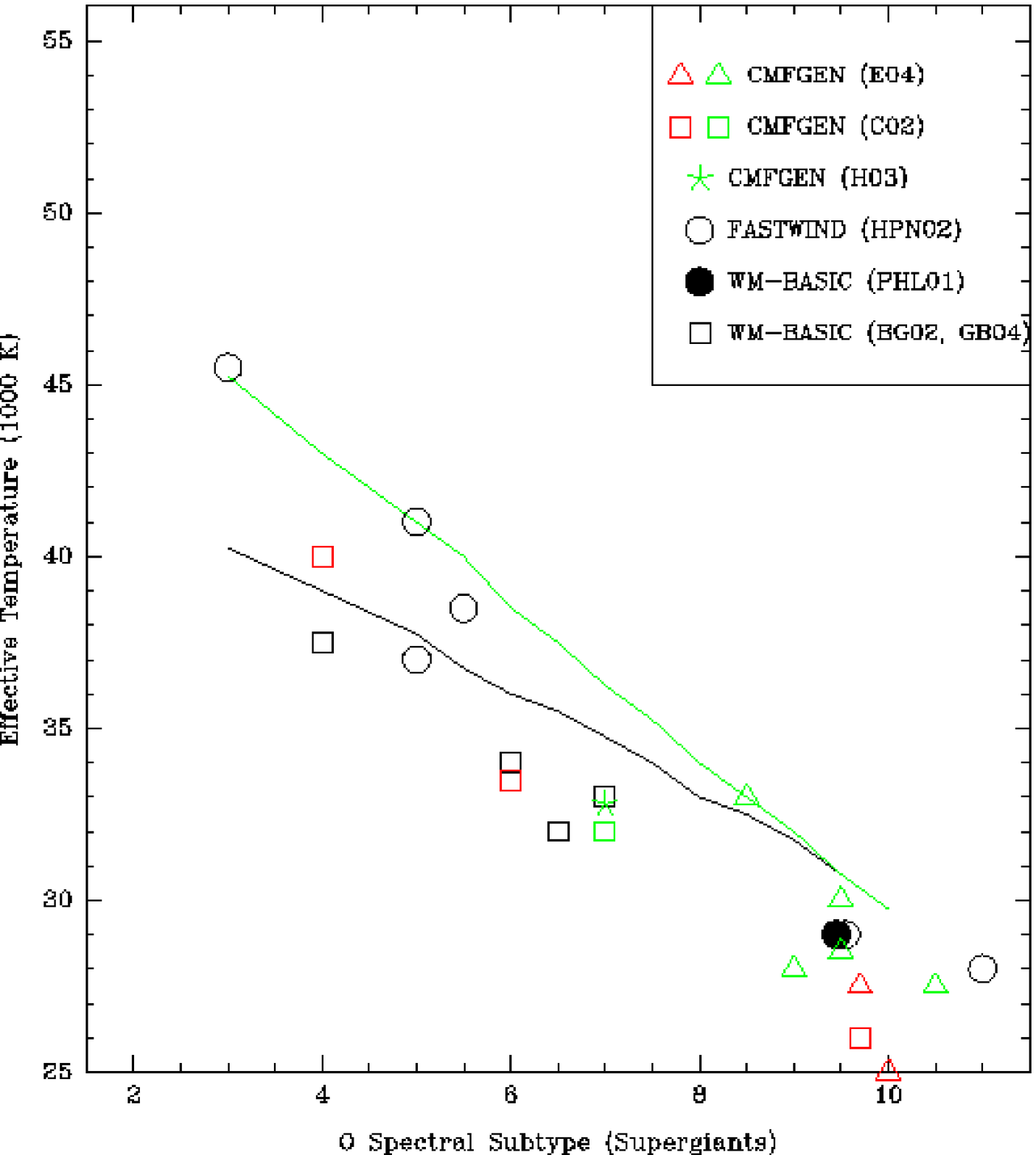}
\plotone{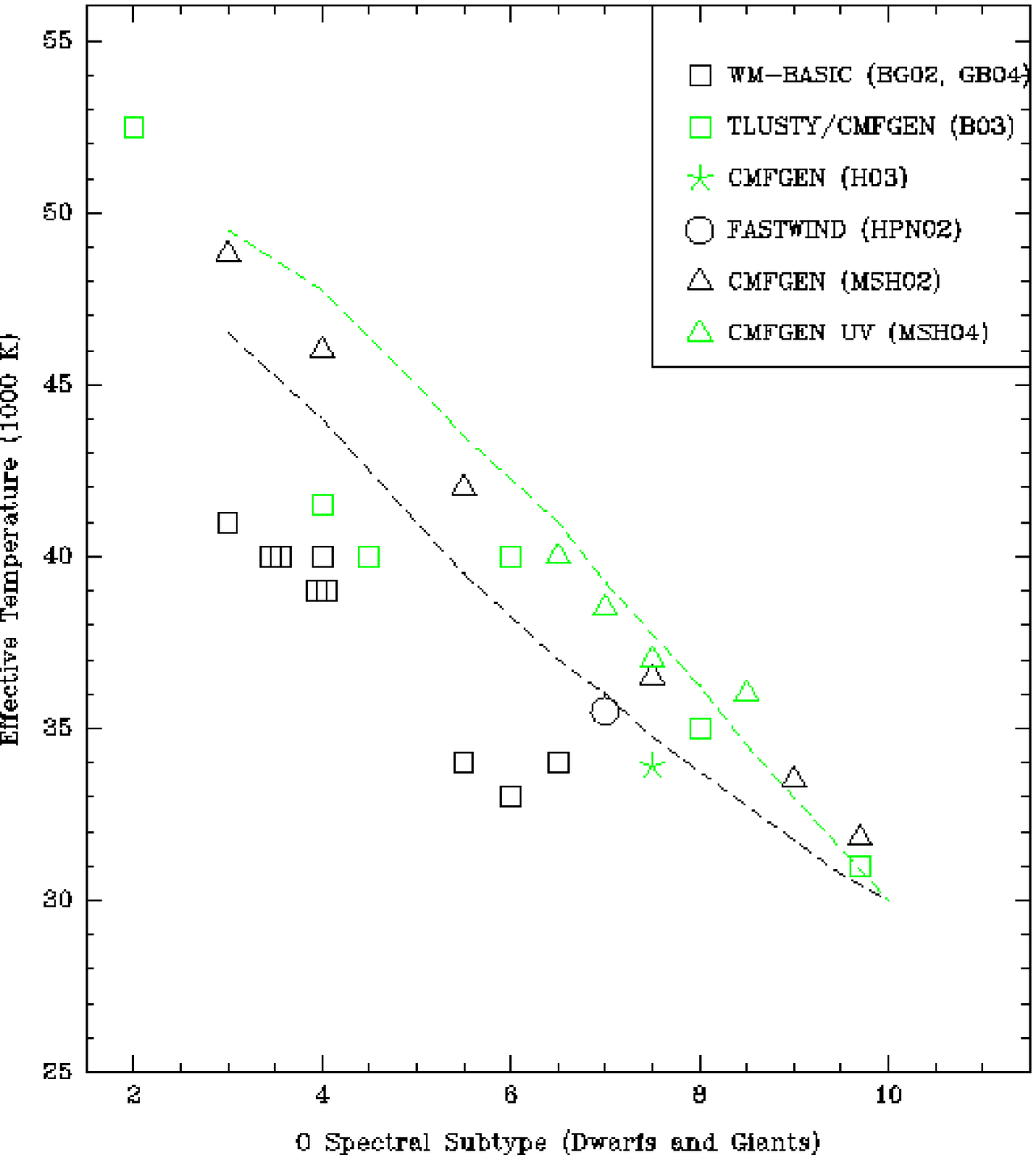}
\caption{\label{fig:others} Our effective temperature scales (dashed or
solid curves) are compared
to recent data from others.   Black symbols represent the Milky Way;
red, the LMC; and green, the SMC. 
The different symbols are indicated in the key, where the data
come from Bianchi \& Garcia (2002), [BG02]; Garcia \& Bianchi (2004), [GB04]; 
Crowther et al.\ (2002), [C02]; Hillier et al.\ (2003), [H03]; and
Herrero et al.\ (2002), [HPN02], Bouret et al.\ (2003), [B03], 
Martins et al.\ (2002), [MSH02], Martins et al.\ (2004), [MSH04],
and Evans et al. (2004) [E04].
}

\end{figure}

\begin{figure}
\epsscale{0.65}
\plotone{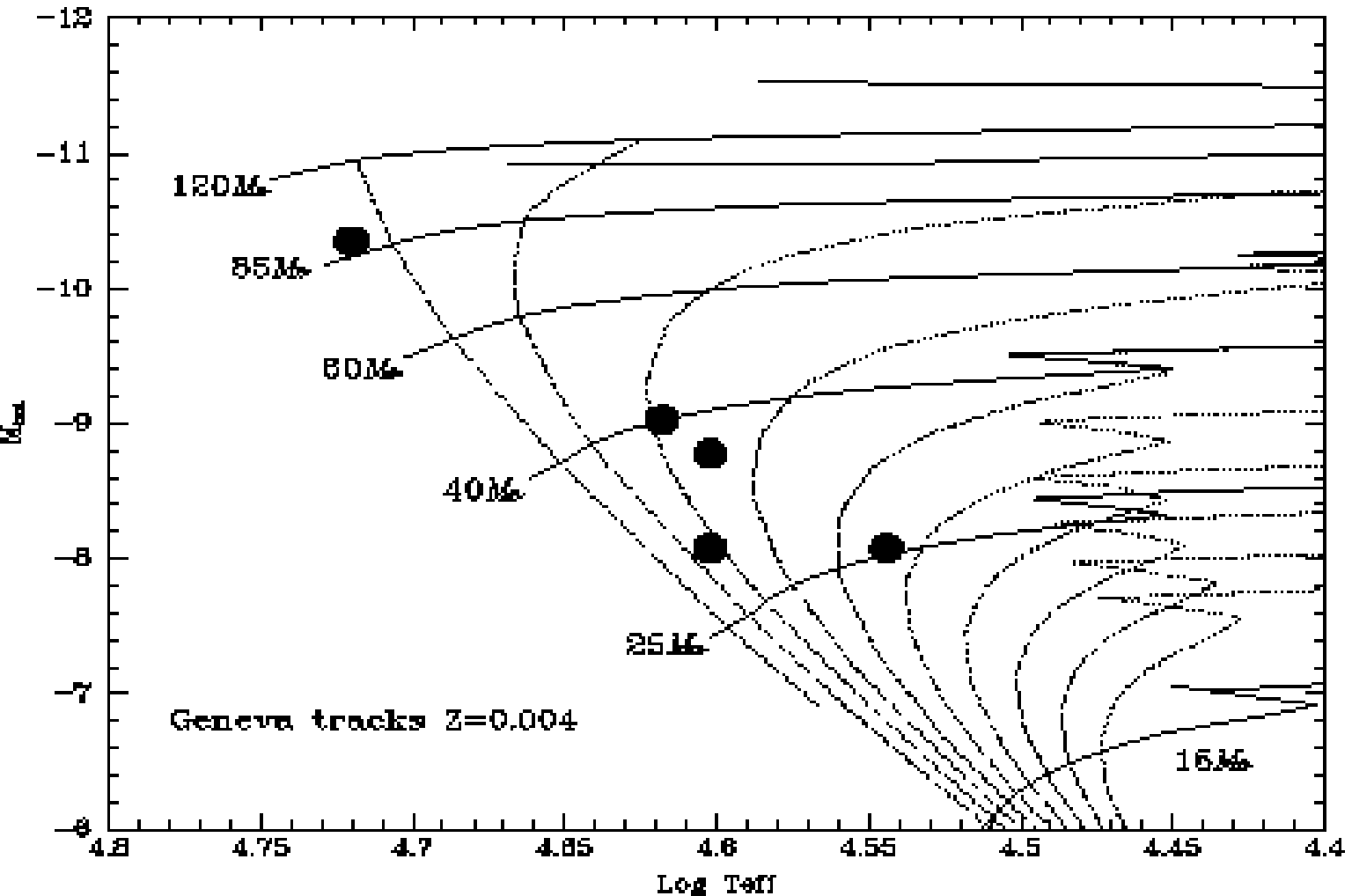}
\plotone{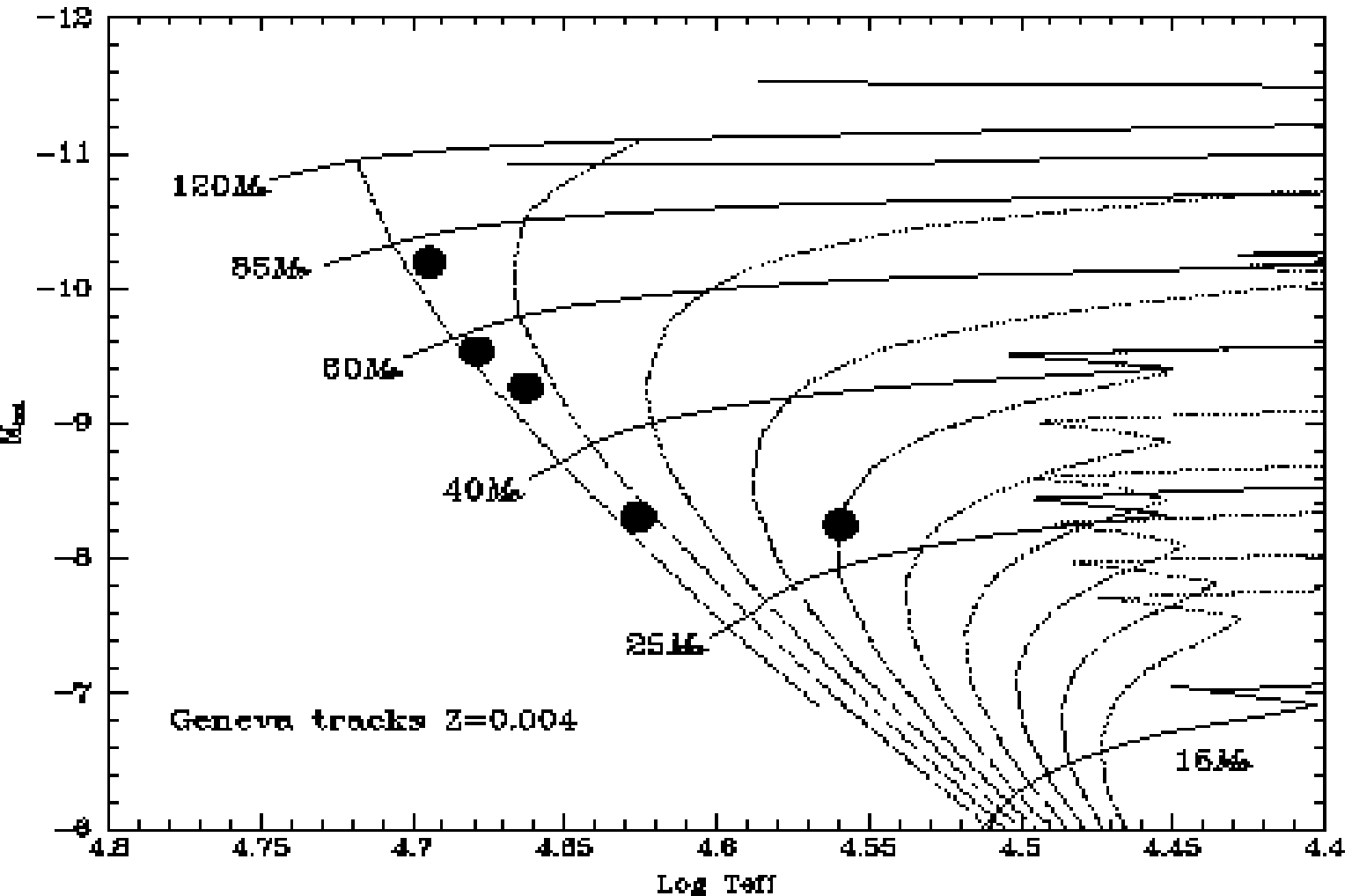}
\caption{\label{fig:N346}.  The H-R diagram for the NGC~346 stars studied
by Bouret et al.\ (2003), excluding NGC~346-MPG12, which
Walborn et al.\ (2000) argue is not part of the cluster.
The evolutionary tracks (solid lines) are from
Charbonnel et al.\ (1993) computed for a metallicity of Z=0.004.  The
dotted lines are isochrones computed form these models for ages of
1-10~Myr at intervals of 1~Myr.  (a) Stars placed with the parameters of
Bouret et al.\ (2003).  (b) Stars placed
using the average
spectral type to effective temperature calibration given here.
In both cases the same photometry and color excesses were used, although
a more realistic estimate of the reddening would result in stars about
0.1~mag more luminous. 
}

\end{figure}

\begin{figure}
\epsscale{0.6}
\plotone{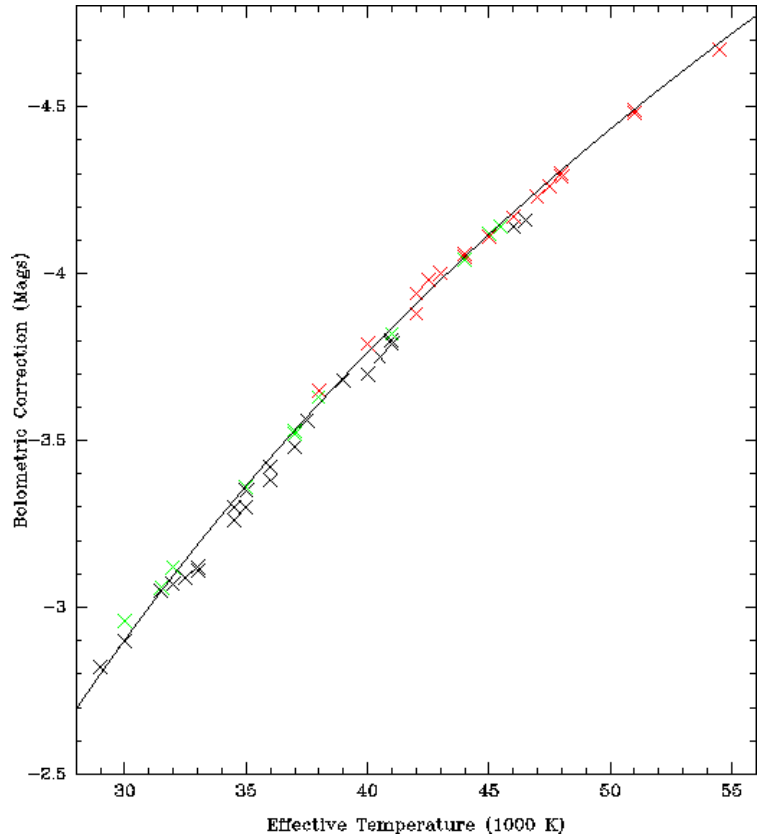}
\caption{\label{fig:bcs} The bolometric corrections are shown as a function
of effective temperature.  The black points are for the Milky Way; the
red points are for the LMC, and the green points are for the SMC.  The
smooth curve denotes the relationship given in Equ.~\ref{equ:bc}.
}
\end{figure}

\begin{figure}
\epsscale{0.49}
\plotone{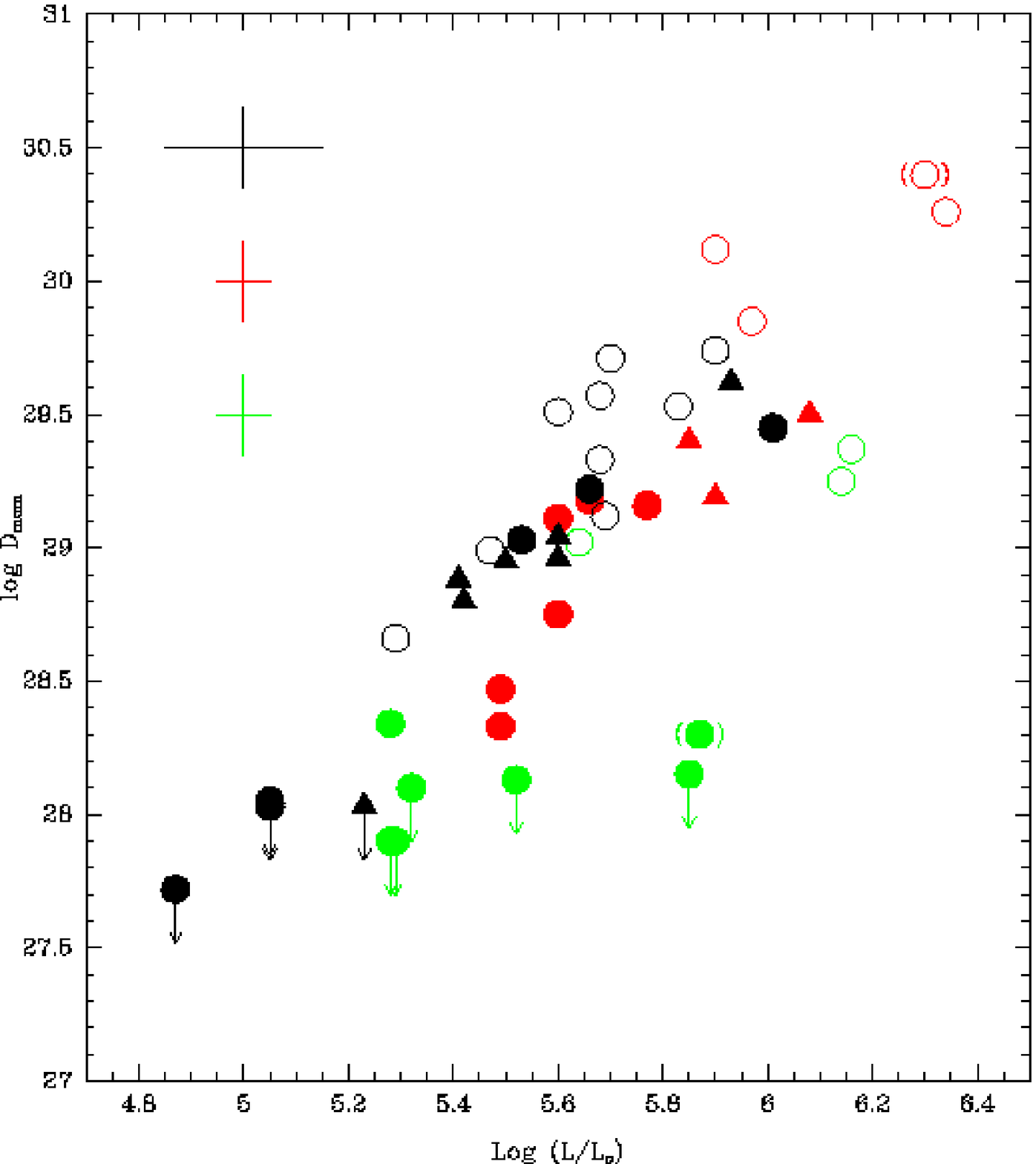}
\plotone{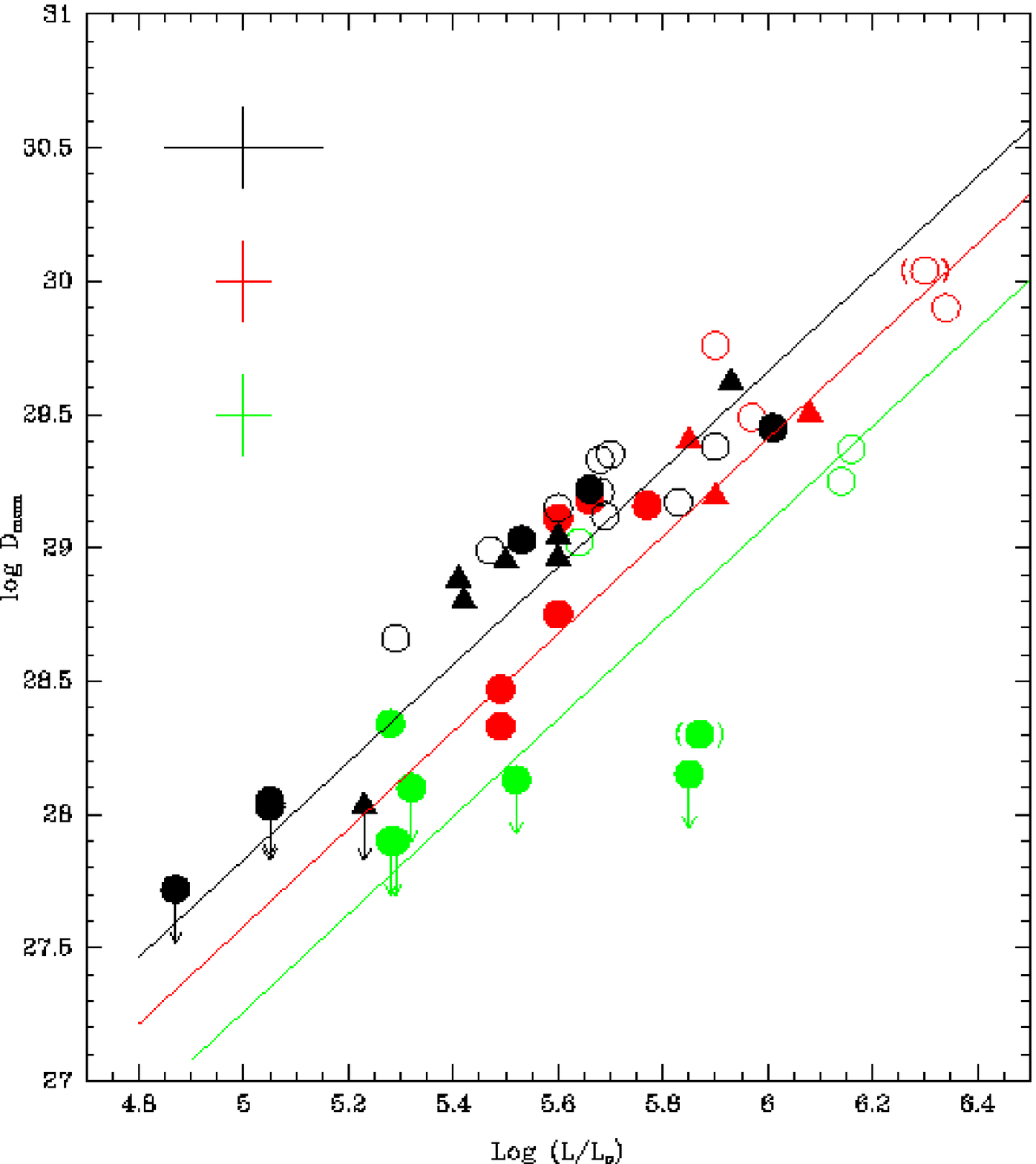}
\caption{\label{fig:wlr} The wind momentum luminosity relationship.  The symbols have the same meaning as in Fig.~\ref{fig:teffs}.
 Stars with only lower limits
on their effective temperatures are not shown, and stars whose values
are particularly uncertain are indicated by parenthesis.  Typical error
bars are shown in the upper left.
(a) The data are shown 
uncorrected for wind clumping.  (b) The stars showing H$\alpha$ emission
have been corrected by -0.36~dex in $D_{\rm mom}$ to correct for the
effects of wind clumping on the deduced mass-loss rates. The three lines
are {\it not} fits to the data; instead, these are the theoretical expectations from radiatively driven wind theory.
}

\end{figure}

\begin{figure}
\epsscale{0.6}
\plotone{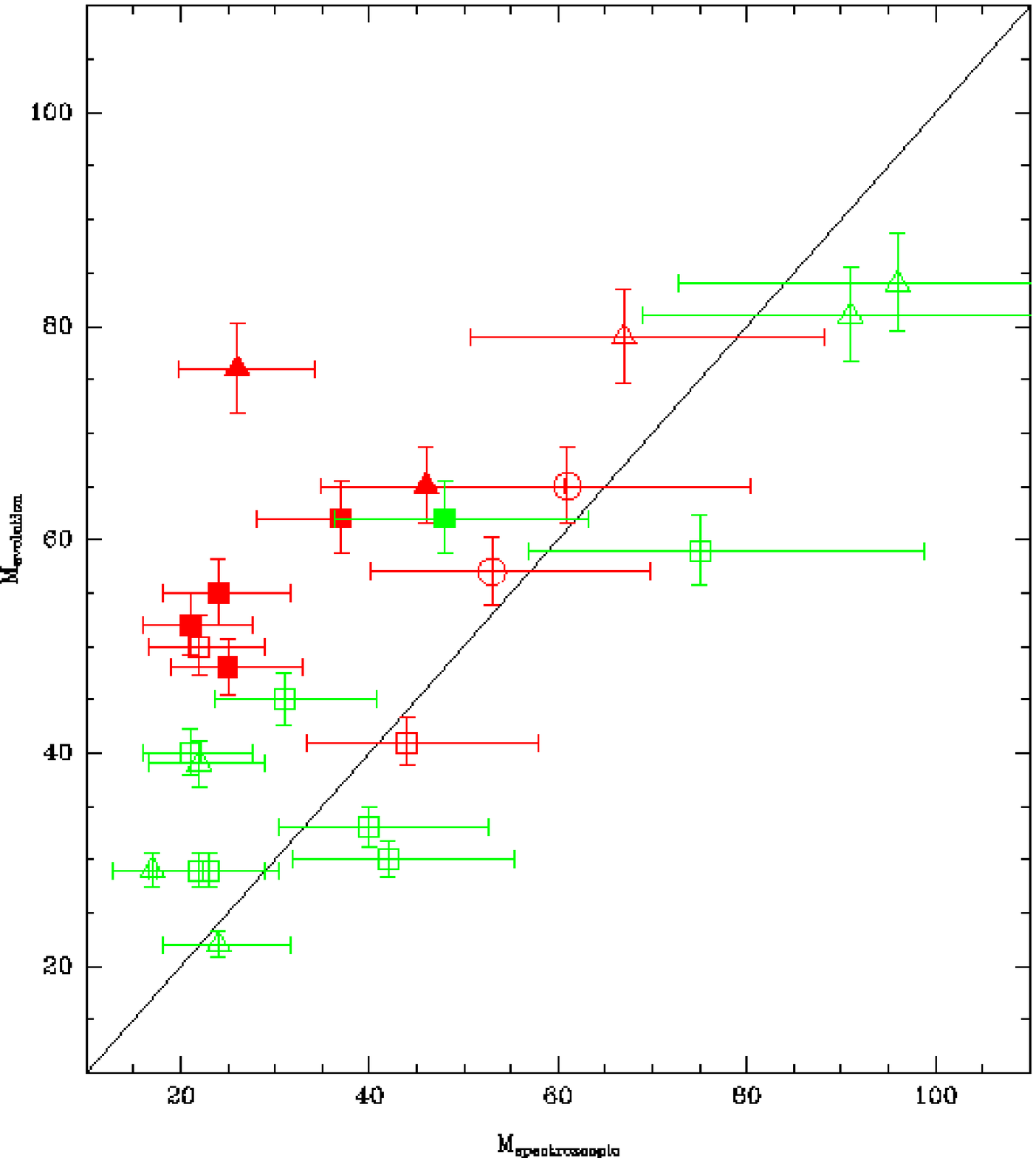}
\caption{\label{fig:massdis} The evolutionary masses are compared to
the spectroscopic masses.  The circles represent supergiants,
the triangles represent giants, and the squares represent dwarfs.
Green shows the data for the SMC, and red for the LMC.  We have
distinguished the hottest stars ($T_{\rm eff}>$45,000) by filled
symbols; all of those stars show a significant mass discrepancy.}
\end{figure}

We have analyzed 40 O-type stars in the Magellanic Clouds, including
many stars of the earliest type.   Modeling was successful for
33 of these stars, with the other 7 showing the spectroscopic signature
of unresolved companions.  This study, in combination with an analysis of
24 Galactic stars by Repolust et al.\ (2004), using similar techniques
and the same model atmosphere code, allow us to obtain the following results:

\begin{enumerate}
\item The effective temperatures of O3-O7 dwarfs and giants in the SMC is
about 4000 K hotter than for stars
of the same spectral type in the Milky Way.  The differences decrease
as one approaches B0~V.  This is readily understood in terms of the
{\it decreased importance} of line- and wind-blanketing at the lower
metallicity that characterizes the SMC.  The results for the LMC appear
to be intermediate between the two galaxies.  A similar effect
is seen for the supergiants, although the differences decrease more
rapidly with increasing spectral type (i.e., Table~\ref{tab:teff} and
Fig.~\ref{fig:teffs}).  For each galaxy, there is no difference in the
effective temperatures of dwarfs and giants of the same spectral type,
while the supergiants are about 4000 K cooler than the dwarfs for the
hottest types in the SMC, and about 6000K cooler than the dwarfs for
the hottest types in the Milky Way. The differences in effective
temperatures between supergiants and dwarfs decrease for the
later O-types stars.  Our effective temperature scale
for dwarfs is in accord with some CMFGEN studies but not for others, and
is significantly hotter than that indicated by WM-Basic modeling the UV.
It is also hotter than the TLUSTY/CMFGEN modeling of NGC~346 stars by
Bouret at al.\ (2003) and Heap et al.\ (2005).  However, 
our scale leads to  more consistent results with
stellar evolutionary theory as evidenced by comparisons of the
degree of coevality of stars in the NGC~346 cluster (i.e, Fig.~\ref{fig:N346}).

\item Our data suggest that the wind momentum ($D_{\rm
mom}\equiv v_\infty\dot{M}R^{0.5}$) scales with luminosity in the
expected way with metallicity as predicted by radiatively driven wind
theory (Kudritzki 2002, Vink et al.\ 2001):  $$D_{\rm mom} \propto
L^{1/\alpha_{\rm eff}},$$ with $\alpha_{\rm eff}\approx0.55$ fairly insensitive
to $T_{\rm eff}$ or $Z$, and with the constant of proportionality
scaling with $Z^{0.82}$.  Two of the SMC stars with low
mass-loss rates does not fit this relationship well, but most of the
others do; a detailed analysis of the UV spectra (to be done in the future)
might provide better constraints on the wind momenta in those cases.

\item Most of the stars in our sample show a reasonable match between
the spectroscopic mass and the evolutionary
mass.
However, stars with $T_{\rm eff} > 45000$K show a
systematic difference, with the spectroscopic mass 
significantly less (by a factor of 2 or more) than the evolutionary
mass.  This is similar to the long-standing ``mass discrepancy" discussed
by Herrero et al.\ (1992) for Galactic supergiants, but which has now
been mostly resolved due to the lower effective temperatures of 
the Galactic models with the effects of line-blanketing included.
Use of newer evolutionary tracks (which contain improved opacities, better
treatment of mass-loss, and the effects of rotation) will tend to
decrease the discrepancy, but such improvement is likely to be only
of order 10\%, and will not account for the factors of 2 discrepancies.
We are
left with the conclusion that the surface gravities or stellar radii
may be underestimated in our models for stars of the highest effective
temperatures.

\item We find that there is little correlation in the {\it physical
properties} (such as $T_{\rm eff}$) with the new spectral types O2-3.5.
Although stars in this group contain the hottest stars, neither the
NIII/NIV emission line ratio nor even the He~I/He~II absorption line
ratio provides a clear indication of the star's effective temperature.
For these extreme spectral types such line ratios appear to be sensitive
to $\log g$ and $\dot{M}$ as well as $T_{\rm eff}$.
Thus although we can predict a star's effective temperature from its
two-dimension spectral type and metallicity for O4's and later, it requires
a detailed analysis of the entire spectrum to derive a reliable effective
temperature (i.e., the mass-loss rate inferred from H$\alpha$ and the
terminal velocity of the wind from UV measurements are needed in addition
to the blue optical spectrum.  

\item Two of the stars with the largest radial velocities with respect
to the LMC are O3 stars listed by Massey et al.\ (1995) as ``field"
stars i.e., they are found far from the nearest OB association.
These provide the first compelling evidence for the origin of this
field population.

\end{enumerate}

Here we wish to emphasize again the point raised in the introduction that
although we have used the best available data and (we believe) the best
models and methodology, that the absolute effective temperature scale of
O stars is likely to undergo revision in the future as the physics
is improved.  Indeed, our study here underscores that other ``best" methods,
such as fitting the UV lines via WM-BASIC, yield temperatures for Galactic
stars that are significant cooler than that found by 
FASTWIND and some CMFGEN modeling of
the optical spectra of an admittedly different sample.  
Reality is that different lines (and different spectral domains) seem to give
different answers at this time, very likely independent of the model code
used. That tells us that the physics used in our models is still imperfect.
It is hard to tell which method is closer to the truth. Both have their
merits.  However, the strength of the current study is that we have been
able to do a strictly differential comparison based on one and the same
method applied to a large sample of objects in three galaxies of different
metallicity. While the absolute scale might still
be uncertain, we believe that the relative differences of the
scales at the different metallicities obtained in our approach are accurate
and significant.

Future work should attempt to identify where improvements are needed; the 
logical place to start is by modeling the UV spectra of the stars studied
here, and by modeling optical data on samples that have been analyzed only
in the UV.  In some cases we believe that better data of the same stars
(such as the NGC 346 sample studied by Bouret et al.\ 2003) may help
resolve differences seen between our effective temperature scale and others,
in much the same way as our better (nebular-subtracted)
data on the R136 stars led to much better model fits (Paper~I).  We keenly 
anticipate the results ahead.

\acknowledgements
We are grateful to the staffs of CTIO, {\it HST}, and {\it FUSE} for providing
first-rate spectroscopic capabilities on well-functioning telescopes; we note
with sadness the loss of STIS, which provided much of the crucial data
in this series.  We also thank Luciana Bianchi and Miriam Garcia for help
in our efforts to analyze the {\it FUSE} data. 
We also acknowledge the strong encouragement of several of our
colleagues over the years for this study, and in particular Peter Conti and
Margaret Hanson.  
Support for {\it HST} programs GO-6417, GO-7739, and GO-9412 were provided by
NASA through a grant from the Space Telescope Science Institute,
which is operated by the Association of Universities for Research in
Astronomy, Inc., under NASA contract NAS 5-26555; support for {\it FUSE}
program C002 was similarly provided through NASA.  An anonymous referee
offered many detailed comments, which led to improvements in the paper.
Chris Evans called our attention to an error in the draft version,
and provided useful information on the issue of sky subtraction of
the NGC~346 data used by Bouret et al.\ (2003) and Heap et al.\ (2005);
Artemio Herrero also offered userful comments.

\clearpage
\begin{deluxetable}{l l l l c c c c c l}
\pagestyle{empty}
\rotate
\tabletypesize{\scriptsize}
\tablewidth{0pc}
\tablenum{1}
\tablecolumns{10}
\tablecaption{\label{tab:stars}Program Stars\tablenotemark{a}}
\tablehead{
\colhead{Name\tablenotemark{b}}
&\colhead{Cat ID\tablenotemark{c}}
&\colhead{$\alpha_{2000}$}
&\colhead{$\delta_{2000}$}
&\colhead{$V$}
&\colhead{$B-V$}
&\colhead{$U-B$}
&\colhead{$E(B-V)$\tablenotemark{d}}
&\colhead{M$_V$\tablenotemark{e}}
&\colhead{Spectral Type\tablenotemark{f}}
}
\startdata 
AV 177                        &SMC-038024&00 56 44.17& -72 03 31.3&14.53&-0.21&-1.05&0.12&-4.78&O4~V((f)) \\
AV 435                        &SMC-067670&01 08 17.88& -71 59 54.3&14.00&-0.06&-0.98&0.28&-5.81&O3~V((f*)) \\
AV 440                        &SMC-068756&01 08 56.01& -71 52 46.5&14.48&-0.18&-1.00&0.15&-4.93&O8 V \\
AV 446                        &SMC-069555&01 09 25.46& -73 09 29.7&14.59&-0.24&-1.06&0.10&-4.66&O6.5~V \\
AV 476                        &SMC-074608&01 13 42.41& -73 17 29.3&13.52&-0.09&-0.93&0.28&-6.29&O2-3 V + comp.\\

Sk$-67^\circ$22=BAT99-12      &LMC-034056&04 57 27.47& -67 39 03.3&13.44&-0.18&-1.05&0.16&-5.53&O2 If* \\
Sk$-65^\circ47$=LH43-18       &\nodata   &05 20 54.67& -65 27 18.3&12.68&-0.13&-0.93:&0.19&-6.39&O4 If \\
LH58-496=LH58-10a\tablenotemark{g} &\nodata   &05 26 44.21& -68 48 42.1&13.73&-0.23&-1.09&0.11&-5.09&O5 V(f) \\
LH81:W28-23                   &\nodata   &05 34 50.11& -69 46 32.3&13.81&-0.16&-1.13&0.15&-5.14 &O3.5~V((f+))\\
LH90:Br58=BAT99-68\tablenotemark{h}   &\nodata&05 35 42.02& -69 11 54.2&14.13 & 0.49&-0.48 & 0.85 & -6.98 & O3If/WN6 \\
LH90:ST2-22                   &\nodata   &05 35 45.26& -69 11 35.1&13.93\tablenotemark{i}&0.18\tablenotemark{i}&-0.73\tablenotemark{i}&0.60&-6.41&O3.5~III(f+) \\
BI~237                        &LMC-164942&05 36 14.68& -67 39 19.3&13.89&-0.12&-0.97&0.25&-5.38&O2 V((f*))\\
BI~253                        &LMC-168644&05 37 34.48& -69 01 10.4&13.76&-0.09&-1.02&0.25&-5.50&O2 V((f*))\\
LH101:W3-14=ST5-52\tablenotemark{j}            &\nodata   &05 39 05.41& -69 29 20.7&13.41&-0.15&-0.89&0.17&-5.60&O3 V + O V Composite\\
LH101:W3-19=ST5-31            &\nodata   &05 39 12.20& -69 30 37.6&12.37\tablenotemark{i}&-0.06\tablenotemark{i}&-0.92\tablenotemark{i}&0.30&-7.04&O2 If* \\
LMC2-675                      &\nodata   &05 43 13.00& -67 51 16.0&13.66&-0.25&-1.14&0.07&-5.04 &O2~III(f*) +  O V Composite \\
R136-007=Mk39\tablenotemark{k}      &  \nodata &05 38 40.3186&-69 06 00.172 & 13.01&\nodata&\nodata&0.47 & -6.9 & O2 If* Composite\\
R136-014=Mk37\tablenotemark{k}      &  \nodata &05 38 42.4986&-69 06 15.396 & 13.57&\nodata&\nodata&0.48 & -6.4 & O3.5 If*\\
R136-018\tablenotemark{k}           &  \nodata &05 38 44.2211&-69 05 56.954 & 13.91&\nodata&\nodata&0.42 & -5.9 & O3 III \\
R136-033\tablenotemark{k}           &  \nodata &05 38 42.2106&-69 06 00.988 & 14.43&\nodata&\nodata&0.35 & -5.1 & O3~V\\
\enddata
\tablenotetext{a}{Coordinates and photometry 
are from Massey 2002 or Massey, Waterhouse, \& DeGioia-Eastwood 2002 unless otherwise noted.}
\tablenotetext{b}{Identifications are as follows: 
``AV" from Azzopardi \& Vigneau 1982;
``BI" from Brunet et al.\ 1975;
``BAT" from Breysacher, Azzopardi, \& Testor 1999;
``Br" from Breysacher 1981;
``LH" from Lucke 1972 except as noted, 
``LMC2" from Massey et al.\ 1995 and Massey 2002;
``Sk" from Sanduleak 1969;
``ST" from Testor \& Niemela 1998;
``W" from Westerlund 1961;
``Mk" from Melnick 1985;
``R136-NNN" from Hunter et al.\ 1997 and Massey \& Hunter 1998}
\tablenotetext{c}{Designations from the catalog of Massey 2002.}
\tablenotetext{d}{From averaging the color excesses in 
$B-V$ and $U-B$ based upon the spectral type.  See Massey 1998b.}
\tablenotetext{e}{Computed using $A_V=3.1\times E(B-V)$, 
with assumed distance moduli for the SMC and LMC of 18.9 and 18.5,
respectively (Westerlund 1997, van den Bergh 2000).}
\tablenotetext{f}{New to this paper.}
\tablenotetext{g}{LH58-10A identification is from Lucke 1972; LH58-496 
identification is from Garmany at al.\ 1994.}
\tablenotetext{h}{Misidentified with a neighboring bright star in Table~4
of Massey 2002.}
\tablenotetext{i}{Photometry new to this paper, based upon the CCD material
described by Massey 2002.}
\tablenotetext{j}{Coordinates and photometry from Testor \& Niemela 1998.}
\tablenotetext{k}{Coordinates and photometry from Hunter et al.~1997
and Massey \& Hunter 1998.}
\end{deluxetable}

\begin{deluxetable}{l c c c c c c c l}
\pagestyle{empty}
\tabletypesize{\tiny}
\tablewidth{0pc}
\tablenum{2}
\tablecolumns{9}
\tablecaption{\label{tab:journal1}Sources of Data Used in This Study}
\tablehead{
\colhead{Spectral Region}
&\colhead{Telescope}
&\colhead{Instrument}
&\colhead{Aperture}
&\colhead{Grating}
&\colhead{Wavelength}
&\colhead{Resolution} 
&\colhead{S/N\tablenotemark{a}}
&\colhead{Use} \\
\colhead{}
&\colhead{}
&\colhead{}
&\colhead{(arcsec x arcsec)}
&\colhead{}
&\colhead{(\AA)}
&\colhead{(\AA)}
&\colhead{}
&\colhead{} \\

}

\startdata

Far-UV        &FUSE/C002&\nodata     &30x30  &\nodata& 905-1187&0.1&40&Terminal velocities\\
UV            &HST/9412 & STIS& 0.2x0.2 &G140L  &1150-1740&0.9 & 30 &Terminal velocities\\
Near-UV       &HST/9412 & STIS & 0.2x52  &G430M  &3020-3300&0.4&10-80&Check on the modeling \\
Blue-optical  &CTIO 4-m & RC Spec & 1.3x330  &KPGLD  &3750-4900&1.4&400-500&Modeling\\
              &HST/7739,9412 & STIS/CCD & 0.2x52  &G430M  &4310-4590&0.4&100&Modeling \\
              &HST/6417      & FOS & 0.26(circ)  &G400M  &3250-4820&3.0 & 40-60 & Spectral class R136 stars\\
H$\alpha$     &CTIO 4-m & RC Spec & 1.3x330  &KPGLD  &5400-7800&2.8 &150 & Modeling\\
              &HST/7739,9412 & STIS/CCD & 0.2x52  &G750M  &6300-6850&0.8& 60 & Modeling\\
\enddata
\tablenotetext{a}{Signal to noise per spectral resolution element.}
\end{deluxetable}

\begin{deluxetable}{l c c c c c}
\pagestyle{empty}
\tabletypesize{\scriptsize}
\tablewidth{0pc}
\tablenum{3}
\tablecolumns{6}
\tablecaption{\label{tab:journal2}Spectral Regions Observed}
\tablehead{
\colhead{Star}
&\colhead{FUV}
&\colhead{UV}
&\colhead{NUV}
&\colhead{Blue-optical}
&\colhead{H$\alpha$}
}

\startdata

AV 177            &FUSE      &HST/9412   & \nodata  & CTIO 4-m & CTIO 4-m, HST/9412 \\     
AV 435            &\nodata   &HST/9412   & \nodata  & CTIO 4-m & CTIO 4-m, HST/9412 \\  
AV 440            &\nodata   &HST/9412   & \nodata  & CTIO 4-m & HST/9412 \\  
AV 446            &\nodata   &HST/9412   & \nodata  & CTIO 4-m & CTIO 4-m, HST/9412\\   
AV 476            &FUSE      &HST/9412
                                        & \nodata  & CTIO 4-m & CTIO 4-m, HST/9412 \\    
Sk$-67^\circ22$   &FUSE      &HST/9412   & \nodata  & CTIO 4-m & HST/9412 \\
Sk$-65^\circ47$   &FUSE      &HST/9412   & HST/9412 & CTIO 4-m, HST/9412 & HST/9412 \\  
LH58-496          &\nodata   &HST/9412   & HST/9412 & CTIO 4-m & HST/9412 \\       
LH81:W28-23       &\nodata   &HST/9412   & HST/9412 & CTIO 4-m & CTIO 4-m, HST/9412 \\            
LH90:Br58         &\nodata   &HST/9412   & \nodata  & CTIO 4-m & CTIO 4-m, HST/9412 \\        
LH90:ST2-22       &\nodata   &HST/9412   & \nodata  & CTIO 4-m & CTIO 4-m, HST/9412 \\           
BI~237            &\nodata   &HST/9412   & \nodata  & CTIO 4-m & CTIO 4-m, HST/9412 \\           
BI~253            &FUSE      &HST/9412   & \nodata  & CTIO 4-m & CTIO 4-m, HST/9412 \\           
LH101:W3-14       &\nodata   &HST/9412   & HST/9412 & CTIO 4-m & CTIO 4-m, HST/9412 \\
LH101:W3-19       &\nodata   &HST/9412   & HST/9412 & CTIO 4-m, HST/9412 & CTIO 4-m, HST/9412 \\
LMC2-675          &\nodata   &HST/9412   & HST/9412 & CTIO 4-m & CTIO 4-m, HST/9412 \\
R136-007          &\nodata   &HST/9412   & \nodata  & HST/7739, HST/6417 & HST/7739 \\
R136-014          &\nodata   &HST/9412   & \nodata  & HST/7739, HST/6417 & HST/7739 \\
R136-018          &\nodata   &HST/9412   & \nodata  & HST/7739, HST/6417 & HST/7739 \\
R136-033          &\nodata   &HST/9412   & \nodata  & HST/7739, HST/6417 & HST/7739 \\

\enddata
\end{deluxetable}

\begin{deluxetable}{l c l l}
\pagestyle{empty}
\tabletypesize{\footnotesize}
\tablewidth{0pc}
\tablenum{4}
\tablecolumns{5}
\tablecaption{\label{tab:termvels}Terminal Velocities in km s$^{-1}$}

\tablehead{
\colhead{Star}
&\colhead{$v_\infty$}
&\colhead{Lines Used}
&\colhead{Comments}
}

\startdata

AV 177		&2650		&CIV, NV	&	              \\
AV 435		&1500::		&CIV, SiIV	&Weak CIV\\
AV 440		&1300:		&CIV		&Weak CIV \\
AV 446		&1400:		&CIV		&Weak CIV\\
AV 476		&2670		&CIV		&  \\
Sk$-67^\circ 22$	&2650		&CIV		& \\
Sk$-65^\circ 47$		&2100		&CIV&\\
LH58-496	&2400::		&CIV		&Wide CIV, plus abs. contamination \\
LH81:W28-23		&3050		&CIV&\\
LH90:Br58	&1900:		&SiIV		&Could not use CIV\\
LH90:ST2-22	&2560		&CIV&\\
BI 237		&3400		&CIV	&\\
BI 253		&3180		&CIV&\\
LH101:W3-14	&3100		&CIV&\\
LH101:W3-19	&2850		&CIV            &\\
LMC2-675	&3200		&CIV&\\
R136-007	&2100		&CIV, SiIV&\\
R136-014	&2000		&CIV, SiIV	&Velocity scale uncertain\\
R136-018	&3200		&CIV&  \\
R136-033	&3250		&CIV&  \\
\enddata
\end{deluxetable}

\begin{deluxetable}{l l c c c c c c c r c c c r l}
\pagestyle{empty}
\rotate
\tabletypesize{\tiny}
\tablewidth{0pc}
\tablenum{5}
\tablecolumns{15}
\tablecaption{\label{tab:results}Results of Model Fits}
\tablehead{
\colhead{Name}
&\colhead{Spectral}
&\colhead{$T_{\rm eff}$}
&\colhead{$\log g_{\rm eff}$}
&\colhead{$\log g_{\rm true}$\tablenotemark{a}}
&\colhead{$R$}
&\colhead{M$_V$}
&\colhead{BC}
&\colhead{M$_{\rm bol}$}
&\colhead{Mass}
&\colhead{$\dot{M}$}
&\colhead{$\beta$}
&\colhead{$v_\infty$}
&\colhead{He/H\tablenotemark{b}}
&\colhead{Comments} \\
\colhead{}
&\colhead{Type}
&\colhead{($^\circ$K)}
&\colhead{[cgs]}
&\colhead{[cgs]}
&\colhead{($R_\odot$)}
&\colhead{mags}
&\colhead{mags}
&\colhead{mags}
&\colhead{$M_\odot$}
&\colhead{($10^{-6} M_\odot$ yr$^{-1}$)}
&\colhead{}
&\colhead{(km s$^{-1}$)}
&\colhead{}
&\colhead{}
}
\startdata 
AV 177      &O4~V((f)) & 44000 &    3.80&3.85&   8.9& -4.78 & -4.04 & -8.82  & 21  & 0.3&   0.8&   2650&    0.15& $\dot{M}<~0.5$   \\
AV 435      &O3~V((f*))& 45000 &    3.80&3.81&  14.2&-5.81 & -4.12 & -9.93 & 48&   0.5 &    0.8 &  1500:: &   0.10& \\
AV 440      &O8 V      & 37000 &    4.00&4.01&  10.6&-4.93 & -3.52 & -8.45 & 42 &  0.1 &   0.8  & 1300:   & 0.12 & $\dot{M}<~0.3$   \\
AV 446      &O6.5 V    & 41000 &    4.15&4.15 & 8.8& -4.66 & -3.82 & -8.48 & 40   &  0.1 &    0.8 &  1400 &   0.15 & $\dot{M}<~0.3$  \\
AV 476      &O2-3 V+comp. &\nodata &\nodata &\nodata &-6.29 &\nodata &\nodata &\nodata  &\nodata &\nodata&\nodata&2670& \nodata & Composite \\
Sk $-67^\circ$22&O2 If* & $>42000$& $\approx$3.5&$\approx$3.56&$\approx$13.2& -5.53&$<-3.94$&-9.47 & $\approx$23 &  15  & 0.8 & 2650 & 0.30 & $T_{\rm eff}$ lower limit\\
Sk $-65^\circ$47&O4 If & 40000 & 3.60&3.62 & 20.1 & -6.39 & -3.79 & -10.18 & 61 & 12 &   0.8 & 2100 & 0.10 & \\
LH58-496 & O5 V & 42000 & 4.00&4.04 & 10.5 & -5.09 & -3.88 & -8.97 & 44  & 0.6 & 0.8 & 2400 & 0.10 & \\
LH81:W28-23 & O3.5 V((f+)) & 47500 & 3.80&3.81 & 10.0 & -5.14 & -4.26 & -9.40 & 24 & 2.5 & 0.8 & 3050 & 0.25 & \\
LH90:Br58 & O3If/WN6 & 40000-42000: & 3.5: & 3.5: & 30: & -6.98 & -4.1:: & -11.1: & 100:: & 40: & 3.0 & 1900 & 0.1:: & See text\\ 
LH90:ST2-22 & O3.5~III(f+) & 44000 & 3.70&3.71& 18.9 & -6.41 & -4.05 & -10.46 & 67 & 4.5 & 0.8 & 2560 & 0.20 & \\
BI~237 & O2~V((f*)) & 48000 & 3.90&3.92 & 11.1 & -5.38 & -4.29 & -9.67 & 37 & 2.0 & 0.8 & 3400 & 0.10\\
BI~253 & O2~V((f*)) & $>48000$ & $\approx$3.90&$\approx$3.93 & $\approx$11.8 & -5.50 & $<$-4.30 & $<$-9.80 & $\approx$43 & 3.5 & 0.8 & 3180 & 0.10 & $T_{\rm eff}$ lower limit  \\
LH101:W3-14  & O3~V + O~V & \nodata & \nodata & \nodata & -5.60 & \nodata & \nodata & \nodata & \nodata & \nodata &\nodata  &3100 &  \nodata& Composite \\
LH101:W3-19 & O2 If* & $>44000$ & $\approx$3.90 &$\approx$3.91& 
$\approx25.5$ & -7.04 & $<$-4.06 & $<$-11.10 & $\approx$193& 20 & 0.8& 2850 & 0.10 & $T_{\rm eff}$ lower limit\\
LMC2-675 & O2 III(f*) + O~V & \nodata & \nodata & \nodata & -5.04 & \nodata & \nodata & \nodata & \nodata & \nodata &\nodata & 3200 & \nodata & Composite \\
R136-007 & O2 If* Composite & \nodata & \nodata & \nodata & -6.9 & \nodata & \nodata & \nodata & \nodata & \nodata &\nodata &  2100 & \nodata & Composite \\
R136-014 & O3.5 If* & 38000 & 3.50&3.51 & 21.1 & -6.4 & -3.65 & -10.0 & 53 & 23 & 0.8 & 2000 & 0.10 & \\
R136-018 & O3 III(f*) & 45000 & 3.75&3.77 & 14.7 & -5.9 & -4.11 & -10.0 & 46 & 2.0 & 0.8 & 3200 & 0.10 & \\
R136-033 & O2-3.5 V & 47000 & 3.75 &3.77& 9.8 & -5.1 & -4.23 & -9.3 & 21 & 2.0 & 0.8 & 3250 & 0.10 & \\
\enddata
\tablenotetext{a}{By number.}
\end{deluxetable}

\begin{deluxetable}{l c c c}
\pagestyle{empty}
\tabletypesize{\footnotesize}
\tablewidth{0pc}
\tablenum{6}
\tablecolumns{4}
\tablecaption{\label{tab:ovi}OVI vs.\ CIV Terminal Velocities (km s$^{-1}$)}
\tablehead{
\colhead{Star}
&\colhead{Sp.Type}
&\colhead{OVI}
&\colhead{CIV}
}
\startdata
AV 177          &O4 V((f))& 2500 & 2650 \\
Sk$-67^\circ$22 &O2 If*   & 2650 & 2650 \\
Sk$-65^\circ$47 &O4 If    & 2000 & 2100 \\
BI 253          &O2 V((f*))& 3200 & 3180 \\
\enddata
\end{deluxetable}

\clearpage
\begin{deluxetable}{l l c c c c c c c c c c l}
\pagestyle{empty}
\rotate
\tabletypesize{\tiny}
\tablewidth{0pc}
\tablenum{7}
\tablecolumns{13}
\tablecaption{\label{tab:o3s}Line Strengths in the Earliest O-type Stars}
\tablehead{
\colhead{Star}
&\colhead{Type}
&\colhead{$T_{\rm eff}$}
&\colhead{$\log g_{\rm true}$ }
&\colhead{$\dot{M}$ }
&\multicolumn{2}{c}{EWs [m\AA]\tablenotemark{a}}
&\colhead{$\log\frac{{\rm EW(HeI)}}{{\rm EW(HeII)}}$}
&\colhead{}
&\multicolumn{2}{c}{EWs [m\AA]\tablenotemark{b}}
&\colhead{$\log\frac{{\rm EW(NIII)}}{{\rm EW(NIV)}}$}
&\colhead{Comments}  \\  \cline{6-7} \cline{10-11}
 & & \colhead{(1000 K)} & \colhead{(cgs)} 
&\colhead{($10^{-6} M_\odot$ yr$^{-1}$)} 
&\colhead{He~I $\lambda 4471$}  
 & \colhead{He~II $\lambda 4542$}
 & &
 & \colhead{N IV $\lambda 4058$}
 & \colhead{N III $\lambda 4634,42$} \\

}
\startdata
\sidehead{DWARFS}
BI 237      &O2 V((f*))& 48.0  & 3.92          & 2.0 &$ 20\pm10$ &$750\pm10$ &$ -1.6\pm0.2$  &&  $-165\pm20$  & $-90\pm20$   &$ -0.3\pm0.1$  & C~IV $\lambda$ 4658 \\
BI 253      &O2 V((f*))&$>$48.0& $\approx3.93$ & 3.5 &$<25\pm10$ &$710\pm10$ &$<-1.5\pm0.2$  &&  $-450\pm20$  &$>-60\pm20$   &$<-0.9\pm0.1$  & C~IV $\lambda$ 4658 \\
\\
AV 435\tablenotemark{a}&O3 V((f*))& 45.0& 3.81 & 0.5 &$125\pm10$ &$750\pm10$ &$ -0.8\pm0.0$  &&  $ -80\pm20$  &$>-60\pm20$   &$<-0.1\pm0.2$ \\ 
\\
LH101:W3-24 &O3.5 V((f+))&  48.0 & 4.01 &0.5         &$120\pm10$ &$790\pm10$ &$ -0.8\pm0.0$  &&  $>-20\pm20$  &$-200\pm20$   &$>1.0\pm0.4$  &  O3 V((f)) in Pap I. \\
LH81:W28-23 &O3.5 V((f+))&  47.5 & 3.81 &2.5         &$ 80\pm10$ &$830\pm10$ &$ -1.0\pm0.1$  &&  $-120\pm20$  &$-500\pm20$   &$ 0.6\pm0.1$  \\
\\
AV 177\tablenotemark{a}&O4 V((f)) &44.0 & 3.85 &0.3  &$170\pm10$ &$950\pm10$ &$ -0.8\pm0.0$  &&  $>-50\pm20$  &$-250\pm20$   &$>0.7\pm0.2$  \\
LH81:W28-5             &O4 V((f+))&46.0 & 3.81 &1.2  &$200\pm10$ &$890\pm10$ &$ -0.7\pm0.0$  &&  $ -50\pm20$  &$-480\pm20$   &$ 1.0\pm0.2$  \\
\\
R136-055    &O2-3.5 V   &47.5 &3.81 &0.9             &$ 35\pm20$ &$650\pm20$ &$ -1.3\pm0.2$  &&  $>-200\pm100$&$>-200\pm100$  &\nodata &  O3 V in Paper I \\
R136-033    &O2-3.5 V   &47.0 &3.77 &2.0             &$ 50\pm20$ &$790\pm20$ &$<-1.2\pm0.2$  &&  $>-200\pm100$&$>-200\pm100$  &\nodata  & O3 V in this paper \\
R136-040    &O2-3.5 V  &$>$51.0 &$\approx$3.81  &2.0 &$<100\pm50$&$760\pm20$ &$<-0.9\pm0.2$  &&  $>-200\pm100$&$>-200\pm100$  &\nodata & O3 V in Paper I \\

\sidehead{GIANTS}
LH64-16    &ON2 III(f*) &  54.5 & 3.91 &4.0          &$100\pm10$&$1090\pm10$ &$ -1.0\pm0.0$  &&  $ -660\pm20$ &$ -120\pm20$   & $-0.7\pm0.1$ &  C~III $\lambda$ 4658\\
R136-047   &O2 III(f*)  & $>$51.0 & $\approx$3.91&6.0&$<30\pm20$&$ 450\pm20$ &$<-1.2\pm0.3$  &&  $ -700\pm100$&$>-250\pm100$  &$<-0.5\pm0.2$ \\
\\
R136-018   &O3 III(f*)  & 45.0 & 3.77 &2.0           &$ 90\pm20$&$ 670\pm20$ &$ -0.9\pm0.1$  &&  $ -750\pm100$&$>-150\pm100$  &$<-0.7\pm0.3$  \\
\\
LH90:ST2-22&O3.5 III(f+)& 44.0 & 3.71  &4.5          &$140\pm10$&$ 880\pm10$ &$ -0.9\pm0.0$  &&  $ -230\pm20$  & $-750\pm20$  &$0.5\pm0.0$ \\

\sidehead{SUPERGIANTS}
Sk$-$67 22  &O2 If*    & $>$42.0 & $\approx$3.56&15   &$<50\pm20$&$650\pm20$     &$<-1.1\pm0.2$  &&  $-1610\pm20$ & $-350\pm20$  & $-0.7\pm0.0$  & C~IV $\lambda 4658$ \\
LH101:W3-19&O2 If*     & $>$44.0 & $\approx$3.91&$<$20&$<\sim50\pm20$ &$570\pm20$&$<-1.1\pm0.2$ &&  $-300\pm20$ & $-230\pm20$  & $-0.1\pm0.1$  & C~IV $\lambda 4658$ \\
R136-020   &O2 If*     & $>$42.5 & $\approx$3.61& 23      & $<30\pm20$ &$510\pm20$&$<-1.2\pm0.3$ &&  $-2570\pm50$ & $>-100\pm50$  &$<-1.4\pm0.2$   \\
R136-036   &O2 If*     & $>$43.0 & $\approx$3.71& 14      & $<50\pm20$ &$550\pm20$&$<-1.0\pm0.2$ &&  $-2020\pm100$ & $>-200\pm100$&$<-1.0\pm0.2$  \\
\\
LH90:Br58\tablenotemark{b}  &O3 If/WN6  & 40$-$42& 3.5:&40:&$<50\pm20$ &$400\pm20$&$<-0.9\pm0.2$ &&  $-1010\pm20$  & $-480\pm20$  & $-0.3\pm0.0$ \\
\\
R136-014   &O3.5 If*  &   38.0 & 3.51 & 23&        $40\pm20$   &$400\pm20$   &  $-1.0\pm0.2$     &&  $-1360\pm100$  & $-1400\pm100$ & $0.0\pm0.0$ \\
\\
Sk$-$65$^\circ$ 47   &O4 If & 40.0 & 3.62 &12&     $115\pm10$   &$550\pm10$  &  $-0.7\pm0.0$     &&  $ -175\pm20$  & $-1540\pm20$ &  $0.9\pm0.1$  \\
\enddata
\tablenotetext{a}{SMC star.}
\tablenotetext{b}{This star is excluded from consideration in 
Fig.~\ref{fig:o3teff} due to the uncertainity in its effective temperature.
}
\end{deluxetable}

\begin{deluxetable}{l c c c c c}
\pagestyle{empty}
\tablewidth{0pc}
\tablenum{8}
\tablecolumns{6}
\tablecaption{\label{tab:teff}Effective Temperature Scale [K]}
\tablehead{
\colhead{Type}
&\multicolumn{2}{c}{Milky Way}
&
&\multicolumn{2}{c}{SMC} \\ \cline{2-3} \cline{5-6}
\colhead{}
&\colhead{V+III}
&\colhead{I}
&
&\colhead{V+III}
&\colhead{I}
}
\startdata
O3  &46500 &40250 &&49500 &45250 \\
O4  &44000 &39000 &&47750 &43000 \\
O5  &41000 &37750 &&45000 &41000 \\
O5.5&39500 &36750 &&43500 &40000 \\
O6  &38250 &36000 &&42250 &38500 \\
O6.5&37000 &35500 &&41000 &37500 \\
O7  &36000 &34750 &&39250 &36250 \\
O7.5&34750 &34000 &&37750 &35250 \\
O8  &33750 &33000 &&36250 &34000 \\
O8.5&32750 &32500 &&34500 &33000 \\
O9  &31750 &31750 &&33000 &32000 \\
O9.5&30750 &30750 &&31500 &30750 \\
B0  &30000 &29750 &&30000 &29750 \\
\enddata
\end{deluxetable}

\clearpage

\begin{deluxetable}{l l c c c c c c c c c c c}
\pagestyle{empty}
\tabletypesize{\tiny}
\tablecolumns{13}
\tablewidth{0pc}
\rotate
\tablenum{9}
\tablecaption{\label{tab:finalresults}Parameters for the Complete Sample}
\tablehead{
\colhead{Star}
&\colhead{Spectral}
&\colhead{$v \sin i$}
&\colhead{$T_{\rm eff}$}
&\colhead{$\log g_{\rm true}$}
&\colhead{$\dot{M} [10^{-6}$}
&\colhead{$v_\infty$}
&\colhead{$R/R_\odot$}
&\colhead{He/H}
&\colhead{$\log D_{\rm mom}$}
&\colhead{$\log L/L_\odot$}
&\colhead{$M_{\rm spect}$}
&\colhead{$M_{\rm evol}$\tablenotemark{a}}   \\
\colhead{}
&\colhead{Type}
&\colhead{km s$^{-1}$}
&\colhead{[1000 K]}
&\colhead{[cgs]}
&\colhead{$M_\odot$ yr$^{-1}$]}
&\colhead{[km s$^{-1}$]}
&\colhead{}
&\colhead{[num.]}
&\colhead{}
&\colhead{}
&\colhead{}
&\colhead{} \\
}
\startdata
\sidehead{DWARFS} 
R136-040    &O2-3.5 V&120&     $>$51.0 &$\approx$3.81   &2.0 &3400 &$\approx$10.3 &0.10& $\approx$29.14 &$>$5.82  &$\approx$25  &$>$70\\
BI 253      &O2 V((f*)) &200&  $>$48.0 & $\approx$3.93  & 3.5 &3180& $\approx$11.8&0.10& $\approx$29.38 &$>$5.82 & $\approx$43  &$>$64\\
BI 237      &O2 V((f*))&150   & 48.0 & 3.92  & 2.0 &3400 & 11.1 &0.10 &29.16 & 5.77 & 37&  62\\
R136-033    &O2-3.5 V &120     &47.0 & 3.77 & 2.0& 3250 &  9.8& 0.10 & 29.11 & 5.6  & 21&  52\\
R136-055    &O2-3.5 V &120    & 47.5 & 3.81  & 0.9 &3250  & 9.4 & 0.10& 28.75 & 5.6  & 21&  52\\
LH101:W3-24 &O3 V((f))&120    & 48.0  &4.01  & 0.5 &2400  & 8.1 & 0.15& 28.33 & 5.49  & 25  & 48 \\
AV 435\tablenotemark{b} &O3 V((f*))&110   & 45.0 & 3.81  & 0.5 &1500:: &14.2&0.10&  28.3::& 5.87  & 48  & 62\\
LH81:W28-23 &O3 V((f+)) &120   & 47.5 & 3.81  & 2.5 &3050 & 10.0 & 0.20 & 29.18 & 5.66  & 24  & 55\\
AV 177\tablenotemark{b}   &O4 V((f))&220    & 44.0 & 3.85  & 0.3 &2650 &  8.9&0.15&  28.18 & 5.43  & 21  & 40\\
LH81:W28-5  &O4 V((f+))&120   & 46.0&  3.81 &  1.2 &2700 &  9.6 &0.20& 28.80  &5.58 &  22 &  50\\
AV 377\tablenotemark{b}    &O5 V((f))&120    & 45.5 & 4.01  &$<$0.3 &2350 &  9.1&0.35& $<$28.13 & 5.52  & 31  & 45\\
AV 14\tablenotemark{b}      &O5 V &150       & 44.0 & 4.01 & $<$0.3 &2000 & 14.2&0.10& $<$28.15 & 5.85 &  75  & 59\\
LH58-496    &O5 V &250        & 42.0 & 4.04 &  0.6 &2400 & 10.5 &0.10& 28.47 & 5.49 &  44  & 41\\
AV 446\tablenotemark{b}     &O6.5 V&95       & 41.0 & 4.15 &$<$0.3& 1400 &  8.8 &0.15&$<$27.90 & 5.29 &  40 &  33\\
AV 207\tablenotemark{b}     &O7.5 V((f))&120   &37.0 & 3.72 & $<$0.3& 2000 & 11.0 &0.10 &$<$28.10 & 5.32  & 23 &  29\\
AV 296\tablenotemark{b}     &O7.5 V((f)) &300 & 35.0 & 3.63 &  0.5 & 2000  &11.9& 0.10& 28.34 & 5.28  & 22  & 29\\
AV 440\tablenotemark{b}     &O8 V&100         & 37.0 & 4.01 & $<$0.3 &1300: &10.6 &0.12&$<$27.9:&  5.28 &  42 &  30\\
\sidehead{GIANTS}
LH64-16     &ON2 III(f*)&120  & 54.5 & 3.91 &  4.0 &3250 &  9.4& 1.0& 29.40 & 5.85  & 26 &   76\\
R136-047    &O2 III(f*) &120  & $>$51.0 & $\approx$3.91 & 6.0 &3500& 
$\approx$10.4 &0.10
& $\approx$28.64 &$>$5.82  &$\approx$32  & $>$70\\
R136-018    &O3 III(f*) &180  & 45.0 & 3.77&  2.0& 3200 & 14.7 &0.10& 29.19 & 5.9  &  46 &   65\\
LH90:ST2-22 &O3.5 III(f+)&120 & 44.0 & 3.71  & 4.5 &2560 & 18.9&0.20 & 29.50 & 6.08  & 67  &  79\\
AV 378\tablenotemark{b}     &O9.5 III&110     & 31.5 & 3.27 & \nodata&\nodata   &15.4 &0.15& \nodata   & 5.34  & 17 &   29\\
AV 396\tablenotemark{b}     &B0 III &120      & 30.0 & 3.52 &  \nodata& \nodata & 14.1 &0.15& \nodata  & 5.17  & 24 &   22\\
\sidehead{SUPERGIANTS}
LH101:W3-19 &O2 If*  &180      & $>$44.0& $\approx$3.91 &  20\tablenotemark{c} & 2850& $\approx$25.5&0.10 &$\approx$30.26 &$>$6.34 &$\approx$193 & $>$109\\
R136-036    &O2 If* &120      & $>$43.0&$\approx$3.71  & 14\tablenotemark{c} & 3700 &$\approx$12.8&0.10 &$\approx$30.07 &$>$5.7   &$\approx$31 & $>$50\\
R136-020    &O2 If* &120       & $>$42.5& $\approx$3.61  & 23\tablenotemark{c} & 3400 &$\approx$16.4&0.20 &$\approx$30.30& $>$5.9  & $>$40 & $>$62\\
Sk -67 22   &O2 If* &200     &  $>$42.0& $\approx$3.56  & 15\tablenotemark{c} & 2650 &$\approx$13.2&0.30 &$\approx$29.96 &$\approx$5.69 & $>$23 & $>$49\\
LH90:Br58   &O3If/WN6 &\nodata &40.0-42.0& 3.5:&  40:\tablenotemark{c} &1900 & 30: &  30.4&0.1:: &  6.3:  & 40:& 101:\\
R136-014    &O3.5 If* &120     & 38.0&  3.51 & 23\tablenotemark{c} & 2000 & 21.1&0.10 & 30.12 & 5.9  &  53  & 57\\
Sk -65 47   &O4 If &160         & 40.0&  3.62 & 12\tablenotemark{c}  &2100 & 20.1 &0.10& 29.85  &5.97  & 61  & 65\\
AV 75\tablenotemark{b}      &O5.5 I(f)&120     & 40.0&  3.61& 3.5  &2100 & 25.4 &0.10 &29.37 & 6.16  & 96  &  84\\
AV 26\tablenotemark{b}      &O6 I(f) &150      & 38.0 & 3.52& 2.5  &2150&  27.5&0.10 & 29.25 & 6.14 &  91  &  81\\
AV 469\tablenotemark{b}     &O8.5 I(f) &120    & 32.0 & 3.13 &1.8  &2000& 21.2&0.20  &29.02 & 5.64  & 22  &  39\\
\enddata
\tablenotetext{a}{From the non-rotating models of Charbonnel et al.\ 1993 (SMC)
and Schaerer et al.\ 1993 (LMC).}
\tablenotetext{b}{SMC}
\tablenotetext{c}{H$\alpha$ in emission; these values for $\dot{M}$ should be 
reduced by a factor of 0.44 to allow
for the effects of clumping in the stellar wind.  See text.}
\end{deluxetable}

\clearpage
\clearpage

\clearpage

\clearpage

\clearpage

\clearpage

\clearpage

\clearpage

\clearpage

\clearpage

\clearpage

\clearpage

\clearpage

\clearpage

\clearpage

\clearpage

\clearpage

\clearpage

\clearpage

\clearpage

\clearpage

\clearpage

\clearpage

\clearpage

\clearpage

\clearpage

\clearpage

\clearpage

\clearpage

\clearpage

\clearpage

\clearpage

\clearpage


\begin{references}

\reference {} Abbott, D. C., \& Hummer, D. G. 1985, ApJ, 294, 286

\reference {} Auer, L. H., \& Mihalas, D. 1972, ApJS, 24, 193

\reference {} Azzopardi, M., \& Vigneau, J. 1982, A\&AS, 30, 261

\reference {} Bianchi, L., \& Garcia, M. 2002, ApJ, 581, 610

\reference {} Bouret, J.-C., Lanz, T., Hillier, D. J., Heap, S. R.,
Hubeny, I., Lennon, D. J., Smith, L. J., \& Evans, C. J. 2003,
ApJ, 595, 1182

\reference {} Breysacher, J. 1981, A\&AS, 43, 20

\reference {} Breysacher, J., Azzopardi, M., \& Testor, G. 1999, A\&AS, 137, 117

\reference {} Brunet, J. P., Imbert, N., Martin, N., Mianes, P.,
Prevot, L., Rebeirot, E., \& Rousseau, J. 1975, A\&AS, 21, 109

\reference {} Burkholder, V., Massey, P., \& Morrell, N. 1997, ApJ 490, 328

\reference {} Charbonnel, C., Meynet, G., Maeder, A., Schaller, G., \& Schaerer, 
D. 1993, A\&AS, 101, 415

\reference {} Conti, P. S. 1988, in O Stars and Wolf-Rayet Stars,
ed.\ P. S. Conti \& A. B. Underhill, NASA SP-497

\reference {} Conti, P. S., \& Alschuler, W. R. 1971, ApJ, 170, 325

\reference {} Conti, P. S., \& Bohannan, B. 1989, in Physics of Luminous
Blue Variables, ed.\ K. Davidson, A. F. J. Moffat, \& H. J. G. L. M. Lamers
(Dorrecht: Kluwer), 297

\reference {} Conti, P. S., \& Frost, S. A. 1977, ApJ, 212, 728
 
\reference {} Conti, P. S., Garmany, C. D., \& Massey, P. 1986, AJ, 92, 48

\reference {} Crampton, D. C., \& Greasley, J. 1982, PASP, 94, 31

\reference {} Crowther, P. A., Hillier, D. J., Evans, C. J., Fullerton, A. W.,
De Marco, O., \& Willis, A. J. 2002, ApJ, 579, 774

\reference {} Evans, C. J., Crowther, P. A., Fullerton, A. W., \& Hillier, D. J. 
2004, ApJ 610, 102

\reference {} Evans, C. J. et al.\ 2005, A\&A, submitted

\reference {} Garcia, M., \& Bianchi, L. 2004, ApJ, 606, 497

\reference {} Garmany, C. D., Conti, P. S., \& Massey, P. 1987, AJ, 93, 1070

\reference {} Garmany, C. D., Massey, P., \& Parker, J. W. 1994, AJ, 108, 1256

\reference {} Groenewegen, M. A. T., Lamers, H. J. G. L. M., \& Pauldrach, A. W. A. 
1989, A\&A, 221, 78

\reference {} Hanson, M. M. 2003, ApJ, 597, 957

\reference {} Haser, S. M., 1995, Ph.D. thesis, Ludwig-Maximilians Univ.

\reference {} Haser, S. M., Lennon, D. J., Kudritzki, R.-P., Puls, J.,
Pauldrach, A. W. A., Bianchi, L., \& Hutchings, J. B. 1995, A\&A, 295,
136

\reference {} Heap, S. R., Lanz, T., \& Hubeny, I. 2005, ApJ, submitted
(astro-ph/0412345)

\reference {} Herrero, A. 2003, in A Massive Star Odyssey: From
Main Sequence to Supernova, IAU Symp.~212, ed. K. A. van der Hucht, A.
Herrero, \& C. Esteban (San Francisco: ASP), 3

\reference {} Herrero, A., Kudritzki, R. P., Vilchez, J. M.,Kunze, D.,
Butler, K., \& Haser, S. 1992, A\&A, 261, 209

\reference {} Herrero, A., Puls, J., \& Najarro, F. 2002, A\&A, 396, 949

\reference {} Hillier, D., Lanz, T., Heap, S. R., Hubeny, I., Smith, L. J.,
Evans, C. J., Lennon, D. J., \& Bouret, J. C. 2003, ApJ, 588, 1039

\reference {} Hummer, D. G. 1982, ApJ< 257, 724

\reference {} Hunter, D. A., Vacca,W. D., Massey, P., Lynds, R.,
\& Oneil, E. J. 1997, AJ, 113, 1691

\reference {} Kim, S., Staveley-Smith, L., Dopita, M. A., Freeman, K. C.,
Sault, R. J., Kesteven, M. J., \& McConnell, D. 1998, ApJ, 503, 674


\reference {} Kudritzki, R. P. 1980, A\&A, 85, 174

\reference {} Kudritzki, R. P. 1998, in Stellar Astrophysics for the Local
Group, ed. A. Aparicio, A. Herrero., \& F. Sanchez (Cambridge: Cambridge
Univ.\ Press), 149

\reference {} Kudritzki, R. P. 2002, ApJ, 577, 389

\reference {} Kudritzki, R. P., Lennon, D. J., \& Puls, J. 1995, in
Science with the VLT, ed. J. R. Walsh \& I. J. Danziger (Berlin: Springer-Verlag), 246

\reference {} Kudritzki, R. P., Pauldrach, A., Pul, J., \& Abbott, D. C. 1989,
A\&A, 219, 205


\reference {} Lamers, H. J. G. L. M., Cerruti-Sola, M., \& Perinotto, M. 1987,
ApJ, 314, 726

\reference {} Larsen, S. S., Clausen, J. V., \& Storm, J. 2000, A\&A, 364, 455

\reference {} Leitherer, C., Robert, C., \& Drissen, L. 1992, ApJ, 401, 596

\reference {} Lucke, P. 1972, Ph.D. thesis, Univ.~Washington

\reference {} Maeder, A., \& Meynet, G. 2000, ARA\&A, 38, 143


\reference {} Martins, F., Schaerer, D., \& Heydari-Malayeri 2004, A\&A, 420, 1087

\reference {} Martins, F., Schaerer, D., \& Hillier, D. J. 2002, A\&A, 382, 999

\reference {} Massey, P. 1998, in The Stellar Initial Mass Function,
38th Herstmonceux Conference, ed. G. Gilmore \& D. Howell (San Francisco,
ASP), 17

\reference {} Massey, P. 2002, ApJS, 141, 81

\reference {} Massey, P. 2003, ARA\&A, 41, 15

\reference {} Massey, P., \& Conti, P. S. 1977, ApJ, 218, 431

\reference {} Massey, P., Bresolin, F., Kudritzki, R. P., Puls, J.,
\& Pauldrach, A. W. A. 2004, ApJ, 608, 1001 (Paper I)

\reference {} Massey, P., \& Hunter, D. A. 1998, ApJ, 493, 180

\reference {} Massey, P., Lang, C. C., DeGioia-Eastwood, K., \&
Garmany, C. D. 1995, ApJ, 438, 188

\reference {} Massey, P., Parker, J. W., \& Garmany, C. D. 1989, AJ, 98, 1305

\reference {} Massey, P., Penny, L. R., \& Vukovich, J. 2002, ApJ, 565, 982

\reference {} Massey, P., Waterhouse, E., \& DeGioia-Eastwood, K. 2000,
AJ, 119, 2214

\reference {} Melnick, J. 1985, A\&A, 153, 235

\reference {} Meynet, G., \& Maeder, A. 2004, A\&A, in press

\reference {} Mihalas, D., \& Hummer, D. G. 1973, ApJ, 179, 827

\reference {} Moos, H. W. et al.\ 2000, ApJ, 538, L1

\reference {} Morgan, W. W., Keenan, P. C., \& Kellman, E. 1943,
An Atlas of Stellar Spectra, (Chicago: Univ.~Chicago Press)

\reference {} Morrison, N. D. 1975, ApJ, 202, 433

\reference {} Nelan, E. P., Walborn, N. R., Wallace, D. J., Moffat, A. F. J.,
Makidon, R. B., Gies, D. R., \& Panagia, N. 2004, AJ, 128, 323

\reference {} Oey, M. S. 2004, in The Local Group as an Astrophysical
Laboratory, ed. M. Livio (Cambridge, Cambrdige University Press), in 
press, astro-ph/0307131

\reference {} Oey, M. S., \& Kennicutt, R. C., Jr. 1997,
MNRAS, 291, 827

\reference {} Owocki, S. P., Gayley, I. G., Cranmer, S. R. 1998,
in Boulder-Munich II: Properties of Hot, Luminous Stars, ed. I. D. Howarth
(San Francisco: ASP), 237

\reference {} Pauldrach, A. W. A., Hoffmann, T. L., \& Lennon, M. 2001, A\&A, 375, 161

\reference {} Pauldrach, A. W. A., Kudritzki, R. P., Puls, J., \& Butler, K.
1990, A\&A, 228, 125


\reference {} Pauldrach, A. W. A., Kudritzki, R. P., Puls, J., Butler, K.,
\& Hunsinger, J. 1994, A\&A, 283, 525

\reference {} Puls, J., Repolust, T., Hoffmann, T. L., Jokuthy, A.
2003, in A Massive Star Odyssey, from Main Sequence to Supernova,
IAU Symp~212, ed. K. A. van der Hucht, A. Herrero, \& C. Esteban
(San Francisco: ASP), 61

\reference {} Puls, J., Springmann, U., \& Lennon, M. 2000 A\&AS, 141, 23

\reference {} Puls, J., Urbaneja, M., Springmann, U., Venero, R., Repolust, T.,
\& Jokothy, A. 2005, A\&A, in press;  astro-ph/0411398

\reference {} Repolust, T., Puls, J., \& Herrero, A. 2004, A\&A, 415, 349

\reference {} Sahnow, D. J., et al.  2000, \apj, 538, L7

\reference {} Sanduleak, N. 1969, CTIO Publ.~89

\reference {} Schaerer, D., Meynet, G., Maeder, A., \& Schaller, G. 1993, A\&AS,
98, 523

\reference {} Schild, H. \& Testor, G. 1992, A\&AS, 92, 729

\reference {} Schwering, P. B. V., \& Israel, F. P. 1991, A\&A, 246, 231

\reference {} Sellmaier, F., Puls, J., Kudritzki, R. P.,
Gabler, A., Gabler, R., \& Voels, S. A.  1993 A\&A 273, 533

\reference {} Simon K. P., Kudritzki, R. P., Jonas, G., \& Rathe, J. 1983, A\&A, 125, 34

\reference {} Slesnick, C. L., Hillenbrand, L. A., \& Massey, P. 2002, ApJ, 576, 880

\reference {} Swings, P. 1948, AnAp, 11, 228

\reference {} Taresch, G., Kudritzki, R. P., Hurwitz, M., Bowyer, S.,
Pauldrach, A. W. A., Puls, J., Butler, K., Lennon, D. J., \& Haser, S. M. 1997,
A\&A, 321, 531

\reference {} Testor, G., \& Niemela, V. 1998, A\&AS, 130, 527

\reference {} Testor, G., Schild, H., \& Lortet, M. C. 1993,
A\&A, 280, 426

\reference {} Vacca, W. D., Garmany, C. D., \& Schull, J. M. 1996, ApJ, 460,
914

\reference {} van den Bergh, S. 2000, The Galaxies of the Local Group
(Cambridge: Cambridge Univ.\ Press)

\reference {} Vink, J. S., de Koter, A., \& Lamers, H. J. G. L. M. 2000,
A\&A, 362, 295

\reference {} Vink, J. S., de Koter, A., \& Lamers, H. J. G. L. M. 2001,
A\&A, 369, 574

\reference {} Voels, S. A., Bohannan, B., \& Abbott, D. C. 1989, ApJ, 340, 1073

\reference {} Walborn, N. R. 1971a, ApJ, 167, L31

\reference {} Walborn, N. R. 1971b, ApJS, 23, 257

\reference {} Walborn, N. R. 1973a, ApJ, 186, 611

\reference {} Walborn, N. R. 1973b, ApJ, 180, L35

\reference {} Walborn, N. R. 1982, ApJ, 254, L15

\reference {} Walborn, N. R. 1983, ApJ, 265, 716

\reference {} Walborn, N. R. et al.\ 2002, AJ, 123, 2754

\reference {} Walborn, N. R., \& Blades, J. D. 1997, ApJS, 112, 457

\reference {} Walborn, N. R., \& Fitzpatrick, E. L. 1990, PASP, 102, 379

\reference {} Walborn, N. R., Lennon, D. J., Heap, S. R., Lindler, D. J.,
Smith, L. J., Evans, C. J., \& Parker, J. W. 2000, PASP, 112, 1243

\reference {} Walborn, N. R., Morrell, N. I., Howarth, I. D., Crowther,
P. A., Lennon, D. J., Massey, P., \& Arias, J. I. 2004, ApJ, 608, 1028

\reference {} Westerlund, B. E. 1961, Uppsala Astron.\ Obs.\ Ann., 5, 1

\reference {} Westerlund, B. E. 1997, The Magellanic Clouds (Cambridge: 
Cambridge Univ.\ Press)

\end{references}
\end{document}